%% file: FullPaper.tex
\documentclass[11pt]{article}

\usepackage{stat260_format}
\usepackage{stat260_commands}
\usepackage{arxiv}

\usepackage{algorithm}
\usepackage{algorithmicx}
\usepackage{algpseudocode}
\usepackage[titletoc,toc,page]{appendix}

\usepackage[utf8]{inputenc} % allow utf-8 input
\usepackage[T1]{fontenc}    % use 8-bit T1 fonts
\usepackage{hyperref}       % hyperlinks
\usepackage{url}            % simple URL typesetting
\usepackage{booktabs}       % professional-quality tables
\usepackage{amsfonts}       % blackboard math symbols
\usepackage{nicefrac}       % compact symbols for 1/2, etc.

\usepackage{natbib}
\usepackage{graphicx}

\usepackage{doi}

\newtheorem{condition}{Condition}

\title{Bayesian inference on Cox regression models using catalytic prior distributions}

\author{
 Weihao Li \\
  National University of Singapore
   \And
   Dongming Huang \\
   National University of Singapore
}

\renewcommand{\headeright}{}
\renewcommand{\undertitle}{}
\renewcommand{\shorttitle}{Cox catalytic prior}

\begin{document}
\maketitle

\begin{abstract}
\input{abstract}

\end{abstract}


\input{maintext}

\section{Acknowledgements}
D. Huang was partially supported by NUS Start-up Grant
A-0004824-00-0 and Singapore Ministry of Education AcRF Tier 1 Grant A-8000466-00-00.

\bibliographystyle{agsm}

\bibliography{all-bibliography}

 \appendix

\input{appendix}

\end{document}

%% file: abstract.tex
The Cox proportional hazards model (Cox model) is a popular model for survival data analysis. 
When the sample size is small relative to the dimension of the model, the standard maximum partial likelihood inference is often problematic. 
In this work, we propose the Cox catalytic prior distributions for Bayesian inference on Cox models, which is an extension of a general class of prior distributions originally designed for stabilizing complex parametric models. 
The Cox catalytic prior is formulated as a weighted likelihood of the regression coefficients based on synthetic data and a surrogate baseline hazard constant. 
This surrogate hazard can be either provided by the user or estimated from the data, and the synthetic data are generated from the predictive distribution of a fitted simpler model. 
For point estimation, we derive an approximation of the marginal posterior mode, which can be computed conveniently as a regularized log partial likelihood estimator. 
We prove that our prior distribution is proper and the resulting estimator is consistent under mild conditions. 
In simulation studies, our proposed method outperforms standard maximum partial likelihood inference and is on par with existing shrinkage methods. 
We further illustrate the application of our method to a real dataset.

%% file: maintext.tex
\section{Introduction}
\label{sec:introduction}

The Cox proportional hazards model (Cox model) \citep{cox1972regression} is a widely used model for analyzing the relationship between survival time and risk factors (covariates). It is composed of two parts: a nonparametric baseline hazard function and a parametric component that contains regression coefficients for covariates. 
The parametric component, which captures the association between covariates and hazards, is often the primary focus of research.

Existing estimation methods for the parametric component in the Cox model are not fully satisfactory when the number of observations $n$ is not sufficiently large relative to the number of covariates $p$, a situation often seen in clinical studies and other survival analyses \citep{beca2021impact}.  
The maximum partial likelihood estimator (MPLE) \citep{cox1972regression, cox1975partial}, though widely used, becomes unstable and biased when $p$ increases to be comparable to $n$ \citep{Bryson1981,zhang2022modern}.

Bayesian estimation offers a general alternative to the MPLE in small samples. 
However, its application to the Cox model faces several challenges, especially in choosing the prior. Throughout the paper, we use the compact term ``priors'' (and ``posterior'') in place of the long term ``prior distributions'' (and ``posterior distribution'') for convenience. 
In cases where the MPLE fails, using a simple flat prior distribution may not result in a well-behaved posterior. 
In addition, commonly used parametric families, such as normal or Cauchy distributions, require subjective choices of hyperparameters, which can be difficult to justify or interpret. 
Furthermore, the Jeffreys prior may not exist \citep{wu2018assessing}, and its implementation often requires numerically intensive evaluations of the Hessian matrix.
Informative priors, such as power priors  \citep{ibrahim1998prior}, can incorporate additional information or historical data \citep{neuenschwander2010summarizing}, but their applicability is limited to certain scenarios \citep{brard2017bayesian,van2018including}.

To address these challenges, we extend a recently proposed class of ``catalytic priors'' \citep{huang_catalytic_2020} to the Cox model. A catalytic prior stabilizes a complex parametric model by adding a small amount of synthetic data generated from the predictive distribution of a simpler model. This prior is formulated as a fractional exponent of the likelihood based on the synthetic data, which leads to improved estimation even when the sample size is not sufficiently large. 
However, the original proposal focuses on parametric models and does not directly apply to Cox models for survival analysis due to the nonparametric components. 
The current paper studies the Bayesian inference for the Cox model with an extension of the catalytic prior method. 

The main contributions are summarized as follows:

(1) We introduce Cox catalytic priors, an extension of the original idea of catalytic priors for coefficients in the semiparametric Cox model. This class of priors seamlessly integrates with existing Bayesian samplers for Cox model inference. We prove that this prior is proper under mild assumptions, a necessary property for many Bayesian methods.

(2) We introduce a regularized estimator based on the approximate marginal posterior of the coefficients. We prove its stability against the randomness of synthetic data and prove its frequentist consistency as the dimension $p$ grows with the sample size $n$.

 (3) We use synthetic data to define a joint prior for both the coefficients and the baseline hazard function. 
 This prior results in a stable estimator for the coefficients that can be easily computed using standard software. We also prove its frequentist consistency as $p$ diverges.

Through simulation studies, we demonstrate that our proposed estimation methods outperform the classical maximum partial likelihood estimation in challenging situations where the sample size is not sufficiently large relative to the dimension of covariates. 
Furthermore, their performance is comparable and sometimes superior to existing regularized methods.

The paper is organized as follows. Section~\ref{sec:review_Methodology} reviews the catalytic prior and the Cox model. In Section~\ref{sec:Methodology}, we introduce Cox catalytic priors and propose two associated estimators. Section~\ref{sec:theory} presents the theoretical guarantees for these priors and estimators. We evaluate the numerical performance of our methods through simulations in Section~\ref{sec:simulation} and apply them to a real-world dataset in Section~\ref{sec:real_data}. Finally, Section~\ref{sec:discussion} concludes the article with a discussion.

\section{Review of existing methods}
\label{sec:review_Methodology}

This section provides a succinct introduction to the Cox model, the MPLE and its variants, and the original catalytic prior method. Detailed discussions on related literature are in Appendix A.

 \subsection{Proportional hazards model and partial likelihood}

% For a given subject $i$, denoted by $T_i$ the survival time which may be right-censored by a censoring time $C_i$. The observed data is represented by a triplet $\left(\boldsymbol{X}_i, Y_i, \delta_i\right)$, where $\boldsymbol{X}_i \in \mathbb R^{p}$ is the covariate vector, $\delta_i$ is the binary event indicator, and $Y_i$ is the observed time, satisfying that $\delta_{i}=\mathbf{1}[T_i\leq C_i]$ and $Y_i=\min(T_i, C_i)$. 
% We assume censoring is non-informative, i.e., $T_i \perp C_i \mid \boldsymbol{X}_i$. 

For subject $i$, let $T_i$ denote the survival time, which may be right-censored by a censoring time $C_i$. The observed data is represented by a triplet $\left(\boldsymbol{X}_i, Y_i, \delta_i\right)$, where $\boldsymbol{X}_i \in \mathbb R^{p}$ is the covariate vector, $\delta_{i}=\mathbf{1}[T_i\leq C_i]$ is the binary event indicator, and $Y_i$ is the observed time with  $Y_i=\min(T_i, C_i)$. 
We assume censoring is non-informative, i.e., $T_i \perp C_i \mid \boldsymbol{X}_i$. 
 
In the Cox model, the hazard for subject $i$ at time $t$ is given by
$
h(t \mid \boldsymbol{X_i})=h_{0}(t) \exp \left(\boldsymbol{X}_i^{\prime} \boldsymbol{\beta}\right)
$,
where $h_0(\cdot)$ is an unknown baseline hazard function and treated as a nuisance parameter in this study. 
Given $n$ independent subjects $\left\{\left(\boldsymbol{X}_i, Y_i, \delta_i\right)\right\}_{i=1}^n$, the likelihood is given by:
\begin{equation}
\label{full_likelihood_cox}
	L\left(\boldsymbol{\beta}, h_0 \mid \left\{\left(\boldsymbol{X}_i, Y_i, \delta_i\right)\right\}_{i=1}^n\right) = \prod_{i=1}^n\left\{\exp \left(\boldsymbol{X}_i^{\prime} \boldsymbol{\beta}\right) h_0\left(Y_i\right)\right\}^{\delta_i}\exp \left\{-\exp \left(\boldsymbol{X}_i^{\prime} \boldsymbol{\beta}\right) H_0\left(Y_i\right)\right\},
\end{equation}
where $H_0(t)=\int_0^{t} h_0(u) d u$ is the cumulative baseline hazard function.

To estimate $\boldsymbol{\beta}$, \citet{cox1975partial} introduced the partial likelihood, which estimates the parametric components without accounting for the baseline hazard.
The partial likelihood  is given by:
\begin{equation}
    \label{PL}
PL(\boldsymbol{\beta})=\prod_{i =1}^n \left\{ \frac{\exp \left(\boldsymbol{X}_{i}^{\prime} \boldsymbol{\beta}\right)}{\sum_{j \in \mathbf{R}_{i}} \exp \left(\boldsymbol{X}_{j}^{\prime} \boldsymbol{\beta}\right)}\right\}^{\delta_i},
\end{equation}
where  $\mathbf{R}_{i}:=\{j:Y_j\geq Y_i \}$ is the collection of subjects who are at risk at time $Y_i$. 
 The estimator obtained by maximizing \eqref{PL} is also referred to as MPLE, when the context is clear. 
 
MPLE can be regularized with an $\ell_1$ \citep{tibshirani1997lasso} or $\ell_2$ \citep{hoerl1970ridge} penalty on the negative log partial likelihood, corresponding to Lasso and ridge regression. 
 
With fixed $p$ and large $n$, 
the asymptotic properties of the MPLE are well-studied  \citep{tsiatis1981large,andersen1982cox,wong1986theory,murphy2000profile}. 
However, when $n$ is small, the MPLE becomes unstable, with some estimates diverging to infinity \citep{Bryson1981}. 
In addition, when $n$ is large but $p$ scales with $n$, the estimated coefficients may have inflated magnitudes and be biased away from zero \citep{zhang2022modern}. 
We illustrate the biased behavior of the MPLE in  Appendix F.

\subsection{Catalytic prior: specification and property}
\label{subsec:catalytic_introduce}

The motivation behind the catalytic prior is the understanding that data are real while models are imperfect tools designed to analyze data. 
From this data-centric perspective, specifying a prior distribution is analogous to adding some hypothetical ``prior data'' to the observed data, which will be elaborated in the following review. 
 
Let $\mathcal{D}=\left\{\left(Y_i, \boldsymbol{X}_i\right)\right\}_{i=1}^n$ be $n$ independent data pairs, where $Y_i$ is the response and $\boldsymbol{X}_i$ the covariates. We aim to use $\mathcal{D}$ to fit a complex model for $Y_i$ given $\boldsymbol{X}_i$, assuming the data are sampled as 
$	Y_i \mid \boldsymbol{X}_i, \boldsymbol{\theta} \sim f\left(y \mid \boldsymbol{X}_i, \boldsymbol{\theta}\right)$, 
where $f$ is the corresponding density function, and $\boldsymbol{\theta}$ is the unknown target parameter that is hard to estimate with a small sample.

The basic idea of catalytic priors is to generate synthetic data from a simpler model, and use the synthetic data as prior information for fitting the complex target model $f$. 
Suppose a simpler model $g(y \mid \boldsymbol{X}, \boldsymbol{\psi})$ with unknown parameter $\boldsymbol{\psi}$ is stably fitted from $\mathcal{D}$ and results in a synthetic data-generating distribution $g_*(y \mid \boldsymbol{x}, \left\{\left(Y_i, \boldsymbol{X}_i\right)\right\}_{i=1}^n)$. 
Then $M$ synthetic data points $\mathcal{D}^*=\left\{\left(Y_i^*, \boldsymbol{X}_i^*\right)\right\}_{i=1}^M$ can be generated as follows: 
\begin{equation}\label{generate_X_star}
\boldsymbol{X}_i^* \stackrel{i . i . d .}{\sim} Q(\boldsymbol{x}), \quad Y_i^* \mid \boldsymbol{X}_i^* \sim g_*\left(y \mid \boldsymbol{X}_i^*, \mathcal{D}\right), 
\end{equation}
where  $Q$ denotes the distribution from which the synthetic covariates $\boldsymbol{X}^*$ are generated. 
The supplement of \cite{huang_catalytic_2020} discusses various strategies to generate synthetic data. 
Examples of the simpler model $g$ include the intercept-only model, a model with only the first principal component of covariates, and sub-models nested in the complex target model. 
It should be emphasized that the simpler model $g$ does not have to be a sub-model of the complex target model $f$; for example, $f$ could be a tree model, where $g$ could be a linear model with a handful of predictors.

Since synthetic data are not actually observed, we reduce the impact of synthetic data by down-weighting with a positive tuning parameter $\tau$, which controls their total weight so that each synthetic data point receives weight $\tau / M$. 
The catalytic prior is defined as
\begin{equation}
\label{Mcat}
\pi_{c a t, M}(\boldsymbol{\theta} \mid \tau) \propto\left\{\prod_{i=1}^M f\left(Y_i^* \mid \boldsymbol{X}_i^*, \boldsymbol{\theta}\right)\right\}^{\frac{\tau}{M}}=\exp \left\{\frac{\tau}{M} \sum_{i=1}^M \log \left(f\left(Y_i^* \mid \boldsymbol{X}_i^*, \boldsymbol{\theta}\right)\right)\right\}.
\end{equation}
We can treat the catalytic prior as a down-weighted version of likelihood based on synthetic data.  
Based on \eqref{Mcat}, the posterior distribution is given by
\begin{equation}
\label{posterior_general}
	\begin{aligned}
		\pi(\boldsymbol{\theta} \mid \mathcal{D})&\propto \left(\prod_i^n f(Y_i\mid \boldsymbol{X}_i,\boldsymbol{\theta}) \right)\pi_{c a t, M}(\boldsymbol{\theta} \mid \tau)\\
		&\propto \exp \left\{\sum_{i=1}^n \log \left(f\left(Y_i \mid \boldsymbol{X}_i, \boldsymbol{\theta}\right)\right)+\frac{\tau}{M} \sum_{i=1}^M \log \left(f\left(Y_i^* \mid \boldsymbol{X}_i^*, \boldsymbol{\theta}\right)\right)\right\}. 
	\end{aligned}
\end{equation} 
As a data-dependent construction, this approach can be seen as an empirical Bayes method, which can be traced back to the classical Box-Cox transformation \citep{Box:1964ev}.

If we take $M \to \infty$, the prior in \eqref{Mcat} converges to the population catalytic prior defined as
$$
\pi_{c a t, \infty}(\boldsymbol{\beta} \mid \tau) \propto \exp \left\{\tau \mathbb{E}_{Q, g_*}\left(\log f\left(Y^* \mid \boldsymbol{X}^*, \boldsymbol{\beta}\right)\right)\right\}
$$
where the expectation $\mathbb{E}_{Q, g_*}$ averages over both $\boldsymbol{X}^*$ and $Y^*$. 
There is an interpretation of this population catalytic prior from information theory: the resultant posterior mode can be identified as the solution to the following optimization problem:
\begin{equation}\label{KLcat_optima}
    \min _{\boldsymbol{\beta}}\left\{-\sum_{i=1}^n \log \left(f\left(Y_i \mid \boldsymbol{X}_i, \boldsymbol{\theta}\right)\right)+\tau \mathbb{E}_{Q}\left(\mathrm{KL}\left(g_*\left(\cdot \mid \boldsymbol{X}^*, \mathcal{D}\right), f\left(\cdot \mid \boldsymbol{X}^*, \boldsymbol{\beta}\right)\right)\right)\right\},
\end{equation}
where the expectation $\mathbb{E}_{Q}$ averages over $\boldsymbol{X}^*$ and $\text{KL}(\cdot,\cdot)$ denotes the usual Kullback-Leibler divergence (KL divergence). 
In \eqref{KLcat_optima}, the expected KL divergence effectively regularizes the complex model by pushing the estimate toward the simpler model $g_*$.

The class of catalytic priors in \eqref{Mcat} offers several significant benefits. First, since the posterior has the same form as the likelihood, Bayesian inference based on the posterior is no more difficult than any likelihood-based inference.  
For example,  the posterior mode can be conveniently computed as the maximum weighted likelihood estimate using standard statistical software.
Second, unlike some other priors such as Gaussian priors or Cauchy priors, this class does not focus on the parameter space and remains invariant under any invertible affine transformation of model parameters. 
Third, catalytic priors provide an intuitive interpretation of the information included by the prior through the use of synthetic data. 
Lastly, this class of priors is generic and applicable to any parametric model, making it a versatile tool in various applications of Bayesian methods. 

\section{Catalytic prior distributions for Cox models}
\label{sec:Methodology}

While the catalytic prior introduced in Section~\ref{subsec:catalytic_introduce} is effective for fully parametric models, it does not directly apply to semiparametric models like the Cox model. 
In the Cox model, our primary interest lies in inference on the regression parameter $\boldsymbol{\beta}$, but the presence of the unknown nuisance component $h_0(\cdot)$ makes it impossible to write down the catalytic prior for $\boldsymbol{\beta}$ directly from the likelihood function in \eqref{full_likelihood_cox}.  
In this section, we propose a new catalytic prior designed specifically for the regression coefficient $\boldsymbol{\beta}$ in the Cox model and derive the corresponding point estimator. 
We also explore an alternative formulation of catalytic priors jointly on both $\boldsymbol{\beta}$ and the baseline hazard function.

\subsection{Prior formulation}\label{sec:formulate}

Assume that the synthetic data have already been generated, with the details of the generation discussed in the next subsection. 
Denote the synthetic data by $\left\{\left(\boldsymbol{X}_i^*, Y_i^*\right)\right\}_{i=1}^M$, where $\boldsymbol{X}_i^*$ is a synthetic covariate vector and $Y_i^*$ is a synthetic survival time. 
We assume these synthetic data are uncensored for two reasons: (1) we have full control over the synthetic data generation and (2) as discussed in Section~\ref{subsec:catalytic_introduce}, these synthetic data can be generated using any simpler model, even if it is misspecified. 
Therefore, the uncensored nature of synthetic data does not conflict with the non-informative censoring mechanism for the observed data.

Given the synthetic data, we consider the strategy that defines the catalytic prior in \eqref{Mcat}: we view the likelihood based on the synthetic data as the prior for the Cox model and control the influence of synthetic data by down-weighting with a positive tuning parameter $\tau$. 
However, since our main focus is on the parameter $\boldsymbol{\beta}$, we should avoid involving the nuisance component $h_0(\cdot)$ of the Cox model in \eqref{full_likelihood_cox}. 
For this purpose, we substitute the nuisance component $h_0(\cdot)$ by a simple user-specific baseline hazard constant $h^+_0$ so that the likelihood based on the synthetic data will depend only on the coefficient $\boldsymbol{\beta}$. 
This substitution leads us to define the \textit{Cox catalytic prior for $\boldsymbol{\beta}$} as follows: 
\begin{equation}
    \label{cat_beta_cox}
    \begin{aligned}
    	\pi_{cox,cat}(\boldsymbol{\beta}\mid \tau) \propto &   L\left(\boldsymbol{\beta}, h_0^+ \mid \left\{\left(\boldsymbol{X}_i^*, Y_i^*\right)\right\}_{i=1}^M\right)^{\tau/M}\\
    	=& \left[\prod_{i=1}^M\left\{\exp \left(\boldsymbol{X}_i^{*\prime} \boldsymbol{\beta}\right) h_0^+\right\}\exp \left\{-\exp \left(\boldsymbol{X}_i^{*\prime} \boldsymbol{\beta}\right) Y_i^* h^+_0\right\}\right]^{\tau/M} ,
    \end{aligned}
\end{equation}
where $ h^+_0$ and $\tau$ are positive constants. 
Similar to the definition in \eqref{Mcat}, since the synthetic data are not actually observed, we use the parameter $\tau$ to control the total weight of the $M$ synthetic units, so that each of them has a relatively small weight $\tau / M$. 
The scalar constant $h_0^+$ acts merely as a surrogate for the nuisance component to facilitate the construction of our catalytic prior for $\boldsymbol{\beta}$; it does not need to be correctly specified. 
In practice, $h_0^+$ can be either specified by the user or computed in a data-driven way.
Similar to the catalytic prior in \ref{Mcat}, Cox catalytic prior is also an empirical Bayes method.

The separable form of the Cox catalytic prior enables us to evaluate it efficiently. 
Furthermore, one can see that the Cox catalytic prior is log-concave, so its tails are typically well-behaved. 
The log-concavity of the Cox catalytic prior may also be useful in the posterior sampling. 
In particular, we provide an example in  Appendix 
 C,  
where the update step of $\boldsymbol{\beta}$ in a Gibbs iteration involves a log-concave distribution, from which efficient sampling is possible \citep{lovasz2006fast}. For a comprehensive review on log-concave densities, see \citep{saumard2014log}. 
In addition, we prove in Section~\ref{sec:theory} that the Cox catalytic prior is proper for any $\tau>0$ under some mild conditions.

The Cox catalytic prior in \eqref{cat_beta_cox} offers a compelling alternative to the default prior for $\boldsymbol{\beta}$ in the standard Bayesian Cox model. 
The Cox catalytic prior is not only simple to implement but also provides an intuitive interpretation through the use of synthetic data. 

\subsection{Specification of components}\label{sec:specification}

To define the Cox catalytic prior, one needs to generate synthetic survival data and specify the synthetic sample size $M$ and the total weight $\tau$.

\paragraph{Generating synthetic survival data}
Similarly to the introduction in Section~\ref{subsec:catalytic_introduce}, the procedure for generating the synthetic survival data $\left\{\left(\boldsymbol{X}_i^*, Y_i^*\right)\right\}_{i=1}^M$ can be summarized as:
 
(1) Synthetic covariates: Use any appropriate methods in \citep{huang_catalytic_2020} to generate $\left\{\boldsymbol{X}_i^*\right\}_{i=1}^M$. For example, each coordinate $\boldsymbol{X}_{i,j}^*$ is independently resampled from the corresponding $j$-th marginal empirical distribution of the observed covariates. 

(2) Synthetic survival time: Consider a simpler model $g(y \mid \boldsymbol{X},\boldsymbol{\delta}, \boldsymbol{\psi})$ with a low-dimensional unknown parameter $\boldsymbol{\psi}$, which can be stably fitted from the observed data $\left\{\left(\boldsymbol{X}_i, Y_i, \delta_i\right)\right\}_{i=1}^n$. Using either a point estimate or a Bayesian posterior, one can obtain a predictive distribution from this simpler model, denoted by $g_*(y \mid \boldsymbol{x}, \left\{\left(\boldsymbol{X}_i, Y_i, \delta_i\right)\right\}_{i=1}^n)$. Synthetic survival times can then be generated as
 $Y_i^* \mid \boldsymbol{X}_i^* \sim g_*\left(y \mid \boldsymbol{X}_i^*, \delta_{i}^*=1, \left\{\left(\boldsymbol{X}_i, Y_i, \delta_i\right)\right\}_{i=1}^n\right)$, where $\delta_{i}^*=1$ denotes the uncensored status of the synthetic unit.

There are many choices for the simpler models $g(y \mid \boldsymbol{X},\boldsymbol{\delta}, \boldsymbol{\psi})$. 
For example, one could use a parametric survival model for the survival time $T$ using exponential distributions or Weibull distributions. 
Alternatively, one could use a nonparametric model for the survival function without using any covariates and fit it using the Kaplan-Meier estimator. 
Similar to the discussion in Section~\ref{subsec:catalytic_introduce}, the simpler model does not have to be a sub-model of the Cox model. 
In view of this, one might even choose a linear model as the simpler model.

As a simple example, we generate synthetic survival times from an estimated exponential distribution, i.e., $Y_i^*\sim \text{Exp}(\hat{\psi})$, where $\hat{\psi}$ is estimated stably based on the observed data by maximizing $L\left( \psi \mid \left\{\left(\boldsymbol{X}_i, Y_i, \delta_i\right)\right\}_{i=1}^n\right) = \prod_{i=1}^n \psi^{\delta_i}\exp \left(- \psi Y_i \right)$ over $ \psi>0.$
This likelihood corresponds to a simple nested model of \eqref{full_likelihood_cox}, where the model parameters are restricted to be $\boldsymbol{\beta}=0$ and $h_0(t)=\psi$ for all $t>0$. 
The estimate $\hat{\psi}$ can also be used to specify the surrogate $h_0^+=\hat{\psi}$ in the expression \eqref{cat_beta_cox}.

\paragraph{Synthetic sample size}

The synthetic sample size $M$ should generally be as large as possible to reduce the randomness of synthetic data. 
However, the marginal gains vanish when $M$ exceeds a certain point, as the Cox catalytic prior density in \eqref{cat_beta_cox} will converge to the following as $M\to\infty$:
\begin{equation}
\label{cat_cox_infinity}
\pi_{cox,cat}^{\infty}(\boldsymbol{\beta}\mid \tau) \propto  \exp\left\{
     \tau \mathbb{E}\left(\boldsymbol{X}^{*\prime} \boldsymbol{\beta}
-\exp \left(\boldsymbol{X}^{*\prime} \boldsymbol{\beta}\right) Y^* h^+_0\right)
     \right\} ,
\end{equation}
where the expectation is taken over the synthetic data $(\boldsymbol{X}^*,Y^*)$.
We refer to the prior \eqref{cat_cox_infinity} as the \textit{population Cox catalytic prior}. 
This convergence suggests that the randomness of synthetic data would eventually have a negligible impact on the prior and the posterior. 
With no universal value of $M$ that suits all problems, we advise users to experiment with different values of $M$ based on the problem's scale and available computational resources. 
To ensure the properness of the Cox catalytic prior with $p$ covariates, $M$ should be at least $4p$, as supported by the theory in Section~\ref{sec:theory}, particularly Remark~\ref{rem:norm-recov}.

\paragraph{Total weight}

The total weight $\tau$ is an important hyperparameter that controls the influence of the synthetic data in the catalytic prior and, consequently, the impact on posterior inference. 
When $\tau\to 0$, the Cox catalytic prior degenerates to the improper flat prior, which provides no regularization for posterior inference.  
As $\tau$ increases, the Cox catalytic prior density becomes more concentrated around its mode, which is given by 
\begin{equation}
    \label{cat_beta_cox_mode}
    \begin{aligned}
    \boldsymbol{\beta}_{cat,0}	= &\arg\max_{\boldsymbol{\beta}}\pi_{cox,cat}(\boldsymbol{\beta}\mid \tau)
        = & \arg\max_{\boldsymbol{\beta}} \frac{1}{M}\sum_{i=1}^M\left\{ \boldsymbol{X}_i^{*\prime} \boldsymbol{\beta} -  \exp \left(\boldsymbol{X}_i^{*\prime} \boldsymbol{\beta}\right) Y_i^* h^+_0\right\}. 
    \end{aligned}
\end{equation}
Consequently, the resultant posterior inference will have more bias but less variance. 

The choice of $\tau$ involves balancing this trade-off between bias and variance in the resultant inference. 
Although no single choice for $\tau$ works the best for all scenarios, 
a simple practical choice is to set $\tau$ equal to $p$, the dimension of $\boldsymbol{\beta}$, following the recommendation by \cite{huang2022catalytic} for logistic regression. 
Another option is to adopt frequentist methods to fine-tune $\tau$ by optimizing a given criterion function, such as the cross-validated partial log-likelihood outlined in  Appendix I.3.

If a fully Bayesian perspective on $\tau$ is adopted, we can specify a joint prior on $(\tau, \boldsymbol{\beta})$ that allows  $\tau$ to be adaptively determined by the data. 
Given any two positive scalar hyperparameters $\alpha$ and $\gamma$, we define the \textit{Cox adaptive catalytic prior} for $(\tau, \boldsymbol{\beta})$ as
\begin{equation}\label{joint_prior}
\pi_{\alpha, \gamma}(\tau, \boldsymbol{\beta}) \propto \Gamma_{\alpha, \gamma}(\tau)\cdot  L\left(\boldsymbol{\beta}, h_0^+ \mid \left\{\left(\boldsymbol{X}_i^*, Y_i^*\right)\right\}_{i=1}^M\right)^{\tau/M} ,
\end{equation}
where $\Gamma_{\alpha, \gamma}(\tau)$ is a function defined as 
\begin{equation}\label{prior_tau}
\Gamma_{\alpha, \gamma}(\tau)=\tau^{p+\alpha-1} e^{-\tau\left(\kappa+\gamma^{-1}\right)} ,
\end{equation} and $ \kappa:=\sup _{\boldsymbol{\beta} \in \mathbb{R}^p} \frac{1}{M}  \log L\left(\boldsymbol{\beta}, h_0^+ \mid \left\{\left(\boldsymbol{X}_i^*, Y_i^* \right)\right\}_{i=1}^M\right) .$
This form of $\Gamma_{\alpha, \gamma}(\tau)$ is chosen so that the conditional posterior of $\tau$ given $\boldsymbol{\beta}$ is a Gamma distribution, which is easy to sample from. This simplifies the update step of $\tau$ in the iteration of the Gibbs algorithm. Furthermore, Theorem~\ref{Thm:proper} shows that the Cox adaptive catalytic prior is proper. 

The hyperparameters $\alpha$ and $\gamma$ in the Cox adaptive catalytic prior for $(\tau, \boldsymbol{\beta})$ control the tail behavior of $\tau$; larger values of $\alpha$ or $\gamma$ encourage stochastically larger values of $\tau$, thereby shrinking the working model closer to the simpler model. 
The values of $\alpha$ and $\gamma$ are user-specified. 
While there is no universal optimal choice across all scenarios, posterior inference is generally robust to these hyperparameters.
Based on our numerical experiments, the choice $(\alpha, \gamma) = (2, 1)$ performs well in practice.

\subsection{Bayesian inference and approximate MAP estimator}  \label{sec:Bayes cox cata}

The full Bayesian analysis of a Cox model using a catalytic prior in \eqref{cat_beta_cox} can be conducted using standard techniques such as posterior sampling. 
This usually involves placing a nonparametric prior process to model the baseline hazard $h_0(t)$ and deriving the joint posterior of $(\boldsymbol{\beta}, h_0)$. 
Monte Carlo methods for Bayesian inference on Cox models have been well-established in the literature; see for example, \citet{chen2012monte}. 
In a nutshell, denoting the prior measure for $h_0(\cdot)$ by $\Pi(h_0)$, the posterior of $(\boldsymbol{\beta}, h_0)$ w.r.t. the product of the Lebesgue measure on $\boldsymbol{\beta}$ and $\Pi(\cdot)$ is proportional to  
\begin{equation}\label{eq:cox-cat-posterior}
L\left(\boldsymbol{\beta}, h_0 \mid \left\{\left(\boldsymbol{X}_i, Y_i, \delta_i\right)\right\}_{i=1}^n\right) \pi_{cox,cat}(\boldsymbol{\beta}\mid \tau).
\end{equation}
We provide a detailed description of the full Bayesian inference using the Cox catalytic prior and a Gamma process for modeling the cumulative hazard function in  Appendix B.
Source code for implementing the posterior sampling using \texttt{R} and \texttt{Stan}, which are open-source software and interfaces for  Bayesian analysis, can be found in  Appendix L.

When point estimation is of the primary interest, maximum a posteriori (MAP) estimation is often favored. 
The MAP estimate is equal to the mode of the posterior distribution, and it is often much simpler to compute than other posterior estimates, especially when sampling from the posterior distribution is difficult. 
This is particularly the case for Cox models because the nonparametric baseline hazard function may make the Bayesian computation expensive and time-consuming.

To mitigate the computational burden of point estimation using the Cox catalytic prior, we propose the following approximation to the mode of the marginal posterior for $\boldsymbol{\beta}$: 
\begin{equation}\label{estimator_penalty}
\hat{\boldsymbol{\beta}}_{CR,\tau}=\arg \max_{\boldsymbol{\beta}} \left\{\log PL(\boldsymbol{\beta})+ \log \pi_{cox,cat}(\boldsymbol{\beta}\mid \tau)  \right\},    
 \end{equation}
 which involves maximizing a regularized log partial likelihood. 
 Here, the regularization term is the log density of the Cox catalytic prior for $\boldsymbol{\beta}$. 
 For simplicity, we refer to the estimator $\hat{\boldsymbol{\beta}}_{CR,\tau}$ as the \textit{catalytic-regularized estimator} (CRE) throughout the paper.
 Since both $\log(PL(\boldsymbol{\beta}))$ and $\log(\pi_{cox,cat}(\boldsymbol{\beta}\mid \tau))$ are concave functions of $\boldsymbol{\beta}$, we can use any efficient convex optimization method, such as the Newton-Raphson method, to find $\hat{\boldsymbol{\beta}}_{CR,\tau}$.

The justification for using the approximation in \eqref{estimator_penalty} is based on the approximation of the marginal posterior of $\boldsymbol{\beta}$ when we place an independent diffused prior on the baseline hazard function. 
A detailed explanation is provided in  Appendix D.

\subsection{Estimation utilizing synthetic data}\label{sec: estimation mix}

In addition to the Cox catalytic prior on $\boldsymbol{\beta}$ defined in Section~\ref{sec:formulate}, we can construct a joint prior on both the coefficients $\boldsymbol{\beta}$ and the baseline hazard function $h_0(\cdot)$. 
This alternative construction will lead to an estimator for $\boldsymbol{\beta}$ that can be intuitively interpreted as an estimation method based on a combination of the actual observed data and the weighted synthetic data.

In the context of parametric modeling, the catalytic prior in Eq.~\eqref{Mcat} is defined as a down-weighted version of the likelihood based on synthetic data. 
As a result, the corresponding posterior can be regarded as a ``likelihood'' based on a mixture of observed data (each with weight 1) and synthetic data (each with weight $\tau/M$). 
This perspective on the posterior can be extended to the Cox model. 
As an analogy, we consider a ``posterior distribution'' $\tilde{L}$ on both $\boldsymbol{\beta}$ and $h_0(\cdot)$ whose density (w.r.t. some base measure) can be expressed as the following weighted likelihood function:  
 \begin{equation}
 \label{mix_likelihood}
 \tilde{L}(\boldsymbol{\beta},h_0(\cdot)\mid \boldsymbol{D},\boldsymbol{D}^*)=L(\boldsymbol{\beta},h_0(\cdot)\mid \boldsymbol{D})\cdot L(\boldsymbol{\beta},h_0(\cdot)\mid \boldsymbol{D}^*)^{\frac{\tau}{M}} ,
\end{equation}
where $\boldsymbol{D}$ denotes the actual data $\{\left(\boldsymbol{X}_i, Y_i, \delta_i\right)\}_{i=1}^n$ , $\boldsymbol{D}^*$ the synthetic data $\left\{\left(\boldsymbol{X}_i^*, Y_i^*\right)\right\}_{i=1}^M$,  and $L(\cdot)$ the likelihood defined in \eqref{full_likelihood_cox}. 
The posterior in \eqref{mix_likelihood} is associated with a prior on  $(\boldsymbol{\beta},h_0(\cdot))$ whose density (w.r.t. the same base measure as before) is $L(\boldsymbol{\beta},h_0(\cdot)\mid \boldsymbol{D}^*)^{\frac{\tau}{M}}$. 

We highlight the difference between the joint prior $L(\boldsymbol{\beta},h_0(\cdot)\mid \boldsymbol{D}^*)^{\frac{\tau}{M}}$ and the Cox catalytic prior for $\boldsymbol{\beta}$ defined in Section~\ref{sec:formulate}: 
the Cox catalytic prior for $\boldsymbol{\beta}$ does not involve $h_0(\cdot)$ but instead uses the surrogate baseline $h_0^+$, whereas the joint prior defined here depends on $h_0(\cdot)$ but does not need the surrogate $h_0^+$. 
Consequently, the posterior under the Cox catalytic prior in \eqref{eq:cox-cat-posterior} use $h_0(\cdot)$ and $h_0^+$ separately for observed data and synthetic data, whereas in \eqref{mix_likelihood} the likelihoods for both observed data and synthetic data share the same $h_0(\cdot)$.

The joint prior $L(\boldsymbol{\beta},h_0(\cdot)\mid \boldsymbol{D}^*)^{\frac{\tau}{M}}$ is a natural extension of the original catalytic prior, but we will not delve into the full Bayesian inference with \eqref{mix_likelihood} because  $\boldsymbol{\beta}$ and $h_0(\cdot)$ are not prior-independent, which diverges from typical Bayesian inference on Cox models. 
Nonetheless, if the primary interest is in estimating $\boldsymbol{\beta}$, \eqref{mix_likelihood} leads to an efficient and stable estimation method for $\boldsymbol{\beta}$ that combines observed data with weighted synthetic data.

To estimate $\boldsymbol{\beta}$ from \eqref{mix_likelihood}, we leverage the profile likelihood method to remove the nuisance parameter $h_0(\cdot)$. 
Following the approach employed by \citep{murphy2000profile}, we reduce the weighted likelihood in \eqref{full_likelihood_cox} to a new partial likelihood $P\tilde{L}(\boldsymbol{\beta})$: 
\begin{align}
    P\tilde{L}(\boldsymbol{\beta})  =  & \prod_{i\in \tilde{\boldsymbol{I}} } \left( \frac{\theta_{i} }{\sum_{j\in \tilde{\mathcal{R}}_i } w_j \theta_j}  \right)^{\tilde{\delta}_i w_i},
    \label{wm partial likelihood}
\end{align}
where the notation is described as follows: 
$\tilde{\boldsymbol{I}} = \boldsymbol{I}_{D} \cup \boldsymbol{I}_{D^*}$ combines observed data indices $\boldsymbol{I}_D = \{1, 2, \ldots, n\}$ and synthetic data indices $\boldsymbol{I}_{D^*} = \{n+1, n+2, \ldots, n+M\}$. 
The combination of observed and synthetic data is denoted as $\{\tilde{Y}_i\}_{i=1}^{n+M}$, where $\tilde{Y}_i = Y_i$ for $i \in \boldsymbol{I}_D$ and $\tilde{Y}_i = Y^*_{i-n}$ for $i \in \boldsymbol{I}_{D^*}$. The risk set at time $\tilde{Y}_i$ is $\tilde{\mathcal{R}}_i = \{j: \tilde{Y}_j \geq \tilde{Y}_i, j \in \tilde{\boldsymbol{I}}\}$. Denote $\theta_i = \exp(\boldsymbol{X}_i^\top\boldsymbol{\beta})$ for $i \in \boldsymbol{I}_D$ and $\theta_i^* = \exp({\boldsymbol{X}_{i-n}^*}^\top\boldsymbol{\beta})$ for $i \in \boldsymbol{I}_{D^*}$. The weights are $w_i = 1$ for $i \in \boldsymbol{I}_D$ and $w_i = \tau / M$ for $i \in \boldsymbol{I}_{D^*}$, while the censoring status is $\tilde{\delta}_i = \delta_i$ for $i \in \boldsymbol{I}_D$ and $\tilde{\delta}_i = 1$ for $i \in \boldsymbol{I}_{D^*}$.

This new partial likelihood is based on the combined dataset that mixes the actual observed data (each with weight 1) and the synthetic data (each with weight $\tau/M$). 
Subsequently, we define the following estimator by maximizing the new partial likelihood:
\begin{equation}\label{estimator_wm}
\hat{\boldsymbol{\beta}}_{WM,\tau}=\arg\max_{\boldsymbol{\beta}} P\tilde{L}(\boldsymbol{\beta}). 
\end{equation}
We term this estimator the \textit{weighted mixture estimator} (WME), since it is derived from the weighting and mixing of observed and synthetic data. 
An advantage of the WME is its computational simplicity, allowing estimates to be easily computed with standard statistical software. See  Appendix E for more details.

The WME has a unique form of regularization that sets it apart from other regularized estimators. 
Unlike the CRE, whose objective function is separable into a standard term and an additive regularization term, the objective function for the WME does not permit such separation. 
Nonetheless, the WME should still be regarded as a regularized estimator, where the regularization arises from adding the synthetic with weight $\frac{\tau}{M}$. 
When the total weight parameter $\tau$ approaches 0, the influence of the synthetic data vanishes and the WME becomes equivalent to the MPLE based solely on the observed data.
Conversely, as $\tau$ increases toward infinity, the WME converges to the MPLE based solely on the synthetic data. 
Another difference between the WME and the CRE is that the definition of WME only requires the generation of synthetic data and does not depend on the surrogate baseline hazard constant $h_0^{+}$. This makes the WME simpler to formulate than the CRE.

\section{Theoretical properties}
\label{sec:theory}

In this section, we first prove the properness of the Cox catalytic prior, then we study frequentist properties of the corresponding estimators, CRE and WME. 
 
\subsection{Properness}
Many Bayesian inferences, such as model comparison using Bayes factors, require a proper prior. 
In addition, a proper prior guarantees a proper posterior regardless of the sample size so that the posterior inference is always valid. 
To establish the properness of the Cox catalytic prior, we start with a mild assumption regarding the synthetic covariates. 

\begin{assumption}[Norm-recoverability.]\label{cond:norm-recov}
The synthetic covariates are said to be \textit{norm-recoverable} if there exists $c_1>0$ such that $\frac{1}{M} \sum_{i=1}^M |\boldsymbol{X}_i^{*\top}\boldsymbol{\beta}| \geq c_1\|\boldsymbol{\beta}\|_2$ for all $\boldsymbol{\beta} \in \mathbb{R}^p$.
\end{assumption}

The norm-recoverability assumption is a mild condition on the synthetic covariates and is generally satisfied when these covariates are generated in specific ways (see Remark~\ref{rem:norm-recov}). 
The next result establishes the properness of the Cox catalytic prior under this assumption.

\begin{theorem}[Properness of Cox catalytic priors]
\label{Thm:proper}
	If Assumption~\ref{cond:norm-recov} holds and $\tau>0$, then the Cox catalytic prior defined in \eqref{cat_beta_cox} is proper, that is, $	\int \pi_{cox,cat}(\boldsymbol{\beta}\mid \tau) d\boldsymbol{\beta} <\infty. $

\end{theorem}

The next result ensures that the Cox adaptive catalytic prior on $(\tau, \boldsymbol{\beta})$ defined in \eqref{joint_prior} is also proper under the norm-recoverability assumption. 

\begin{theorem}[Properness of Cox adaptive catalytic priors]
\label{Thm:joint proper}
If Assumption~\ref{cond:norm-recov} holds, then 
(1) the Cox adaptive catalytic prior $\pi_{\alpha, \gamma}(\tau, \boldsymbol{\beta})$ defined in \eqref{joint_prior} is proper, and 
(2) for any $\alpha^{\prime} \in(0, \alpha)$, the $\alpha^{\prime}$-th moment of $\boldsymbol{\beta}$ is finite.
\end{theorem}

\begin{remark}\label{rem:norm-recov}
   Assumption~\ref{cond:norm-recov} holds with high probability if the synthetic sample size $M$ is sufficiently large relative to the dimension $p$ and the synthetic covariates are i.i.d. copies of $\boldsymbol{X}^{*}=\left(X_1^{*}, X_2^{*}, \ldots, X_p^{*}\right)$ such that $X_1^{*}, \ldots, X_p^{*}$ are independent and are bounded. 
    This statement is a consequence of Theorem 5.7 in \cite{huang_catalytic_2020}. 
    Boundedness is automatically satisfied if the synthetic covariates are resampled from the observed covariates. 
    In particular, as recommended in Section~\ref{sec:specification}, setting $M\geq 4p$ ensures that Assumption~\ref{cond:norm-recov} is typically met.
     
\end{remark}

\subsection{Stability}

We investigate the impact of the synthetic sample size $M$ on the stability of the CRE. 
As $M$ increases, the Cox catalytic prior converges to the population Cox catalytic prior as defined in \eqref{cat_cox_infinity}. 
Therefore, for large $M$, the estimate $\widehat{\boldsymbol{\beta}}_{C R, \tau}$ obtained from $M$ synthetic samples is expected to approximate  $\widehat{\boldsymbol{\beta}}_{\infty}:=\arg \max_{\boldsymbol{\beta}} \left\{\log PL(\boldsymbol{\beta})+ \log \pi^{\infty}_{cox,cat}(\boldsymbol{\beta}\mid \tau)  \right\}$.  
The following result provides an upper bound on the discrepancy between $\widehat{\boldsymbol{\beta}}_{C R, \tau}$  and $\widehat{\boldsymbol{\beta}}_{\infty}$ in terms of $M$. 

\begin{theorem}\label{thm:stability_CR}
Under a mild condition on the synthetic data (Condition 4 in Appendix H), there exists a constant $C$ such that if $M>C p^2$, then $\|\widehat{\boldsymbol{\beta}}_{C R, \tau}-\widehat{\boldsymbol{\beta}}_{\infty}\|^2=O_p\left( \frac{p}{M}\right)$. 
\end{theorem}
Theorem \ref{thm:stability_CR} addresses the general case where the dimension $p$ is not assumed to be bounded. The theorem shows that $\|\widehat{\boldsymbol{\beta}}_{C R, \tau}-\widehat{\boldsymbol{\beta}}_{\infty}\|^2$ decays linearly in the synthetic sample size $M$. 
Moreover, the theorem indicates that the influence of the random synthetic data on the CRE can be effectively mitigated by choosing $M$ to be much larger than $p$. 
To obtain the upper bound in Theorem \ref{thm:stability_CR} when $p$ is allowed to grow, we developed a series of results on the convergence of the empirical process involving exponential functions $\exp\left(\boldsymbol{\beta}^\top\boldsymbol{X}_i^*\right)$. These results are novel in the literature and may be of independent interest.

\subsection{Consistency}
The frequentist properties of Bayesian procedures has been an important topic as reviewed in Chapter 4 of \citet{carlin2000}. 
We analyze these properties of the proposed regularized estimators, namely CRE in \eqref{estimator_penalty} and WME in \eqref{estimator_wm} in the regime where the dimension $p$ grow with the sample size $n$ of the observed data. 
The following theorem shows that although CRE and WME rely on synthetic data, they are consistent under standard regularity conditions for the MPLE.

\begin{theorem}[Consistency of CRE and WME] \label{thm:consistency}
Suppose the observed survival data are sampled from a Cox regression model with regression coefficient $\boldsymbol{\beta}_0$.  
Assume the regularity conditions stated in Appendix H hold. 
If  $\tau\leq c_1p$ for some $c_1>0$ and $p^2/n\to 0$, then  $\|\hat{\boldsymbol{\beta}}_{CR,\tau}-\boldsymbol{\beta}\|^2=O_p(p/n)$ and $ \|\hat{\boldsymbol{\beta}}_{WM,\tau}-\boldsymbol{\beta}\|^2=O_p(p/n)$.
\end{theorem}

Theorem~\ref{thm:consistency} rigorously justifies the following intuition:  
as $n$ increases, the regularization term in Eq~\eqref{estimator_penalty} becomes negligible relative to the log partial likelihood, making the CRE behave similarly to the MPLE, and the objective function of WME is asymptotically equivalent to that of the MPLE. 
Although this intuition is straightforward, proving it rigorously is challenging when $p$ grows with $n$.  
To address the regularization terms introduced by synthetic data, we develop a nontrivial extension of the classical analysis of objective functions using a counting processes representation. 
We provide a detailed proof of Theorem~\ref{thm:consistency} in  Appendix H along with two numerical illustrations on the convergence properties.

\section{Simulation study}
\label{sec:simulation}

In this section, we conduct simulation studies to compare the finite-sample performance of our proposed methods against various alternative methods. 
Specifically, we examine our two proposed estimators, CRE and WME, and the posterior mean estimators resulting from using a Cox catalytic prior and a Cox adaptive catalytic prior, respectively. 
These estimators are compared with several standard methods, including the MPLE, the penalized log partial likelihood estimators with $\ell_2$-penalty (ridge estimator) and $\ell_1$-penalty (Lasso estimator), and the posterior mean estimator resulting from Gaussian priors on coefficients.

\subsection{Simulation setting} 
We simulate the observed $p$-dimensional covariate vectors $\boldsymbol{X}_i$ ($i=1,\ldots,n$) with independent entries defined as follows: for $j=1$, $\boldsymbol{X}_{i, j}\sim \text{Bernoulli}(0.1)$; for $j=2, \boldsymbol{X}_{i, j}\sim \chi_1^2$; for $j=3, \boldsymbol{X}_{i, j}\sim \chi_4^2$, and for remaining columns, $\boldsymbol{X}_{i, j} \sim  N(0,1)$, 
where Bernoulli$(q)$ is the Bernoulli distribution with probability $q$,  $\chi_k^2$ is the Chi-square distribution with $k$ degrees of freedom, and $N(0,1)$ is the standard normal distribution. 
The first three entries of a covariate vector are intended to mimic the prevalent characteristics of highly unbalanced categorical variables and skewed continuous variables in real-world datasets. 

The true regression coefficient vector is set to be $\boldsymbol{\beta}_0 = (4, -4, 3, -3, 1, -1, 1, -1, \mathbf{1}_{p-8}) / \sqrt{p}$. Each survival time $T_i$ is independently sampled from an exponential distribution with a rate parameter $0.5 \exp(\boldsymbol{X}_i^\top \boldsymbol{\beta}_0)$. Censoring times are drawn independently from a uniform distribution on $[0, \xi]$, where $\xi$ is selected to achieve a target censoring rate $r$, as described in \citet{lee2011bayesian}. In our simulations, the sample size is fixed at $n=100$, with the dimension $p$ varying across $\{20, 40, 60\}$ and the censoring rate $r$ across $\{10\%, 20\%, 40\%\}$. Implementation details for estimators are deferred to  Appendix I.

\subsection{Performance measurement} 
We assess the estimation performance of our methods and other competing methods using squared error loss $\|\hat{\boldsymbol{\beta}}-\boldsymbol{\beta}_0 \|^2$ and a measure of prediction accuracy defined as follows. 
After simulating a separate test dataset of size $n_{test}=100$, we compute the predictive deviance
$ \operatorname{Dev}(\boldsymbol{\beta}_0,\hat{\boldsymbol{\beta}})=\left\{\ell^{(\text{test})}({\boldsymbol{\beta}}_0)-\ell^{(\text{test})}(\hat{\boldsymbol{\beta}})\right\},$ 
where $\ell^{(\text {test})}({\boldsymbol{\beta}})$ is the log partial likelihood based on the test dataset. 
Smaller values of this predictive deviance indicate more accurate predictions. 

In addition, the performance of Bayesian credible intervals for coordinates of $\boldsymbol{\beta}$ is assessed through coverage rates and average widths. 
Each interval is constructed using the $2.5\%$ and $97.5\%$ quantiles of the marginal posterior. 
For comparison, confidence intervals based on the asymptotic normality of MPLE, as described in Andersen (1982), serve as the baseline.

\subsection{Simulation results} 
Table~\ref{table:censor2} summarizes the estimation and prediction performance of various methods under a $20\%$ censoring rate. 
The reported metrics, average squared error and predictive deviance, are computed over 100 replications.
Among the evaluated methods, the MPLE performs worst when the dimension $p$ reaches 40, where it exhibits the highest estimation error and the lowest prediction accuracy.  
In contrast, the CRE and the WME tuned via cross-validation consistently rank among the best three methods for both metrics.
The superior performance of these methods can be attributed to regularization. When $p$ is large, the Hessian matrix of the partial likelihood is approximately singular, which leads to a flat curvature and causes high variability in the MPLE. The use of synthetic data provides regularization, either through penalty terms or weighted synthetic data, that reduces the variability of the estimators. 
The ridge estimator also demonstrates competitive performance, aligning with findings in the existing literature \citep{huang_penalized_2002, bovelstad2007predicting}. 
Conversely, the Lasso estimator performs poorly in terms of squared error, which may be attributed to the absence of sparsity in the true coefficients.

Additional simulation results for 10\% and 40\% censoring rates can be found in   Appendix J.
The findings there are consistent with those in Table~\ref{table:censor2}. 
Regarding the impact of the censoring rate, we observe that as more observations become censored, the performance of all methods worsens because less information is available for the true survival times. 
Nonetheless, our proposed methods, in particular $\hat{\boldsymbol{\beta}}_{CR,\hat{\tau}_{cv}}$ and $\hat{\boldsymbol{\beta}}_{WM,\hat{\tau}_{cv}}$, appear to be robust and maintain superior performance across all scenarios considered. 
\renewcommand{\arraystretch}{0.5}
\begin{table}
\centering
\small
\caption{Summary of estimation and prediction performance for various methods under a 20\% censoring rate. 
Standard errors are presented in parentheses. 
The best-performing method for each scenario is highlighted in bold. Shorthand notations: 
``CPM'', ``APM'', and ``GPM'' stand for Bayesian \underline{p}osterior \underline{m}ean resulting from using a Cox \underline{c}atalytic prior, a Cox \underline{a}daptive catalytic prior, and  a \underline{G}aussian prior respectively;  
``CV'' in parentheses indicates the selection of tuning parameters through cross-validation.}

$$
\begin{array}{clccc}
\hline
p & \text{Methods} & \|\hat{\boldsymbol{\beta}}-\boldsymbol{\beta}_0 \|^2 & \text { Predictive deviance }  \\
\hline
\hline
 & \text{MPLE} & 0.95(0.06) & 19.73(1.32) \\
& \text{CRE (CV)} & 0.63(0.04) & 12.87(0.80) \\
& \text{WME (CV)} & \textbf{0.51}(0.02) & \textbf{12.44}(0.75) \\
20& \text{CPM (CV)} & 0.79(0.04) & 18.82(0.67) \\
& \text{APM} & 0.74(0.01) & 20.61(0.66) \\
& \text{GPM (CV)} & 0.61(0.03) & 13.72(0.60) \\
& \text{Ridge (CV)} & 0.58(0.03) & 13.07(0.63) \\
& \text{Lasso (CV)} & 0.75(0.04) & 13.42(0.63) \\
\hline\hline
 & \text{MPLE} & 4.25(0.21) & 103.38(4.99) \\
& \text{CRE (CV)} & 0.82(0.02) & 24.09(0.84) \\
& \text{WME (CV)} & \textbf{0.76}(0.02) & \textbf{23.37}(0.82) \\
40& \text{CPM (CV)} & 0.84(0.02) & 24.63(0.76) \\
& \text{APM} & 0.79(0.02) & 23.69(0.74) \\
& \text{GPM (CV)} & 0.95(0.03) & 25.61(0.81) \\
& \text{Ridge (CV)} & 0.94(0.03) & 24.53(0.87) \\
& \text{Lasso (CV)} & 1.19(0.03) & 24.06(0.92) \\
\hline\hline
 & \text{MPLE} & 27.71(1.50) & 461.44(17.76) \\
& \text{CRE (CV)} & 1.02(0.02) & 30.61(0.88) \\
& \text{WME (CV)} & 0.98(0.02) & 29.77(0.86) \\
60& \text{CPM (CV)} & 1.07(0.02) & 29.62(0.80) \\
& \text{APM} & \textbf{0.97}(0.02) & 27.50(0.73) \\
& \text{GPM (CV)} & 1.11(0.02) & 30.58(0.79) \\
& \text{Ridge (CV)} & 1.11(0.02) & 30.57(0.77) \\
& \text{Lasso (CV)} & 1.29(0.02) & \textbf{27.37}(0.82) \\
\hline
\end{array}
$$
\label{table:censor2}
\end{table}

\renewcommand{\arraystretch}{1}

\renewcommand{\arraystretch}{0.5}
\begin{table}[!ht]
    \caption{Average coverage probability (\%) and width of various $95\%$ interval estimation, including credible intervals under the Cox catalytic prior with $\tau=\hat{\tau}_{cv}$, the Cox adaptive catalytic prior, Gaussian prior, and the confidence intervals associated with the MPLE. The best-performing method for each scenario is highlighted in bold.}
    \label{table:coverage_width}
    \small
    \centering
\begin{tabular}{cc|c|cccc}
\toprule
\multicolumn{2}{c|}{Setting} & & \multicolumn{4}{c}{Methods} \\
$\text{censoring}$ & $p$ & & Cat (CV)  & Cat Adaptive & Gaussian & MPLE \\
\hline
\midrule
0.1 & 20 & Coverage & 80.50\%  & 79.30\% & 84.75\% & \textbf{90.30}\% \\
& & Width & 0.48  & 0.46 & 0.53 & 0.61 \\
\midrule
& 40 & Coverage & 88.98\%  & \textbf{90.20}\% & 88.60\% & 82.38\% \\
& & Width & 0.44  & 0.45 & 0.46 & 0.73 \\
\midrule
& 60 & Coverage & 88.30\%  & \textbf{92.75}\% & 87.50\% & 68.83\% \\
& & Width & 0.37 & 0.44 & 0.37 & 1.05 \\
\hline\midrule
0.2 & 20 & Coverage & 83.80\% & 82.05\% & 87.55\% & \textbf{89.75}\% \\
& & Width & 0.51 & 0.49 & 0.56 & 0.66 \\
\midrule
& 40 & Coverage & 88.45\%  & \textbf{90.42}\% & 89.58\% & 80.92\% \\
& & Width & 0.46 & 0.47 & 0.47 & 0.81 \\
\midrule
& 60 & Coverage & 88.62\% & \textbf{93.12}\% & 87.65\% & 66.63\% \\
& & Width & 0.39  & 0.46 & 0.37 & 1.24 \\
\hline\midrule
0.4 & 20 & Coverage & 85.86\%  & 84.14\% & 89.34\% & \textbf{89.24}\% \\
& & Width & 0.58  & 0.56 & 0.64 & 0.80 \\
\midrule
& 40 & Coverage & 88.85\%  & \textbf{90.80}\% & 89.30\% & 76.68\% \\
& & Width & 0.51  & 0.53 & 0.51 & 1.08 \\
\midrule
& 60 & Coverage & 88.92\%  & \textbf{93.00}\% & 89.41\% & 49.57\% \\
& & Width & 0.42  & 0.50 & 0.40 & 3.02 \\
\hline
\end{tabular}

\renewcommand{\arraystretch}{1}
\end{table}

Table~\ref{table:coverage_width} presents the average coverage probability and the average width of $95\%$ interval estimates for $\beta_j$, averaged over $j$. 
The first three columns present results for credible intervals under catalytic priors with different specifications of the weight parameter $\tau$, the fourth for posterior intervals under Gaussian priors, and the last column for confidence intervals based on the asymptotic normality of MPLE. 
It is seen that, as the dimension $p$ increases, the performance of the asymptotic intervals deteriorates, with expanding widths and decreasing coverage rates. 
The credible intervals derived from the Cox catalytic prior with $\tau=\hat{\tau}_{c v}$ and the Gaussian prior perform consistently well across all settings. 
In contrast, those derived from the Cox catalytic prior with $\tau=p$ and the Cox adaptive catalytic prior perform well in the settings with $p=40$ and $p=60$. 
Notably, in these cases, the coverage rates using the Cox adaptive catalytic prior are the closest to the nominal coverage level.

In summary, our proposed methods outperform classical maximum partial likelihood estimation, particularly when the sample size is insufficient for stable estimation. 
Additionally, these methods generally compare favorably with the competitive regularization methods (ridge and Lasso regression). 
For small sample sizes, we recommend using $\tau=p$ for point estimation and an adaptive catalytic prior for the full Bayesian procedure. 
When the ratio of $p$ to $n$ is moderate, we recommend the weighted mixture estimator tuned via cross-validation, which achieves the lowest mean squared error in our simulation and can be computed easily using standard software without extra implementation.

\section{Application to real-world data}
\label{sec:real_data}

In this section, we apply the Cox catalytic prior to analyzing the well-known Primary Biliary Cholangitis (PBC) dataset from the Mayo Clinic, which has been widely used in survival analysis studies \citep{therneau_modeling_2000}. 
A detailed description of this dataset and our preprocessing steps can be found in  Appendix K. 
After removing subjects with missing data, our analysis will be based on the remaining $n=276$ observations.

Our data analysis is divided into two segments: in the first segment, we conduct a full Bayesian analysis of the PBC dataset; in the second segment, we compare different point estimators based on their out-of-sample prediction performance.

\subsection{Bayesian analysis using Cox catalytic priors}

We begin with a Bayesian analysis of the whole PBC dataset using the Cox catalytic prior. 

\paragraph{Generation of synthetic data} 
To implement the catalytic prior methods, we generate a synthetic dataset comprising $M=1,000$ subjects.
The generation of synthetic binary covariates and continuous covariates follows the same procedure stated in  Section \ref{sec:simulation}. 
For the three-level categorical variable edema, half of the synthetic edema covariates are randomly selected from the observed edema values, and the other half are uniformly sampled from the three possible levels. After generating the synthetic edema vector, we convert it into two separate vectors of dummy variables. 
We sample synthetic survival times from an estimated exponential distribution, following the same procedure as in the simulation studies in Section~\ref{sec:simulation}.

\paragraph{Full Bayesian analysis} 
We place a Cox adaptive catalytic prior on the coefficients $\boldsymbol{\beta}$ and weight parameter $\tau$, and independently assign a Gamma process prior to the cumulative baseline hazard function $H_0(\cdot)$. 
Following the full Bayesian approach outlined in  Appendix B, we can obtain posterior samples of $\boldsymbol{\beta}$ using standard MCMC samplers. 

\begin{table}[htbp]
    \centering
    \caption{Summary of estimated model coefficients for PBC Data. The ``Cat\_mean'' column represents the posterior mean estimates and the ``Cred\_Int'' column represents the \(95\%\) credible intervals derived from posterior samples. The last two columns represent MPLE estimates and the corresponding \(95\%\) confidence intervals. If the interval does not contain 0, we annotate it in boldface.}
    \label{table:summary_pbc}
    \scriptsize  
    \setlength{\tabcolsep}{4pt}  
    \begin{tabular}{llp{1.5cm}p{2cm}p{1.5cm}p{2cm}}
    \toprule
    \# & Variable & Cat\_mean & Cred\_Int & MPLE & Conf\_Int  \\
    \midrule
    1 & DPCA Treatment & 0.114 & [-0.25, 0.47] & 0.172 & [-0.26, 0.60] \\
    2 & Age & 0.294 & \textbf{[0.05, 0.54]} & 0.309 & \textbf{[0.07, 0.55]} \\
    3 & Gender & -0.257 & [-0.71, 0.16] & -0.352 & [-0.96, 0.25] \\
     4 & Presence of ascites & -0.039 & [-0.63, 0.57] & 0.016 & [-0.75, 0.78] \\
    5 & Hepatomegaly & -0.032 & [-0.47, 0.39] & 0.058 & [-0.44, 0.55] \\
    6 & Edema1 (despite diuretic therapy) & 0.557 & [-0.13, 1.19] & 1.150 & \textbf{[0.34, 1.96]} \\
    7 & Edema2 (untreated or successfully treated) & 0.073 & [-0.55, 0.65] & 0.256 & [-0.4, 0.91] \\
    8 & Serum bilirubin & 0.435 & \textbf{[0.22, 0.64]} & 0.369 & \textbf{[0.13, 0.61]} \\
    9 & Serum cholesterol & 0.069 & [-0.14, 0.27] & 0.115 & [-0.09, 0.32] \\
    10 & Serum albumin & -0.337 & \textbf{[-0.56, -0.12]} & -0.304 & \textbf{[-0.55, -0.06]} \\
    11 & Urine copper & 0.229 & \textbf{[0.02, 0.43]} & 0.212 & \textbf{[0.01, 0.41]} \\
    12 & Phosphatase & 0.025 & [-0.14, 0.19] & 0.006 & [-0.16, 0.17] \\
    13 & Aspartate aminotransferase & 0.218 & [-0.01, 0.45] & 0.219 & [0.00, 0.44] \\
    14 & Triglycerides & -0.053 & [-0.22, 0.12] & -0.035 & [-0.22, 0.15] \\
    15 & Platelet count & 0.020 & [-0.22, 0.24] & 0.074 & [-0.14, 0.29] \\
    16 & Standardized blood clotting time & 0.265 & \textbf{[0.06, 0.46]} & 0.243 & \textbf{[0.03, 0.45]} \\
    17 & Histologic stage of disease & 0.423 & \textbf{[0.15, 0.72]} & 0.384 & \textbf{[0.09, 0.68]}\\
    18 & Blood vessel malformations & 0.059 & [-0.36, 0.46] & 0.067 & [-0.42, 0.55] \\
    \bottomrule
    \end{tabular}
\end{table}

Table~\ref{table:summary_pbc} summarizes the posterior means and 95\% credible intervals for individual coefficients, along with results from the MPLE for comparison.
Both Bayesian credible intervals and MPLE confidence intervals identify the following variables as significant: 
age, serum bilirubin, serum albumin, urine copper, standardized blood clotting time, and the histologic stage of disease. 
However, the MPLE additionally identifies Edema1 as significant, which Bayesian analysis does not. 
Notably, both the credible and confidence intervals for the DPCA treatment variable contain zero, suggesting insufficient evidence to support the effectiveness of the treatment. 
This observation is consistent with existing literature, where the ineffectiveness of DPCA treatment has been reported \citep{locke1996time}.

\subsection{Comparison of different methods} 

To examine the finite-sample performance of our proposed methods, we consider the estimation of a Cox model based on a subsample (\textit{training set}) and the evaluation of the fitted model on a separate subsample (\textit{test set}). 
Specifically, in each of the 100 replications, a test set of size $136$ is randomly drawn from the full data, and the remaining 140 units form three nested training sets of size $140, 100, 60$: $I_{140}$, $I_{100}$, and $I_{60}$, with $I_{140} \supset I_{100} \supset I_{60}$. 
A training set is used to define Cox catalytic priors and to estimate coefficients.

We analyze the estimators evaluated in the simulation studies (see Section~\ref{sec:simulation}), using 10-fold cross-validation to select tuning parameters.   
To compare their out-of-sample prediction performance, we employ a metric based on the difference in the log partial likelihood of the fitted Cox model and the null model. 
This metric was formulated by \citet{bovelstad2007predicting} and we refer to it as \textit{prediction score}:
$2\left\{\ell^{(\text {test })}\left(\hat{\boldsymbol{\beta}}_{\text {train }}\right)-\ell^{(\text {test })}(\mathbf{0})\right\}, $ 
where $\hat{\boldsymbol{\beta}}_{\text {train }}$ is any estimate and 
$\ell^{(\text {test})}({\boldsymbol{\beta}})$ is the log partial likelihood based on the test dataset. 
A larger prediction score means better predictive performance of $\hat{\boldsymbol{\beta}}_{\text {train }}$.

\begin{table}[!ht]
    \centering
    \caption{Summary of prediction scores for different methods with various training sample sizes. 
    The standard error is in the parentheses following each score. 
    The best-performing method for each setting is highlighted in bold. 
    Shorthand notations: ``CPM'' stands for Bayesian posterior mean resulting from using a Cox catalytic prior, ``GPM'' stands for posterior mean using a Gaussian prior, and ``CV'' in parentheses indicates the selection of tuning parameters through cross-validation.
    }

    {
   \begin{tabular}{lccc}
    \toprule
    Method & \(I_{60}\) & \(I_{100}\) & \(I_{140}\) \\ 
    \midrule
    MPLE & -488.34(59.94) & -113.57(26.99) & -11.17(6.38) \\
    CRE (CV) & 51.66(2.47) & \textbf{57.33}(1.36) & \textbf{59.98}(1.25) \\
    WME (CV) & 52.21(2.49) & 56.83(1.33) & 59.65(1.22) \\
    CPM (CV) & 50.50(2.31) & 56.58(1.35) & 59.31(1.23) \\
    APM & 43.96(2.79) & 47.84(2.31) & 50.55(2.06) \\
    GPM (CV) & 51.47(2.46) & 56.46(1.31) & 59.09(1.25) \\
    Ridge (CV) & 52.51(2.26) & 57.07(1.24) & 59.62(1.21) \\
    Lasso (CV) & 37.61(2.55) & 38.78(6.79) & 52.52(1.46) \\
    \bottomrule
\end{tabular}
}
    \label{table:subsampling}
\end{table}

Table~\ref{table:subsampling} shows that as the training sample size increases, all methods exhibit increased prediction scores.   
However, the classical MPLE method consistently underperforms, which reflects the challenges of analyzing survival data when the sample size is only slightly larger than the number of covariates. 
When the sample size is $60$, both CRE\_p and WME\_p significantly outperform ridge regression. 
For all sample sizes, the performance of CRE\_CV and WME\_CV is comparable to that of ridge regression. 
Overall, our proposed methods either outperform or are on par with the popular ridge and Lasso regularized methods.

\section{Discussion}
\label{sec:discussion}

When a Cox model is applied to analyze survival data where the dimension of covariates is large relative to the sample size, traditional methods like maximum partial likelihood estimation often yield biased and unreliable estimates. 
In this work, we introduce the class of Cox catalytic priors for Bayesian inference in Cox models by leveraging synthetic data. 
This prior is constructed as a weighted likelihood based on synthetic data and a user-specified surrogate hazard.
The influence of synthetic data is governed by a weight parameter, which is specified by users or treated as a hyperparameter. 
Cox catalytic priors are shown to be proper and can seamlessly integrate with standard Bayesian inference.

For estimating regression coefficients, we derive two regularized estimators: WME by approximating the marginal posterior mode, and CRE by combining the synthetic data with the observed data. 
Both estimators can be efficiently computed: the CRE is solvable via convex optimization, and the WME can be conveniently implemented using standard software. 
In terms of frequentist evaluation, 
we prove that both estimators are consistent in the regime where the dimension diverges, which justifies the idea of \textit{regularization through synthetic data}. 
Empirical studies demonstrate that the resulting inferences using the Cox catalytic priors outperform classical maximum partial likelihood-based inferences, and are comparable or superior to other regularization methods.

Cox catalytic priors have better interpretability than many other priors that focus on parameters, as they encode prior information through synthetic data.
This approach emphasizes that data are real while models are imperfect tools for analyzing data. 
Unlike the original construction of catalytic priors for parametric models, our construction applies to a semiparametric model. 
This prior construction scheme may generalize to other inferences that involve partial likelihood or complex nuisance parameters.

%% file: appendix.tex
\section{Related literature}\label{supp:related literature}
Our work is related to three areas of research: (1) specification of prior distributions, (2) regularized estimations, and (3) survival models.

The use of synthetic data for constructing prior distributions has a long-standing tradition in Bayesian statistics \citep{good1983good}. It is well-known that conjugate priors for exponential families can be interpreted as the likelihood of pseudo-observations \citep{raiffa1961applied}. 
Although there are prior formulations that incorporate additional synthetic data from experts' knowledge \citep{bedrick1996new,bedrick1997bayesian}, they often encounter challenges during practical implementation, especially when there are a lot of covariates in the data or when there are a lot of models being considered. 
To address the challenges in Bayesian model selection, a promising alternative is to use the class of the expected-posterior priors \citep{iwaki1997posterior,perez2002expected,neal2001transferring}, where a prior is defined as the average posterior distribution derived from a set of synthetic data sampled from a straightforward predictive model. However, this class of priors has not been adapted to Cox models. 
\citet{wu2018assessing} proposed Jeffreys-type priors to deal with the monotone partial likelihood problem in Cox models. 
Informative priors, such as the power priors \citep{ibrahim2000power,chen2000power, ibrahim1998prior}, effectively incorporate information from historical data using a power of the likelihood function of those data. However, the validity of a power prior requires that the covariate matrix of historical data has full column rank, which may fail to hold.
In addition, historical data might not always be available. 
In contrast, the catalytic prior is based on synthetic data sampled from a simpler model, which serves as a replacement for historical data.

Apart from the Bayesian strategies, regularized partial likelihood methods are also employed to stabilize Cox models with lots of covariates. 
In particular, the $\ell_2$-regularizer is frequently used in practice to shrink regression coefficients towards smaller values, which is often referred to as ridge regression \citep{hoerl1970ridge}.  
\citep{verweij1993cross} adopted the $\ell_2$-penalty to regularize the negative log partial likelihood, and \citep{huang_penalized_2002} demonstrated that ridge regression can yield a lower mean square error and improved prediction accuracy of relative risk compared to the MPLE, especially in small samples. 
\citep{bovelstad2007predicting} compared the prediction performance resulting from different estimations, where  
ridge regression demonstrated the best overall performance for their microarray gene expression datasets.  

Although this paper focuses on the proportional hazards model, there are many other semiparametric survival models \citep{aalen1989linear,wei1992accelerated,murphy1997maximum}.  
We refer the interested reader to \citet{hanson2013surviving} for a review and to the monograph \citet{ibrahim_bayesian_2001} for a comprehensive treatment of Bayesian survival analysis.

\section{Full Bayesian approach}
\label{sec:supp_full_bayes_cox} 
In standard Bayesian Cox modeling for survival data, it is common to use a nonparametric prior process to model the baseline hazard or the cumulative baseline hazard function
\citep[Chapter~3]{ibrahim_bayesian_2001}. 
There are many choices for modeling the baseline hazard function.
Here we use a Gamma process to model the cumulative baseline hazard function and incorporate it with the Cox catalytic prior in Section~\ref{sec:formulate}.

Let $\mathcal{G}(\alpha, \lambda)$ denote the Gamma distribution with shape parameter $\alpha>0$ and scale parameter $\lambda>0$. Let $\alpha(t), t \geq 0$, be an increasing left-continuous function such that $\alpha(0)=0$, and let $\bar{Z}(t), t \geq 0$, be a stochastic process with the properties:

(i) $Z(0)=0$;\\
(ii) $Z(t)$ has independent increments in disjoint intervals; and\\
(iii) for $t>s, Z(t)-Z(s) \sim \mathcal{G}(c(\alpha(t)-\alpha(s)), c)$.

Then the process $\{Z(t): t \geq 0\}$ is called a gamma process and is denoted by $Z(t) \sim \mathcal{G P}(c \alpha(t), c)$.

In addition, we make use of the grouped data likelihood proposed in \citep{urridge1981empirical} to express the full likelihood, which can handle survival data with distinct continuous failure time and survival data with many ties on failure time. 
Furthermore, we impose the catalytic prior defined in \eqref{cat_beta_cox} for the coefficient $\boldsymbol{\beta}$. A similar setting was adopted in \citep{lee2011bayesian}, where they use a Laplace prior on $\boldsymbol{\beta}$ to perform variable selection. 
For clarity, \citep{kalbfleisch1978non} assigns a Gamma process prior to model cumulative baseline hazard:
\begin{equation}
\label{Gamma_process}
H_0 \sim \mathcal{G} \mathcal{P}\left(c_0 H^*, c_0\right),
\end{equation}

where $H^*(t)$ is an increasing function with $H^*(0)=0$, and $c_0$ is a positive constant controlling the weight of $H^*$. In practice, $H^*$ is taken to be a known parametric function, such as exponential or Weibull distribution. 
In our numerical studies, we take $H^*(t)=\eta_0 t^{\kappa_0}$, which corresponds to Weibull distribution, the hyperparameters $\left(\eta_0, \kappa_0\right)$ are determined by the maximum likelihood estimate from an intercept-only Weibull regression model based on the training data. In this case, the increasing function $H^*(t)$ looks similar to the Nelson-Aalen estimator, which provides an initial guess of $H_0(t)$.  A smaller value of $c_0$ corresponds to a more diffuse and less informative prior. We use $c_0=2$ to get reasonably fast convergence in our numerical study.  

In order to construct the grouped data likelihood, we introduce a convenient representation of the observed data. 
Consider a finite partition of the time axis as follows:  $0=s_0<s_1<s_2<\ldots<s_J$, with $s_J> \underset{1\leq i\leq n}{\max}  \left( Y_i \right)$.  Then, we have $J$ disjoint intervals, $I_j=\left(s_{j-1}, s_j\right]$ for $j=1,2, \ldots, J$, and any observed survival time $Y_i$ falls in one of these $J$ intervals.  
We further introduce two notations regarding these intervals. 
Let $\tilde{\mathcal{R}}_j=\{i:Y_i>s_{j-1}\}$ be the risk set
and $\tilde{\mathcal{D}}_j=\{i:\delta_i=1,Y_i\in (s_{j-1},s_j]\}$ be the failure set corresponding to the $j^{\text {th }}$ interval $I_j$. 
We can now represent the observed data as $\tilde{\bD}=\left(\bX, \tilde{\mathcal{R}}_j, \tilde{\mathcal{D}}_j: j=1,2, \ldots, J\right)$ where $\bX$ is the $n \times p$ matrix of covariates with $i^{t h}$ row $\bX_i^{\prime}$ denoting the risk factors.

Let $h_j$ denote the increment in the cumulative baseline hazard in the interval $I_j$ as follows
\begin{equation}
	h_j=H_0\left(s_j\right)-H_0\left(s_{j-1}\right), \quad j=1,2, \ldots, J .
\end{equation}
The gamma process prior in \eqref{Gamma_process} implies that the $h_j$'s follow independent gamma distributions, that is, 
\begin{equation}
	\label{prior_h}
	h_j \sim \mathcal{G}amma\left(\alpha_{0 j}-\alpha_{0, j-1}, c_0\right), 
\end{equation}
where $\alpha_{0 j}=c_0 H^*\left(s_j\right)$. According to \citep[Chapter~3.2.2]{ibrahim_bayesian_2001}, the grouped data likelihood is given by
\begin{equation}
\label{group_likelihood}
L(\boldsymbol{\beta}, \boldsymbol{h} \mid \tilde{\bD}) \propto \prod_{j=1}^J G_j,
\end{equation}
where $\boldsymbol{h}=\left(h_1, h_2, \ldots, h_J\right)^{\prime}$ and 
$$G_j=\exp \left\{-h_j \sum_{k \in \tilde{\mathcal{R}}_j-\tilde{\mathcal{D}}_j} \exp \left(\bX_k^{\prime} \boldsymbol{\beta}\right)\right\} \prod_{l \in \tilde{\mathcal{D}}_j}\left[1-\exp \left\{-h_j \exp \left(\bX_l^{\prime} \boldsymbol{\beta}\right)\right\}\right].$$
We make the typical assumption that $\boldsymbol{\beta}$ and $\bh$ are  independent in their joint prior, i.e., $\pi_{prior}(\boldsymbol{\beta},\bh)=\pi_{prior,\boldsymbol{\beta}}(\boldsymbol{\beta})\cdot\pi_{prior,\bh}(\bh)$. Based on the prior on $\boldsymbol{\beta}$ \eqref{cat_beta_cox}, the prior on $\bh$ \eqref{prior_h}, and the likelihood for $(\boldsymbol{\beta},\bh)$ \eqref{group_likelihood}, we can write down the joint posterior distribution:
\begin{equation}
	\label{posterior}
	\begin{aligned}
		\pi(\boldsymbol{\beta},\bh\mid \tilde{\bD})&\propto L(\boldsymbol{\beta}, \boldsymbol{h} \mid \tilde{\bD})\cdot\pi_{prior,\bh}(\bh)\cdot \pi_{prior,\boldsymbol{\beta}}(\boldsymbol{\beta})\\&\propto \left(\prod_{j=1}^J G_j\right) \times \left(\prod_{j=1}^J\left[h_j^{\left(\alpha_{0 j}-\alpha_{0, j-1}\right)-1} \exp \left(-c_0 h_j\right)\right]\right)\\
  &\quad \times \left(\prod_{i=1}^M\left\{\exp \left(\bX_i^{*\prime} \boldsymbol{\beta}\right) h_0^+\right\}\exp \left\{-\exp \left(\bX_i^{*\prime} \boldsymbol{\beta}\right) Y_i^* h^+_0\right\}\right)^{\tau/M}
	\end{aligned}
\end{equation}

Similar to the construction of the Cox adaptive catalytic prior on $(\tau, \boldsymbol{\beta})$ in \eqref{joint_prior}, we propose a joint posterior distribution for $(\boldsymbol{\beta},\bh,\tau)$ as follows
$$\begin{aligned}
    \pi(\boldsymbol{\beta},\bh,\tau \mid \tilde{\bD})&\propto \left(\prod_{j=1}^J G_j\right) \times \left(\prod_{j=1}^J\left[h_j^{\left(\alpha_{0 j}-\alpha_{0, j-1}\right)-1} \exp \left(-c_0 h_j\right)\right]\right)\times \tau^{p+\alpha-1} e^{-\tau\left(\kappa+\gamma^{-1}\right)} \\
    &\quad \times \left(\prod_{i=1}^M\left\{\exp \left(\bX_i^{*\prime} \boldsymbol{\beta}\right) h_0^+\right\}\exp \left\{-\exp \left(\bX_i^{*\prime} \boldsymbol{\beta}\right) Y_i^* h^+_0\right\}\right)^{\tau/M}
\end{aligned}$$
where $\alpha,\gamma$ are two positive hyperparameters given in (\ref{joint_prior}), and $\kappa$ is defined after \eqref{prior_tau}.

\section{Potential Advantages of Log-Concavity in the Catalytic Prior}
\label{supp:potential_logconcave}
 
There exist multiple approaches to formulate the semiparametric Cox model within a Bayesian framework. In our method, detailed in section \ref{sec:supp_full_bayes_cox}, we employ a Gamma process combined with a Grouped likelihood. Another notable method is the ``Piecewise Constant Hazard Model" as described in \cite[Chapter 3]{ibrahim_bayesian_2001}. 
The model's likelihood is constructed by first establishing a finite partition of the time axis: $0<s_1<s_2<\ldots<s_J$, ensuring $s_J>Y_i$ for all $i=1,2, \ldots, n$. This results in $J$ intervals such as $\left(0, s_1\right],\left(s_1, s_2\right]$, and so on up to $\left(s_{J-1}, s_J\right]$. For the $j^{t h}$ interval, a constant baseline hazard, $h_0(y)=\lambda_j$, is assumed for $y \in I_j=\left(s_{j-1}, s_j\right]$.
The observed data is represented by $\boldsymbol{D}=\{ \boldsymbol{X}_i, Y_i, \delta_i\}_{i=1}^n$, where  $\delta_i=1$ indicates the failure of the $i^{t h}$ subject and 0 otherwise. With $\boldsymbol{\lambda}=\left(\lambda_1, \lambda_2, \ldots, \lambda_J\right)^{\prime}$, the likelihood function of $(\boldsymbol{\beta}, \boldsymbol{\lambda})$ for the $n$ subjects can be expressed as:
$$\begin{gathered}L(\boldsymbol{\beta}, \boldsymbol{\lambda} \mid \boldsymbol{D})=\prod_{i=1}^n \prod_{j=1}^J\left(\lambda_j \exp \left(\boldsymbol{X}_i^{\prime} \boldsymbol{\beta}\right)\right)^{\nu_{i j} \delta_i} \exp \left\{-\nu_{i j}\left[\lambda_j\left(y_i-s_{j-1}\right)\right.\right. \\ \left.\left.+\sum_{g=1}^{j-1} \lambda_g\left(s_g-s_{g-1}\right)\right] \exp \left(\boldsymbol{X}_i^{\prime} \boldsymbol{\beta}\right)\right\}\end{gathered}$$

where $\nu_{i j}=1$ if the $i^{t h}$ subject failed or was censored in the $j^{t h}$ interval, and $0$ otherwise.
This model guarantees a log-concave likelihood for $\boldsymbol{\beta}$ given $\boldsymbol{\lambda}$. Given this likelihood,  alongside our log-concave catalytic prior, the posterior distribution of $\boldsymbol{\beta} \mid \boldsymbol{\lambda}$ is also log-concave. We can exploit the log-concavity to draw samples quickly from the conditional posterior distribution of $\boldsymbol{\beta} \mid \boldsymbol{\lambda}$, the Gibbs sampler can potentially proceed more rapidly. Algorithms like the adaptive rejection sampling \citep{gilks1992adaptive} and hit-and-run algorithms \citep{lovasz2007geometry} can be used to efficiently sample from log-concave distributions.

\section{Justification of catalytic-regularized estimator (CRE)}
\label{supp:justify_penalized}
Generally, posterior modes can be easily computed via some M-estimation. 
For instance, a regression model with Laplace prior or Gaussian prior on the coefficients will correspond to Lasso or Ridge regression, respectively. 
It would be helpful if we could compute the mode of the marginal posterior
 of $\boldsymbol{\beta}$  based on  $\pi(\boldsymbol{\beta},\bh\mid \tilde{\bD})$.  
 We expect the marginal likelihood of $\boldsymbol{\beta}$ will be close to the partial likelihood when we impose a non-informative prior on the baseline hazard function.  For continuous survival data, \citep{kalbfleisch1978non} show that with the gamma process prior for cumulative baseline hazard, the partial likelihood of $\boldsymbol{\beta}$ in \eqref{PL} can be obtained as the limit of the marginal posterior of $\boldsymbol{\beta}$ in the Cox model when $c_0\rightarrow 0$. For grouped data likelihood, \citep{sinha_bayesian_2003} show that for sufficiently small grouping intervals and $c_0\rightarrow 0 $, the marginal posterior of $\boldsymbol{\beta}$ can be approximated by $PL_{group}(\boldsymbol{\beta}\mid \tilde{\bD})\times \pi_{prior,\boldsymbol{\beta}}(\boldsymbol{\beta})$, where $PL_{group}(\boldsymbol{\beta}\mid \tilde{\bD})$ is defined as
 \begin{equation}
 	\label{PL_group}
 	PL_{group}(\boldsymbol{\beta}\mid \tilde{\bD})=\prod_{j=1}^J \frac{\exp \left(\sum_{l \in \tilde{\mathcal{D}}_j} \bX_l^{\prime} \boldsymbol{\beta}\right)}{\left\{\sum_{k \in \tilde{\mathcal{R}}_j} \exp \left(\bX_k^{\prime} \boldsymbol{\beta}\right)\right\}^{d_j}}
 \end{equation}
 and $d_j$ is the number of subjects in $\mathcal{D}_j$. $PL_{group}$ will agree with the ordinary partial likelihood in \eqref{PL} when the grouping intervals are so small that there is at most one distinct failure time in each interval.

The relationship between the partial likelihood and the marginal likelihood of $\boldsymbol{\beta}$ under a non-informative prior on the baseline hazard admits an efficient way to approximate the mode of marginal posterior of $\boldsymbol{\beta}$: we can directly maximize $PL(\boldsymbol{\beta} )\times \pi_{prior,\boldsymbol{\beta}}(\boldsymbol{\beta})$ over $\boldsymbol{\beta} $ to obtain an estimator of $\boldsymbol{\beta}$.  
Specifically, if $\pi_{prior,\boldsymbol{\beta}}(\boldsymbol{\beta})$ is the catalytic prior defined in \eqref{cat_beta_cox}, we can get the following estimator:
 \begin{equation}
 \label{supp:estimator_penalty}
 	 \begin{aligned}
        	\hat{\boldsymbol{\beta}}_{CR,\tau}&:=\arg \max_{\boldsymbol{\beta}}\left\{PL(\boldsymbol{\beta})\cdot \pi_{cox,cat}(\boldsymbol{\beta}\mid \tau) \right\}
        	\\&=\arg \max_{\boldsymbol{\beta}} \left\{\left(\prod_{i =1,\delta_i=1}^n \left\{ \frac{\exp \left(\bX_{i}^{\prime} \boldsymbol{\beta}\right)}{\sum_{l \in \mathbf{R}_{i}} \exp \left(\bX_{l}^{\prime} \boldsymbol{\beta}\right)}\right\}
\right)\times \right.\\
& \quad \quad \quad \left. \left[ \prod_{i=1}^M\left\{\exp \left(\bX_i^{*\prime} \boldsymbol{\beta}\right) h_0^+\right\}\exp \left\{-\exp \left(\bX_i^{*\prime} \boldsymbol{\beta}\right) Y_i^* h^+_0\right\}\right]^{\tau/M}  \right\}\\
        	&=\arg \max_{\boldsymbol{\beta}} \left\{\log PL(\boldsymbol{\beta})+\frac{\tau}{M}\log L\left(\boldsymbol{\beta}, h_0^+ \mid \left\{\left(\bX_i^*, Y_i^*\right)\right\}_{i=1}^M\right) \right\}, 
        \end{aligned}
 \end{equation}
 where $L\left(\boldsymbol{\beta}, h_0^+ \mid \left\{\left(\bX_i^*, Y_i^*\right)\right\}_{i=1}^M\right)$ is the likelihood of ($\boldsymbol{\beta},h_0^+$) based on the synthetic data in the formulation of the catalytic prior. 
 The last equation of \eqref{supp:estimator_penalty} implies that the original log partial likelihood is penalized by the weighted log-likelihood based on synthetic data.
 Due to this reason, the resulting estimator is termed as $\hat{\boldsymbol{\beta}}_{CR,\tau}$.

The property of $\hat{\boldsymbol{\beta}}_{CR,\tau}$ is similar to the property of posterior mode of (\ref{posterior_general}) for parametric model. When the weight of synthetic data $\tau$ converges to zero, we actually recover the MPLE. As the weight of synthetic data increases, it will pull MPLE towards $\hat{\psi}$, where $\hat{\psi}$ is a fitted parameter based on a simpler model that is used to generate synthetic survival time.

 \section{Details of weighted mixture estimator (WME)}
 
 \label{sec:supp_mix_detail} 

 Suppose the observed data set is $\bD=\{\left(\bX_i, Y_i, \delta_i\right)\}_{i=1}^n$ and  the synthetic data set is $\bD^*=\left\{\left(\bX_i^*, Y_i^*\right)\right\}_{i=1}^M$, where $\bX_i$ is  covariate vector, $Y_i$ is  response time, $\delta_i$ denoted whether $i$-th  data is censored or not, $(\bX^*_i,Y_i^*)$ are corresponding synthetic version. The idea of this alternative method is to merge both observed data and synthetic data together and assign a different weight, one benefit of this approach is that we can utilize currently available software to compute the estimate for regression coefficients directly. In the subsequent derivation, we will make a simplifying assumption that there are no censored observations in either the observed or synthetic data. The version incorporating censored observations can be easily extended from this simplified case.

Since synthetic data are not generated from the same distribution as observed data, we need to reduce the impact of the synthetic data when we merge both data together. We may define a modified likelihood: 
 $$ \tilde{L}(\boldsymbol{\beta},h(\cdot)\mid \bD,\bD^*,\tau)=L(\boldsymbol{\beta},h(\cdot)\mid \bD)\cdot L(\boldsymbol{\beta},h(\cdot)\mid \bD^*)^{\frac{\tau}{M}}.$$

This likelihood function still involves the nuisance parameter $h(\cdot)$, and a common strategy to remove a nuisance parameter is profiling them out.
Mimicking the reasoning developed in \citep{murphy2000profile}, we can derive a new partial likelihood, denoted by $P\tilde{L}(\boldsymbol{\beta})$, as follows: 
\begin{equation}
	P\tilde{L}(\boldsymbol{\beta})=\left[\prod_{i \in \bI_D} \left( \frac{\exp(\bX_i^\top \boldsymbol{\beta})}{\sum_{j\in \tilde{\mathcal{R}}_i\cap \bI_D }\theta_j+\frac{\tau}{M}\sum_{j\in \tilde{\mathcal{R}}_i\cap \bI_{D^*}}\theta_j^*}  \right)^{\delta_i}\right]\left[\prod_{i \in \bI_{D^*}}\frac{\exp({\bX_{i-n}^*} ^\top \boldsymbol{\beta})}{\sum_{j\in \tilde{\mathcal{R}}_i\cap \bI_D }\theta_j+\frac{\tau}{M}\sum_{j\in \tilde{\mathcal{R}}_i\cap \bI_{D^*}}\theta_j^*}  \right]^{\tau/M},
\end{equation}

The notation in above equation is defined in the following way: $\bI_D=\{1,2,\ldots n\}$ denote the index set for observed data, $\bI_{D^*}=\{n+1,n+2,\ldots n+M\}$ denote the index set for synthetic data, i.e., we append the synthetic data to observed data, and if we use $\{\tilde{Y}_i\}_{i=1}^{n+M} $ to represent the collection of $\{Y_i\}_{i=1}^{n}$ and $\{Y_i^*\}_{i=1}^{M}$ then we have

$$
\tilde{Y}_i=\begin{cases}Y_i & \text { if } i\in \bI_D \\ Y^*_{i-n} & \text { if } i\in \bI_{D^*}\end{cases}
$$

$\tilde{\mathcal{R}}_i$ is the collection of subjects in $\{\tilde{Y}_i\}_{i=1}^{n+M} $ who is 
 at risk at time $\tilde{Y}_i$, i.e., $\tilde{\mathcal{R}}_i=\{j:\tilde{Y}_j\geq \tilde{Y}_i, j\in \{1,2,\ldots M+n\}\}$  and $\theta_j=\exp(\bX_j^\top\boldsymbol{\beta}),\theta_j^*=\exp({\bX_{j-n}^*}^\top \boldsymbol{\beta})$. 
Consequently, a new estimator can be constructed by maximizing $P\tilde{L}(\boldsymbol{\beta})$ as follows:
\begin{equation}\label{supp:mix_optimization}
	\begin{aligned}
        		\hat{\boldsymbol{\beta}}_{WM,\tau}&:=\arg\max_{\boldsymbol{\beta}}\log P\tilde{L}(\boldsymbol{\beta}) \\
        		&=\arg\max_{\boldsymbol{\beta}}\left\{\sum_{i\in \bI_D}\delta_i \left[ {\bX_i}^\top\boldsymbol{\beta} - \log (\sum_{j\in \tilde{\mathcal{R}}_i\cap \bI_D }\theta_j+\frac{\tau}{M}\sum_{j\in \tilde{\mathcal{R}}_i\cap \bI_{D^*}}\theta_j^*)\right]\right.\\
        		&\quad \quad +\left.\frac{\tau}{M} \sum_{i\in \bI_{D^*}}\left[ {\bX_{i-n}^*}^\top\boldsymbol{\beta} - \log (\sum_{j\in \tilde{\mathcal{R}}_i\cap \bI_D }\theta_j+\frac{\tau}{M}\sum_{j\in \tilde{\mathcal{R}}_i\cap \bI_{D^*}}\theta_j^*)\right]\right\}. 
        	\end{aligned}
\end{equation}

We can express the estimator succinctly using \eqref{wm partial likelihood}. The estimate can be computed easily using standard software. Specifically, one can merge the generated synthetic data with the observed data, and then use the standard Cox model fitting function with a weight vector (weight 1 for each observed data point and weight $\tau/M$ for each synthetic data point), for example, using the function \texttt{coxph()} in the \textbf{R} package \texttt{survival} \citep{survivalpackage}; see also \cite[Ch 7.1]{therneau_modeling_2000}. 
We summarize the procedure in Algorithm~\ref{alg: wme in R}.

\begin{algorithm}
\caption{Compute Weighted Mixture Estimate in \textbf{R} }\label{alg: wme in R}
\begin{algorithmic}
\State \textbf{Require:} \textbf{R} 
 package \texttt{survival}
\State \textbf{Input:} Observed data $ \{\mathbb{X},\bY,\boldsymbol{\delta}\} $ and synthetic data $ \{\mathbb{X}^*, \bY^*\} $ \\ 
\hspace{3em} Weight $ \tau $, number of synthetic data $ M $, sample size $ n $
\State \textbf{Procedure:}
\State \hspace{0.2cm} 1. Combine data:
\State \hspace{0.8cm} $ \tilde{\mathbb{X}} $ <- $ \text{rbind}(\mathbb{X}, \mathbb{X}^*) $
\State \hspace{0.8cm} $ \tilde{\bY} $ <- $ \text{c}(\bY, \bY^*) $
\State \hspace{0.8cm} $ \tilde{\boldsymbol{\delta}} $ <- $ \text{c}(\boldsymbol{\delta}, \boldsymbol{1}) $
\State \hspace{0.2cm} 2. Compute weight vector:
\State \hspace{0.8cm} $ \tilde{\mathbf{w}} $ <- c $(\text{rep}(1, n), \text{rep}(\frac{\tau}{M}, M))$
\State \hspace{0.2cm} 3. Fit proportional hazards regression model with combined data and weight vector: 
\State \hspace{0.8cm} $ \text{fit} $ <- $ \text{coxph(Surv}(\tilde{\bY}, \tilde{\boldsymbol{\delta}}) \sim \tilde{\mathbb{X}}, \text{weights} = \tilde{\mathbf{w}}) $
\State \textbf{Output:} Coefficients of the fitted model 
\State \hspace{0.8cm} $ \text{coef(fit)} $
\end{algorithmic}
\end{algorithm}

\section{Illustration of bias of MPLE}\label{supp:bias_MPLE}

Recently, \citet{zhang2022modern} observed the upward bias for MPLE in a simple simulation study and explained its occurrence in the asymptotic regime where the model dimension $p$ scales with the sample size $n$. 
Here we reproduce the simulation in \citep{zhang2022modern} to illustrate the problem of MPLE when dimension is large. We take $p=200$ and $n=400$. The true regression coefficients are set as follows: $\beta_{i0}=10$ for $i\in \{1,2,\ldots,40\}$. $,\beta_{i0}=-10$ for $i \in \{41,42,\ldots,80\}$,  and all other $\beta_{i0}$'s are zero. 
We assume an independent Gaussian design $X_{ij}\sim N(0,1)/\sqrt{n}$ and we set a constant baseline function $h_0(y)\equiv 1$. 
All subjects are uncensored. 

\begin{figure}[h!]
\centering
	\includegraphics[scale=0.4]{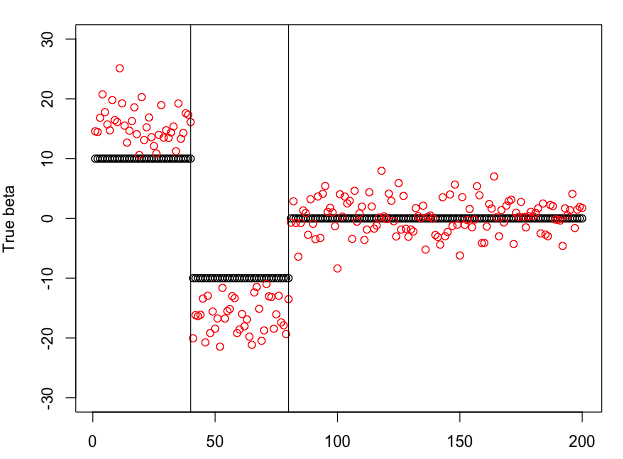}
	\caption{Red points represent MPLE $\hat{\beta_i}$, black points represent corresponding true value $\beta_{i0}$}
	\label{fig:MPLE_bias_plot}
\end{figure}

For one generated dataset, we plot the coordinates of the MPLE against the true values of $\boldsymbol{\beta}_0$ in Figure~\ref{fig:MPLE_bias_plot}. 
It is clear that for any coordinate $i$ with $\beta_{i0}\neq 0$, the MPLE $\hat{\beta}_i$ is biased away from zero. 
This indicates that shrinkage methods should be adopted to reduce both bias and variance in the estimation for large-dimensional Cox models.

\section{Proof of Theorem~\ref{Thm:proper}}\label{supp:proof_thm_one}

We first present a useful lemma, which is proved in \citep[ Lemma 5.1]{huang_catalytic_2020}
\begin{lemma} 
\label{int_lemma}
For $K>0, a>0$
	$$
\int_{\|\bX\|_2>K} \exp (-a\|\bX\|_2) d \bX \leq \frac{\Gamma(p) s_{p-1}}{a^p} \min \left(1, e^{p-a K}\left(\frac{a K}{p}\right)^p\right)
$$
where the constant $s_{p-1}=\frac{2 \pi^{p / 2}}{\Gamma(p / 2)}$ is the surface area of $a(p-1)$-dimension sphere. An immediate consequence is that 
$$\int \exp(-a\|\bX\|_2)d\bX =\int_{\|\bX\|_2\geq K}\exp(-a\|\bX\|_2)d\bX +\int_{\|\bX\|_2< K}\exp(-a\|\bX\|_2)d\bX <\infty$$
\end{lemma}
\subsection{Properness of catalytic prior with fixed weight}
Recall that we have assumed the norm-recoverability condition: $\frac{1}{M} \sum_{i=1}^M |\bX_i^{*\top}\boldsymbol{\beta}| \geq c_1\|\boldsymbol{\beta}\|_2$ for all $\boldsymbol{\beta} \in \mathbb{R}^p$. 
Now we are ready to prove that the integral of the catalytic prior is finite.

We first define some constants that are all positive. 
Let $q=\min_{1\leq i \leq M} Y_i^* h_0^+$, $ C_1:=\prod_{i=1}^M[h_0^+]^\frac{\tau}{M}$, $C_q:=\max\{0,-2+2\log(\frac{2}{q})\}$. 
We immediately have 
 
\begin{equation}\label{eq:int-cox-cat}
    \begin{aligned}
    	\int \pi_{cox,cat}(\boldsymbol{\beta}\mid \tau) d\boldsymbol{\beta} = & \int  \left[\prod_{i=1}^M\left\{\exp \left(\bX_i^{*\prime} \boldsymbol{\beta}\right) h_0^+\right\}\exp \left\{-\exp \left(\bX_i^{*\prime} \boldsymbol{\beta}\right) Y_i^* h_0^+\right\}\right]^{\tau/M}d\boldsymbol{\beta}\\
    	=&C_1 \int \left[\prod_{i=1}^M\exp \left\{\bX_i^{*\prime} \boldsymbol{\beta}-\exp \left(\bX_i^{*\prime} \boldsymbol{\beta}\right) Y_i^* h_0^+\right\}\right]^{\tau/M}d\boldsymbol{\beta} \\
    	\leq & C_1 \int \exp \left\{\frac{\tau}{M}\sum_{i=1}^M \left( \bX_i^{*\prime} \boldsymbol{\beta}-\exp \left(\bX_i^{*\prime} \boldsymbol{\beta}\right) q \right) \right\}d\boldsymbol{\beta} \\
    	\leq&C_1 \int \exp \left\{\frac{\tau}{M}\sum_{i=1}^M \left( -|\bX_i^{*\prime} \boldsymbol{\beta}| +C_q \right) \}\right\}d\boldsymbol{\beta}\\
    	= & C_1 \int \exp\left\{\tau C_q \right\}\exp \left\{-\frac{\tau}{M}\sum_{i=1}^M|\bX_i^{*\prime} \boldsymbol{\beta}| \right\}d\boldsymbol{\beta},
    \end{aligned}
\end{equation}
where the first inequality is due to the definition of the constant $q$, 
the second inequality follows from the elementary identity that 
$$
\max_y \{y-\alpha e^y+|y|\}= \begin{cases} -\alpha, & \text { if } \alpha \geq 2 \\ 
\max(-\alpha,2\log \frac{2}{\alpha}-2), & \text { if }\alpha \in (0,2) 
 \end{cases}
$$
In view of the norm-recoverability condition, we apply Lemma~\ref{int_lemma} with $a=\tau c_1$ to obtain 
\begin{equation*}
\begin{aligned}
    \text{Righ hand side of \eqref{eq:int-cox-cat} }	\leq & C_1 \exp\left\{\tau C_q \right\} \int \exp(-\tau c_1 \|\boldsymbol{\beta}\|_2) d\boldsymbol{\beta}\\
    	< & \infty. 
     \end{aligned}
\end{equation*}

\subsection{Properness of the Cox adaptive catalytic prior}
Denote by $\ell(\boldsymbol{\beta})$ the log-likelihood based on the synthetic data:
$$
\ell(\boldsymbol{\beta})=\frac{1}{M} \sum_{i=1}^M\left[\log(h_0^+)+\bX_i^{*\prime} \boldsymbol{\beta}-Y_i^* h_0^+\exp \left(\bX_i^{*\prime} \boldsymbol{\beta}\right) \right]
$$
and we have $\tau\ell(\boldsymbol{\beta})=\log  \pi_{cox,cat}(\boldsymbol{\beta}\mid \tau)$.
Based on the  proof in the previous section, we have 
\begin{equation}\label{proof:loglikee_upper_bound}
    \exp (\ell(\boldsymbol{\beta})) \leq C_1 \exp(C_q) \exp \left(-c_1\|\boldsymbol{\beta}\|\right), \quad \forall \boldsymbol{\beta} \neq 0
\end{equation}
Recall $\Gamma_{\alpha, \gamma}(\tau)$ is taken as \eqref{prior_tau} of the main text, and we have $\kappa\geq \ell(\boldsymbol{\beta})$. By Tonelli's theorem, for any $\alpha^{\prime} \in(0, \alpha)$,
$$
\int_{\tau>0} \int_{\boldsymbol{\beta} \in \mathbb{R}^p} \|\boldsymbol{\beta}\|^{\alpha^{\prime}}\tau^{p+\alpha-1} \exp \left(-\left(\kappa+\frac{1}{\gamma}\right) \tau+\tau \ell(\boldsymbol{\beta})\right) d \boldsymbol{\beta} d \tau=\int_{\boldsymbol{\beta} \in \mathbb{R}^p} \frac{\|\boldsymbol{\beta}\|^{\alpha^{\prime}}\Gamma(p+\alpha)}{\left(\kappa+\gamma^{-1}-\ell(\boldsymbol{\beta})\right)^{p+\alpha}} d \boldsymbol{\beta}
$$
Split the integral above into two: $\int_{\ell(\boldsymbol{\beta}) \leq \kappa-c_1\|\boldsymbol{\beta}\| / 2}$ and $\int_{\ell(\boldsymbol{\beta})>\kappa-c_1\|\boldsymbol{\beta}\| / 2}$. We will separately bound the two integrals (without the constant term $\Gamma(p+\alpha)$ there).
For the first integral, we have
$$
\begin{aligned}
& \int_{\ell(\boldsymbol{\beta}) \leq \kappa-c_1\|\boldsymbol{\beta}\| / 2}\frac{\|\boldsymbol{\beta}\|^{\alpha^{\prime}}}{(\kappa+1 / \gamma-\ell(\boldsymbol{\beta}))^{p+\alpha}} d \boldsymbol{\beta} \\
\leq & \int_{\ell(\boldsymbol{\beta}) \leq \kappa-c_1\|\boldsymbol{\beta}\| / 2}\frac{\|\boldsymbol{\beta}\|^{\alpha^{\prime}}}{\left(1 / \gamma+c_1\|\boldsymbol{\beta}\| / 2\right)^{p+\alpha}} d \boldsymbol{\beta} \leq \int_{\mathbb{R}^p} \frac{\|\boldsymbol{\beta}\|^{\alpha^{\prime}}}{\left(1 / \gamma+c_1\|\boldsymbol{\beta}\| / 2\right)^{p+\alpha}} d \boldsymbol{\beta},
\end{aligned}
$$
where the last integral is finite by elementary calculus.

For the second integral, we have
$$
\begin{aligned}
& \int_{\ell(\boldsymbol{\beta})>\kappa-c_1\|\boldsymbol{\beta}\| / 2}\frac{\|\boldsymbol{\beta}\|^{\alpha^{\prime}}}{(\kappa+1 / \gamma-\ell(\boldsymbol{\beta}))^{p+\alpha}} d \boldsymbol{\beta} \\
\leq & \int_{\ell(\boldsymbol{\beta})>\kappa-c_1\|\boldsymbol{\beta}\| / 2}\frac{\|\boldsymbol{\beta}\|^{\alpha^{\prime}}}{(1 / \gamma)^{p+\alpha}} d \boldsymbol{\beta} \\
\leq & \int_{\ell(\boldsymbol{\beta})>\kappa-c_1\|\boldsymbol{\beta}\| / 2}\frac{\|\boldsymbol{\beta}\|^{\alpha^{\prime}}}{(1 / \gamma)^{p+\alpha}} \exp \left(\ell(\boldsymbol{\beta})+c_1\|\boldsymbol{\beta}\| / 2-\kappa\right) d \boldsymbol{\beta} \\
\leq & C_1 \exp(C_q)\int_{\ell(\boldsymbol{\beta})>\kappa-c_1\|\boldsymbol{\beta}\| / 2} \gamma^{p+\alpha} \|\boldsymbol{\beta}\|^{\alpha^{\prime}} e^{-\kappa}\exp \left(-c_1\|\boldsymbol{\beta}\|+c_1\|\boldsymbol{\beta}\| / 2\right) d \boldsymbol{\beta} \\
\leq & C_1\exp(C_q) \int_{\boldsymbol{\beta} \in \mathbb{R}^p} \gamma^{p+\alpha} \|\boldsymbol{\beta}\|^{\alpha^{\prime}} e^{-\kappa}\exp \left(-c_1\|\boldsymbol{\beta}\| / 2\right) d \boldsymbol{\beta},
\end{aligned}
$$
where the first inequality is because by definition of $\kappa$ it is no less than $\ell(\boldsymbol{\beta})$, the second inequality is due to the fact that $\ell(\boldsymbol{\beta})+c_1\|\boldsymbol{\beta}\| / 2-\kappa \geq 0$ in the domain of the integral, and the third inequality is due to \eqref{proof:loglikee_upper_bound}. The last expression is finite due to its exponential tail.

\section{Proof of  Theorem ~\ref{thm:consistency} }
\label{supp:consistency_proof}

We first introduce some notations and conditions. 

We define $N_i(t)=I\left\{T_i \leq t, T_i \leq C_i\right\}$, $\bar{N}(t)=\sum_{i=1}^n N_i(t)$ and $\Lambda_i(t)=I\left\{T_i \geq t, C_i \geq t\right\}$.  
 We assume the both survival time and censoring time are bounded by $\xi_U$.
 For a vector $v \in \mathbb{R}^p$, we use $v^{\otimes r}:=v \otimes v \otimes \cdots \otimes v \in \bigotimes^r \mathbb{R}^p$ to denote its $r$-th tensor power.
 In particular, $v^{\otimes 0}=1$, $v^{\otimes 1}=v$, and $v^{\otimes 2}=v v^T$. 
 For $\ell=0,1,2$, we define
 $$
S_n^{(\ell)}(\boldsymbol{\beta}, t):=n^{-1} \sum_{i=1}^n \Lambda_i(t)\left\{\mathbf{X}_i\right\}^{\otimes \ell} \exp \left(\boldsymbol{\beta}^T \mathbf{X}_i\right), ~~ \boldsymbol{\beta}\in \mathbb{R}^p, t\in [0, \xi_{U}].
$$

 In the following, we denote the log partial likelihood by $Q_n(\boldsymbol{\beta})$. 
 Using the counting process notation, we can rewrite it as
\begin{equation}\label{eq: lpl-counting process}
Q_n(\boldsymbol{\beta})=\sum_{i=1}^n \int_0^{\xi_U}\left\{\boldsymbol{\beta}^T \mathbf{X}_i-\log \left(S_n^{(0)}(\boldsymbol{\beta}, t)\right)\right\} d N_i(t), ~~ \boldsymbol{\beta}\in \mathbb{R}^p. 
\end{equation}

We are considering the regime where $p$ may grow along with $n$, and the 
the true regression coefficient $\boldsymbol{\beta}_0$ should be viewed as a sequence of vectors in $\mathbb{R}^p$. 
The cumulative baseline hazard function is fixed and is denoted by $H_0(\cdot)$.

To study the asymptotics, we require the following regularity conditions.

\begin{condition}\label{condition:S_converge_s}
There exists a sequence of compact neighborhood $\mathcal{B}$ of $\boldsymbol{\beta}_0$ such that all the following statements hold:
\begin{enumerate}
	\item   There exist three sequences of scalar, vector, and matrix functions $s^{(j)}$ defined on $\mathcal{B} \times$ $[0, \xi_U]$ such that 
  for $j=0,1,2$,   as $n \rightarrow \infty$, 
$$
\sup_{t \in [0, \xi_U], \boldsymbol{\beta} \in \mathcal{B}} \| S_n^{(j)}(\boldsymbol{\beta}, t) - s^{(j)}(\boldsymbol{\beta}, t) \|_2 \to 0, \text{in probability.}
$$

    \item  There exist a positive constant $c_0$ such that for $j=0,1,2$,   $||s^{(j)}||_2$  is bounded by $c_0$ and $s^{(0)}$ is lower bounded by $1/c_0$ on $\mathcal{B} \times[0, \xi_U]$; for $j=0,1,2$, the family of functions $s^{(j)}(\cdot, t), 0 \leq t \leq \xi_U$, is an equicontinuous family at $\boldsymbol{\beta}_0$.
\item 
For any $\boldsymbol{\beta}\in \mathbb{R}^p$ and $t\in [0, \xi_{U}]$,
let   $\mathbf{e}(\boldsymbol{\beta}, t)=s^{(1)}(\boldsymbol{\beta}, t) / s^{(0)}(\boldsymbol{\beta}, t), \mathbf{v}(\boldsymbol{\beta}, t)=s^{(2)}(\boldsymbol{\beta}, t) / s^{(0)}(\boldsymbol{\beta}, t)-\{\mathbf{e}(\boldsymbol{\beta}$, $t)\}^{\otimes 2}$ and $\boldsymbol{\Sigma}_\bbeta(t)=\int_0^t \mathbf{v}(\boldsymbol{\beta}, u) s^{(0)}\left(\boldsymbol{\beta}_0, u\right) d H_0(u)$.  
Let  $\boldsymbol{\Sigma}_{\bbeta}=\boldsymbol{\Sigma}_{\bbeta}(\xi_U)$. Assume that the $p \times p$ matrix $\boldsymbol{\Sigma}_{\bbeta_0}$ is positive definite for all $n$ and $H_0(\xi_U)<\infty$. 
\item There exists a positive constant $r_0$ such that $\mathcal{B}\supset\left\{\boldsymbol{\beta}:\|\boldsymbol{\beta}-\boldsymbol{\beta}_0\|<r_0\right\}$. 
\end{enumerate}
\end{condition}

Condition \ref{condition:S_converge_s} is a standard condition for deriving asymptotic results in Cox models. 
Similar conditions have been considered in Section 8.2 of \cite{fleming2005counting} for cases with fixed $p$ and in \cite{cai2005variable,bradic2011regularization} for cases with diverging $p$. 

\begin{condition}\label{condition:popu_hessian_eigen}
There exists a positive constant $c_1$ such that operator norm of $\boldsymbol{\Sigma}_{\bbeta_0}$ and $\boldsymbol{\Sigma}_{\bbeta_0}^{-1}$ are bounded by $c_1$.

\end{condition}

Condition \ref{condition:popu_hessian_eigen} is a refinement of the positive definiteness of $\boldsymbol{\Sigma}_{\bbeta_0}$ in Condition~\ref{condition:S_converge_s}. It imposes a bound on the eigenvalues of the information matrix so that the condition number can be controlled. 
This condition is also imposed by Condition A3 of \cite{cai2005variable}. 

\begin{condition}\label{condition:gradient_truth_Op}
    $E\left\{\sup _{0 \leq t \leq \xi_U} \Lambda(t)\|\mathbf{X}\|_2^2 \exp \left(\boldsymbol{\beta}_0^{ \top } \mathbf{X}\right)\right\}=O(p)$.  Furthermore, there exists a positive constant $c_2$ such that 
    $\|\boldsymbol{\beta}_0\|_2\leq c_2$. 
\end{condition}

The first part of Condition \ref{condition:gradient_truth_Op} controls the tail behavior of the covariates in terms of $p$. 
Similar conditions can be seen in \cite{andersen1982cox,cai2005variable,bradic2011regularization}. 
The second part of Condition \ref{condition:gradient_truth_Op} requires as $n,p$ diverges,  $\|\boldsymbol{\beta}_0\|$ remains bounded.

\begin{condition}\label{conditions:synthetic_X_Y}
    The synthetic data are i.i.d. copies of $(\boldsymbol{X}^*, Y^*)$ such that the statements hold: 
	\begin{itemize}
	    \item The synthetic covariate vector $\boldsymbol{X}^*=\left(X_1^*, X_2^*, \cdots, X_p^*\right)$ satisfies (1)  $\mathbb{E} X_j^*=0$, $\operatorname{Var}\left(X_j^*\right)=1$, and $\left|X_j^*\right| \leq B_1$, a.s., for $j=1, \cdots, p$;
     (2) $X_2^*, \cdots, X_p^*$ are independent. 
 
		\item For the synthetic response $Y^*$, there are two positive constants $\xi_L$ and $\xi_U$ such that $\xi_L\leq Y^* \leq \xi_U$.
	\end{itemize}
\end{condition}

Condition~\ref{conditions:synthetic_X_Y}  is mild. The first two requirement on $\boldsymbol{X}^*$ ensures that the coordinates are standardized and bounded, and the second one ensures the coordinates are independent. 
These requirements  are automatically satisfied if the coordinates of $\boldsymbol{X}^*$ are resampled independently from the coordinates of observed covariate data (and historical data if available). 
Since synthetic data generation is fully under our control, Condition~\ref{conditions:synthetic_X_Y} can always be met. A similar assumption is also considered in \cite{huang_catalytic_2020}.

We first present Theorem \ref{thm:est_error_CRE}, which establishes the consistency of CRE. The corresponding result for WME is provided in Theorem \ref{thm:est_error_WME}.

\begin{theorem} \label{thm:est_error_CRE}
Suppose Conditions  \ref{condition:S_converge_s}, \ref{condition:popu_hessian_eigen}, \ref{condition:gradient_truth_Op}, and \ref{conditions:synthetic_X_Y} hold, and suppose $p^2/n\rightarrow 0$. If we choose $\tau\leq c_1p$, then we have 
$$\|\widehat{\boldsymbol{\beta}}_{CR,\tau}-\bbeta_0\|_2^2=O_p\left(\frac{p}{n}\right), $$
where $\widehat{\boldsymbol{\beta}}_{CR,\tau}=\arg\max_{\bbeta}  Q_n(\bbeta)+\frac{\tau}{M}\sum_{i=1}^M \left[\bX_i^{*\prime} \boldsymbol{\beta}-Y_i^* h_0^+\exp \left(\bX_i^{*\prime} \boldsymbol{\beta}\right) \right]$.
\end{theorem}

\begin{remark}
We can also consider an estimator based on an infinite amount of synthetic data:
    $$\widehat{\boldsymbol{\beta}}_{\infty} :=\arg\max_{\bbeta}  Q_n(\bbeta)+ \tau\mathbb{E}\left( \bX^{*\top}\bbeta-Y^*h_0^+\exp(\bX^{*\top}\bbeta)\right).$$
    Follow the same arguments used in proof of Theorem \ref{thm:est_error_CRE}, we can show that $\|\widehat{\boldsymbol{\beta}}_{\infty}-\bbeta_0\|_2^2=O_p\left(\frac{p}{n}\right) $. 
\end{remark}

\begin{proof}[Proof of Theorem \ref{thm:est_error_CRE}]
Using the counting process representation of $Q_n$ in \eqref{eq: lpl-counting process},  the penalized log partial likelihood for CRE can be written as
 $$\mathcal{C}(\boldsymbol{\beta}, \xi_U) \equiv \sum_{i=1}^n \int_0^{\xi_U}\left\{\boldsymbol{\beta}^T \mathbf{X}_i-\log \left(S_n^{(0)}(\boldsymbol{\beta}, t)\right)\right\} d N_i(t)+\frac{\tau}{M}\log L\left(\boldsymbol{\beta}, h_0^+ \mid \left\{\left(\bX_i^*, Y_i^*\right)\right\}_{i=1}^M\right).$$

 It suffices to show that, for any $\varepsilon>0$, there exists a large constant $B$ such that with  $\gamma_n=B \sqrt{\frac{p}{n}}$, it holds that  
\begin{equation}\label{eq: CRE consistency probability}
    P\left\{\sup _{\|\mathbf{u}\|_2=1} \mathcal{C}\left(\boldsymbol{\beta}_0+\gamma_n \mathbf{u}, \xi_U\right)<\mathcal{C}\left(\boldsymbol{\beta}_0, \xi_U\right)\right\} \geq 1-\varepsilon
    \end{equation}
for sufficiently large $n$. 
This is because when the event in the probability holds, there exists a local maximizer $\widehat \bbeta$ of $\mathcal{C}(\boldsymbol{\beta}, \xi_U) $ in $\{\boldsymbol{\beta}:\|\boldsymbol{\beta}-\boldsymbol{\beta}_0\|< \gamma_n\}$, which is also the global maximizer since $\mathcal{C}(\boldsymbol{\beta}, \xi_U)$ is strictly convex. 

By the Taylor expansion of $\mathcal{C}\left(\boldsymbol{\beta}_0+ t \mathbf{u}, \xi_U\right)$ at $t=0$,
$$
\begin{aligned}
\mathcal{C}\left(\boldsymbol{\beta}_0+\right. & \left.\gamma_n \mathbf{u}, \xi_U\right)-\mathcal{C}\left(\boldsymbol{\beta}_0, \xi_U\right) \\
= & {Q}_n(\bbeta_0+\gamma_n \mathbf{u})- {Q}_n(\bbeta_0 ) \\
& \quad+\frac{\tau}{M}\log L\left(\boldsymbol{\beta}_0+\gamma_n \mathbf{u}, h_0^+ \mid \left\{\left(\bX_i^*, Y_i^*\right)\right\}_{i=1}^M\right)-\frac{\tau}{M}\log L\left(\boldsymbol{\beta}_0, h_0^+ \mid \left\{\left(\bX_i^*, Y_i^*\right)\right\}_{i=1}^M\right)\\
= & \mathbf{u}^T U_n\left(\boldsymbol{\beta}_0\right) \gamma_n+\frac{1}{2}\gamma_n^2 \mathbf{u}^T \partial U_n\left(\boldsymbol{\beta}_0\right) \mathbf{u}+r_n\left(\widetilde{\boldsymbol{\beta}} \right) \\
& \quad+\frac{\tau}{M}\log L\left(\boldsymbol{\beta}_0+\gamma_n \mathbf{u}, h_0^+ \mid \left\{\left(\bX_i^*, Y_i^*\right)\right\}_{i=1}^M\right)-\frac{\tau}{M}\log L\left(\boldsymbol{\beta}_0, h_0^+ \mid \left\{\left(\bX_i^*, Y_i^*\right)\right\}_{i=1}^M\right),
\end{aligned}
$$
where 
$
U_n\left(\boldsymbol{\beta}\right)=\frac{\partial Q_n\left(\boldsymbol{\beta}\right)}{\partial \bbeta}=\sum_{i=1}^n \int_0^{\xi_U}\left\{\mathbf{X}_i-\mathbf{e}_n \left(\boldsymbol{\beta}, t\right)\right\} d N_i(t) 
$,
and  the remainder term $r_n\left(\widetilde{\boldsymbol{\beta}}\right)$ is equal to
\begin{equation}\label{eq: CRE consistency PL}
\frac{1}{6} \sum_{j, k,l}\left(\tilde{\beta}_{ j}-\tilde{\beta}_{ j}^*\right)\left(\tilde{\beta}_{ k}-\tilde{\beta}_{ k}^*\right)\left(\tilde{\beta}_{ \ell}-\tilde{\beta}_{\ell}^*\right) \frac{\partial^2 U_{n \ell}\left(\widetilde{\boldsymbol{\beta}}\right)}{\partial \tilde{\beta}_{j} \partial \tilde{\beta}_{ k}}
\end{equation}
with $U_{n \ell}$ being the $\ell$ th component of $U_n$ and $\widetilde{\boldsymbol{\beta}}$ lying between $\boldsymbol{\beta}_0+\gamma_n \mathbf{u}$ and $\bbeta_0$. Under Conditions \ref{condition:S_converge_s}, \ref{condition:popu_hessian_eigen}, and \ref{condition:gradient_truth_Op}, the proof of Theorem 4.2 of  \cite{bradic2011regularization} shows that there exist a constant $c>0$ such that  
\begin{align*}
    \sup _{\|\mathbf{u}\|_2=1} \left[\mathbf{u}^T U_n\left(\boldsymbol{\beta}_0\right) \gamma_n+\frac{1}{2}\gamma_n^2 \mathbf{u}^T \partial U_n\left(\boldsymbol{\beta}_0\right) \mathbf{u}+r_n\left(\widetilde{\boldsymbol{\beta}} \right)\right]
    & 
    < n\gamma_n\left \{O_p(\sqrt{p/n})-c\gamma_n(1+o_p(1)) \right\} \\
    & = Bp\left \{O_p(1)-cB(1+o_p(1)) \right\}, 
\end{align*}
which is negative if $B$ is chosen properly. 

Recall the condition that $\tau\leq c_4 p$. 
In order to establish  $\sup _{\|\mathbf{u}\|_2=1} \left[ \mathcal{C}\left(\boldsymbol{\beta}_0+\gamma_n \mathbf{u}, \xi_U\right)-\mathcal{C}\left(\boldsymbol{\beta}_0, \xi_U\right)\right]<0$ for large $n$, it suffices to bound
\begin{equation}\label{eq: CRE consistency prior}
    \sup _{\|\mathbf{u}\|_2=1}\frac{1}{M}\left[\log L\left(\boldsymbol{\beta}_0+\gamma_n \mathbf{u}, h_0^+ \mid \left\{\left(\bX_i^*, Y_i^*\right)\right\}_{i=1}^M\right)-\frac{1}{M}\log L\left(\boldsymbol{\beta}_0, h_0^+ \mid \left\{\left(\bX_i^*, Y_i^*\right)\right\}_{i=1}^M\right)\right].
\end{equation}

By the condition that $p^2/n\rightarrow 0$, when $n$ is large enough, it holds that  $\sqrt{p}\gamma_n\leq 1$ and we have:
\begin{align*}
  &\quad  \sup _{\|\mathbf{u}\|_2=1} \left[\frac{1}{M}\log L\left(\boldsymbol{\beta}_0+\gamma_n \mathbf{u}, h_0^+ \mid \left\{\left(\bX_i^*, Y_i^*\right)\right\}_{i=1}^M\right)-\frac{1}{M}\log L\left(\boldsymbol{\beta}_0, h_0^+ \mid \left\{\left(\bX_i^*, Y_i^*\right)\right\}_{i=1}^M\right)\right]\\
    &= \sup _{\|\mathbf{u}\|_2=1}\frac{1}{M}\sum_{i=1}^M \left\{ h_0^+ Y_i^*\exp(\bX_i^{*\top}\bbeta_0)[\exp(\gamma_n \bX_i^{*\top} \mathbf{u})-1]  -\gamma_n \bX_i^{*\top} \mathbf{u}\right\}\\
    & \leq   \frac{1}{M}\sum_{i=1}^M \left\{ h_0^+ Y_i^*\exp(\bX_i^{*\top}\bbeta_0)[\exp(\max |X^*_{i,j}|)-1]\right\}+ \gamma_n \left\| \frac{1}{M}\sum_{i=1}^M \bX_i^*\right\|\\
    &\leq \tilde{C} \frac{1}{M}\sum_{i=1}^M    \exp(\bX_i^{*\top}\bbeta_0) +\frac{1}{\sqrt{p}}  \left\| \frac{1}{M}\sum_{i=1}^M \bX_i^*\right\|\\
    &=O_p(1),
\end{align*}
where $O_p(1)$ in the last step does not depend on $B$. 
Given any $\epsilon$, suppose for large $n$, 
(1) with probability $1-\epsilon/2$ , the term $O_p(1)$ in \eqref{eq: CRE consistency PL} is bounded by $C_{\epsilon/2}$ and $o_p(1)$ is bounded by $1/4$, and (2) with probability $1-\epsilon/2$, \eqref{eq: CRE consistency prior} is bounded by $C_{\epsilon/2}^\prime$. 
If we choose a value of $B$ so that $cB>2C_{\epsilon/2}$ and $Bc/c_4>2C_{\epsilon/2}^\prime$, then we have $\sup _{\|\mathbf{u}\|_2=1} \left[ \mathcal{C}\left(\boldsymbol{\beta}_0+\gamma_n \mathbf{u}, \xi_U\right)-\mathcal{C}\left(\boldsymbol{\beta}_0, \xi_U\right)\right]<0$. Therefore,  we complete the proof of \eqref{eq: CRE consistency probability}.

\end{proof}

The next theorem establishes the consistency of WME.

\begin{theorem} \label{thm:est_error_WME}
Suppose Conditions  \ref{condition:S_converge_s}, \ref{condition:popu_hessian_eigen}, \ref{condition:gradient_truth_Op}, and \ref{conditions:synthetic_X_Y} hold, and suppose $p^2/n\rightarrow 0$. If we choose $\tau\leq c_4 p$, then we have  
$$\|\widehat{\boldsymbol{\beta}}_{WM,\tau}-\bbeta_0\|_2^2=O_p\left(\frac{p}{n}\right). $$
\end{theorem}

\begin{proof}[Proof of Theorem \ref{thm:est_error_WME}]

We first extend the counting process notation to the synthetic data. 
Recall that all synthetic data are uncensored. 
Define $N^*_i(t)=I\left\{Y_i^* \leq t\right\}$, 
and $\Lambda_i^*(t)=I\left\{Y_i^* \geq t\right\}$. 
Then the objective function of WME in \eqref{supp:mix_optimization} can be written as:  
\begin{align*}
    \tilde{\mathcal{C}} (\bbeta,\xi_U)&=\tilde{Q}_n(\bbeta)+\tau \tilde{P}_n(\bbeta),
\end{align*}
where
\begin{align*}
    &\tilde{Q}_n(\bbeta):=\sum_{i=1}^n \int_0^{\xi_U}\boldsymbol{\beta}^{\top} \bX_i-\log \left\{ {\frac{1}{n}\sum_{j=1}^n \Lambda_j(t) e^{\boldsymbol{\beta}^{\top} \bX_j}+\frac{\tau}{nM}\sum_{j=n+1}^{n+M} \Lambda_j^*(t) e^{\boldsymbol{\beta}^{\top} \bX_j^*}}\right\} d N_i(t).\\
    &\tilde{P}_n(\bbeta):=\frac{1}{M}\sum_{i=n+1}^{n+M} \int_0^{\xi_U}\boldsymbol{\beta}^{\top} \bX^*_i-\log \left\{ {\frac{1}{n}\sum_{j=1}^n \Lambda_j(t) e^{\boldsymbol{\beta}^{\top} \bX_j}+\frac{\tau}{nM}\sum_{j=n+1}^{n+M} \Lambda_j^*(t) e^{\boldsymbol{\beta}^{\top} \bX_j^*}}\right\} d N_i^*(t).
\end{align*}
Following the same argument in the proof of Theorem~\ref{thm:est_error_CRE}, it suffices to show that for any $\varepsilon>0$, there exists a large constant $B$ and $\gamma_n=B \sqrt{\frac{p}{n}}$ such that
$$
P\left\{\sup _{\|\mathbf{u}\|_2=1} 
 \tilde {\mathcal{C}}\left(\boldsymbol{\beta}_0+\gamma_n \mathbf{u}, \xi_U\right)< \tilde{\mathcal{C}}\left(\boldsymbol{\beta}_0, \xi_U\right)\right\} \geq 1-\varepsilon, 
$$
for large enough $n$. 
To prove this, we can use a similar argument in the proof of Theorem~\ref{thm:est_error_CRE} if we can show 
\begin{enumerate}
    \item [Part (1).] $\sup_{\|\mathbf{u}\|=1}\left\{ \tilde{Q}_n(\bbeta_0+\gamma_n \mathbf{u})-\tilde{Q}_n(\bbeta_0 ) - \left[  {Q}_n(\bbeta_0+\gamma_n \mathbf{u})- {Q}_n(\bbeta_0 )\right]\right\}=O_p(p)$, and
    \item [Part (2).] 
    $\sup_{\|\mathbf{u}\|=1}\left[\tilde{P}_n(\bbeta_0+\gamma_n \mathbf{u})-\tilde{P}_n(\bbeta_0)\right]=O_p(1) $, 
\end{enumerate}
where the constants in the terms $O_p(p)$ and $O_p(1)$ do not depend on the value of $B$.

\textbf{Proof of Part (1):}
In the proof of consistency of CRE, we have showed that 
$$ {Q}_n(\bbeta_0+\gamma_n \mathbf{u})- {Q}_n(\bbeta_0 )=\sum_{i=1}^n\int_0^{\xi_U}\left( \gamma_n\mathbf{u}^{\top} \bX_i-\log \left\{\frac{\frac{1}{n}\sum_{j=1}^n \Lambda_j(t) e^{(\boldsymbol{\beta}_0+\gamma_n\mathbf{u})^{\top} \bX_j} }{\frac{1}{n}\sum_{j=1}^n \Lambda_j(t) e^{\boldsymbol{\beta}_0^{\top} \bX_j} }\right\} \right)d N_i(t).$$
Based on definition of $\tilde{Q}_n$, we have
\begin{align*}
    \tilde{Q}_n(\bbeta_0+\gamma_n \mathbf{u})-\tilde{Q}_n(\bbeta_0 )&=\sum_{i=1}^n  \int_0^{\xi_U}\left( \gamma_n\mathbf{u}^{\top} \bX_i-\right.\\
    &\left.\log \left\{\frac{\frac{1}{n}\sum_{j=1}^n \Lambda_j(t) e^{(\boldsymbol{\beta}_0+\gamma_n\mathbf{u})^{\top} \bX_j}+\frac{\tau}{nM}\sum_{j=n+1}^{n+M} \Lambda_j^*(t) e^{(\boldsymbol{\beta}_0+\gamma_n\mathbf{u})^{\top} \bX_j^*}}{\frac{1}{n}\sum_{j=1}^n \Lambda_j(t) e^{\boldsymbol{\beta}_0^{\top} \bX_j}+\frac{\tau}{nM}\sum_{j=n+1}^{n+M} \Lambda_j^*(t) e^{\boldsymbol{\beta}_0^{\top} \bX_j^*}}\right\} \right)d N_i(t).
\end{align*}

It then follows that 
\begin{align*}
     &\tilde{Q}_n(\bbeta_0+\gamma_n \mathbf{u})-\tilde{Q}_n(\bbeta_0 )= {Q}_n(\bbeta_0+\gamma_n \mathbf{u})- {Q}_n(\bbeta_0 )+\int_0^{\xi_U}\log\left\{\frac{A_{n,u}}{A_n}\frac{(A_n+A^*_{n})}{(A_{n,u}+A^*_{n,u})}\right\}d \bar{N}(t),
\end{align*}
where we have used the following shorthand notation: 
\begin{align}\nonumber
    &A_{n,u}:=A_{n,u}(t)=\frac{1}{n}\sum_{j=1}^n \Lambda_j(t) e^{(\boldsymbol{\beta}_0+\gamma_n\mathbf{u})^{\top} \bX_j},\\
    &A_n:=A_n(t)=\frac{1}{n}\sum_{j=1}^n \Lambda_j(t) e^{\boldsymbol{\beta}_0^{\top} \bX_j}, \nonumber \\
    &A^*_{n,u}:=A^*_{n,u}(t)=\frac{\tau}{nM}\sum_{j=n+1}^{n+M} \Lambda_j^*(t) e^{(\boldsymbol{\beta}_0+\gamma_n\mathbf{u})^{\top} \bX_j^*}, \nonumber\\
    &A^*_{n}:=A^*_{n}(t)=\frac{\tau}{nM}\sum_{j=n+1}^{n+M} \Lambda_j^*(t) e^{\boldsymbol{\beta}_0^{\top} \bX_j^*}. \label{eq: shorthand A}
\end{align}

Since $\|\bbeta_0\|_2\leq c_2$ and all entries of the synthetic covariate vectors are bounded and independent, one can easily use Chebyshev's inequality and Hoeffding's lemma to show that 
\begin{equation}\label{eq: sample mean of synthetic exp beta0}
\frac{1}{M}\sum_{j=n+1}^{n+M} e^{\boldsymbol{\beta}_0^{\top} \bX_j^*}=O_p(1), 
\end{equation}
where the $O_p$-term does not depend on $B$. 
Furthermore, under the condition $\tau\leq c_4 p$, we have
$$\sup_{t\in [0,\xi_U]}A^*_n(t)\leq \frac{\tau}{nM}\sum_{j=n+1}^{n+M} e^{\boldsymbol{\beta}_0^{\top} \bX_j^*}
=O_p\left(\frac{p}{n}\right).
$$

Statement 2 of Condition~\ref{condition:S_converge_s} implies that $\inf_{t\in [0, \xi_U]} A_n(t) \geq (2c_0)^{-1}>0$ holds with probability tending to 1.  
Using the elementary inequality that $\log(t)\leq t-1$ for any $t>0$, we have
 \begin{align*}
    \log \left\{\frac{A_{n,u}}{A_n}\frac{(A_n+A^*_{n})}{(A_{n,u}+A^*_{n,u})} \right\} &\leq \frac{A_{n,u}}{A_n}\frac{(A_n+A^*_{n})}{(A_{n,u}+A^*_{n,u})}-1\leq \frac{A^*_{n}}{A_n}=O_p\left(\frac{p}{n}\right),
 \end{align*}
 where the $O_p$-term does not depend on $B$ or $\mathbf{u}$ because both $A_n$ and $A_n^*$ does not depend on $\gamma_n$ and $\mathbf{u}$. 
 We then conclude that
\begin{equation}
  \sup_{\|\mathbf{u}\|=1}  \int_0^{\xi_U}\log\left\{\frac{A_{n,u}}{A_n}\frac{(A_n+A^*_{n})}{(A_{n,u}+A^*_{n,u})}\right\}d \bar{N}(t)=O_p(p),
\end{equation}
and the proof of Part (1) is completed.

\textbf{Proof of Part (2):} 
Using the shorthand in \eqref{eq: shorthand A}, we can write 
\begin{align*}
 \tilde{P}_n(\bbeta_0+\gamma_n \mathbf{u})-\tilde{P}_n(\bbeta_0)&=\frac{1}{M}\sum_{i=n+1}^{n+M} \int_0^{\xi_U}\left(\gamma_n\mathbf{u}^{\top} \bX^*_i+ \log \left\{ \frac{(A_n+A^*_{n})}{(A_{n,u}+A^*_{n,u})} \right\}\right)dN_i(t).
\end{align*}
By Condition~\ref{conditions:synthetic_X_Y}, $|\mathbf{u}^{\top} \bX^*_i|\leq \sqrt{p}B_1$. 
By the condition that $p^2/n\rightarrow 0$, when $n$ is large enough, it holds that $\sqrt{p}\gamma_n\leq 1$. 
It follows that 
$$\frac{1}{M}\sum_{i=n+1}^{n+M} \int_0^{\xi_U}|\gamma_n\mathbf{u}^{\top} \bX^*_i|dN_i(s)\leq \sqrt{p}\gamma_n B_1\leq B_1.$$

Next, it suffices to show \begin{equation}\label{eq:WME log ratio bound}
\sup_{t\in [0, \xi_U], \|\mathbf{u}\|=1}
\frac{1}{M}\sum_{i=n+1}^{n+M} \int_0^{\xi_U}  \log \left\{ \frac{(A_n+A^*_{n})}{(A_{n,u}+A^*_{n,u})} \right\} dN_i(t)=O_p(1). 
\end{equation}
Following the same argument in the proof of Part (1), we have
\begin{equation}\label{eq:WME log ratio decompose}
\log\left\{ \frac{(A_n+A^*_{n})}{(A_{n,u}+A^*_{n,u})} \right\}\leq  \frac{(A_n+A^*_{n})}{(A_{n,u}+A^*_{n,u})}  -1\leq \frac{\left|A_n-A_{n,u}+A_n^*-A^*_{n,u}\right|}{A_{n,u} }.\end{equation}

When $\gamma_n<r_0$ for $r_0$ in Statement 4 of Condition~\ref{condition:S_converge_s}, we have 
\begin{equation}\label{eq: WME neigh in B}
\{\bbeta_0+\gamma_n\mathbf{u}:\|\mathbf{u}\|=1\}\subset \mathcal{B}. 
\end{equation}
Consequently, Statements 1 and 2 of Condition \ref{condition:S_converge_s} imply that  
\begin{equation}\label{eq: lower bound A_nu}
  \inf_{t\in [0, \xi_U], \|\mathbf{u}\|=1}A_{n,u}(t)\geq 1/(2c_0)
\end{equation}
with probability tending to 1.

In view of \eqref{eq: WME neigh in B}, we can use Statements 1 and 2 of Condition \ref{condition:S_converge_s} to show that 
\begin{equation}\label{eq:An_minus_A_nu}
   \sup_{t\in [0, \xi_{U}],\|\mathbf{u}\|=1}\left| A_{n}-A_{n,u}\right|=\sup_{t\in [0, \xi_{U}\,\|\mathbf{u}\|=1}\left| \frac{1}{n}\sum_{j=1}^n \Lambda_j(t) \left(e^{ \boldsymbol{\beta}_0  ^{\top} \bX_j}-e^{(\boldsymbol{\beta}_0+\gamma_n\mathbf{u})^{\top} \bX_j}\right) \right|\leq 4c_0, 
\end{equation}
with probability tending to 1. 

We rewrite $A_n^*-A^*_{n,u}$ as follows:
\begin{align*}
     A_n^*-A^*_{n,u}&=\frac{\tau}{nM}\sum_{j=n+1}^{n+M} \Lambda_j^*(t) \left(e^{\boldsymbol{\beta}_0 ^{\top} \bX_j^*}-e^{(\boldsymbol{\beta}_0+\gamma_n\mathbf{u})^{\top} \bX_j^*}\right)\nonumber \\
     &= \frac{\tau}{nM}\sum_{j=n+1}^{n+M} \Lambda_j^*(t)e^{\boldsymbol{\beta}_0 ^{\top} \bX_j^*} \left(1-e^{ \gamma_n\mathbf{u}^{\top} \bX_j^*}\right). 
\end{align*}
Recall that for sufficiently large $n$, $\sqrt{p}\gamma_n\leq 1$ and $|\gamma_n\mathbf{u}^{\top} \bX^*_i|\leq B_1$. 
Under the condition $\tau\leq c_4 p$, we have
\begin{align}
      \sup_{t\in [0, \xi_{U}],\|\mathbf{u}\|=1}\left|
      A_n^*-A^*_{n,u}\right| &\leq \frac{c_4 p}{n}\left(e^{ B_1}-1\right) \frac{1}{M}\sum_{j=n+1}^{n+M} e^{\boldsymbol{\beta}_0 ^{\top} \bX_j^*} =O_p\left(\frac{p}{n}\right), \label{eq:An*_minus_Anu*}
\end{align}
where the last step is due to \eqref{eq: sample mean of synthetic exp beta0}.

Putting \eqref{eq: lower bound A_nu}, \eqref{eq:An_minus_A_nu}, and \eqref{eq:An*_minus_Anu*} together, we can conclude \eqref{eq:WME log ratio bound} from \eqref{eq:WME log ratio decompose}. 
Therefore, we have
$$\sup_{\|\mathbf{u}\|_2=1} \left|\tilde{P}_n(\bbeta_0+\gamma_n \mathbf{u})-\tilde{P}_n(\bbeta_0) \right| \leq  B_1+O_p(1)=O_p(1),$$
where the constant in the $O_p$-term does not depend on $B$.  
After proving Part 1 and Part 2, the rest of the proof follows the same argument in the proof of Theorem~\ref{thm:est_error_CRE}. 
\end{proof}

\subsection{A numerical illustration}

We conducted a simple simulation study to illustrate the consistency of the estimators $\hat{\boldsymbol{\beta}}_{CR,\tau}$ and $\hat{\boldsymbol{\beta}}_{WM,\tau}$ for $p=20, p=50$ and $ p=100$. The true parameter $\boldsymbol{\beta}_0\sim 2 \cdot $ Unif($\mathbb{S}^{p-1} $), i.e. $\boldsymbol{\beta}_0$ is uniformly distributed on a sphere with length 2. 
The design matrix $\mathbb{X}$ consisted of independent entries sampled from the standard normal distribution.  

The number of synthetic data was set to $M=400$. The synthetic covariate matrix $\mathbb{X}^*$ was generated via independent resampling from $\mathbb{X}$, and the synthetic response vector $\bY^*$ was generated in the same manner described in Section~\ref{sec:simulation}. 

\begin{figure}[h!]
\centering
	\includegraphics[scale=0.3]{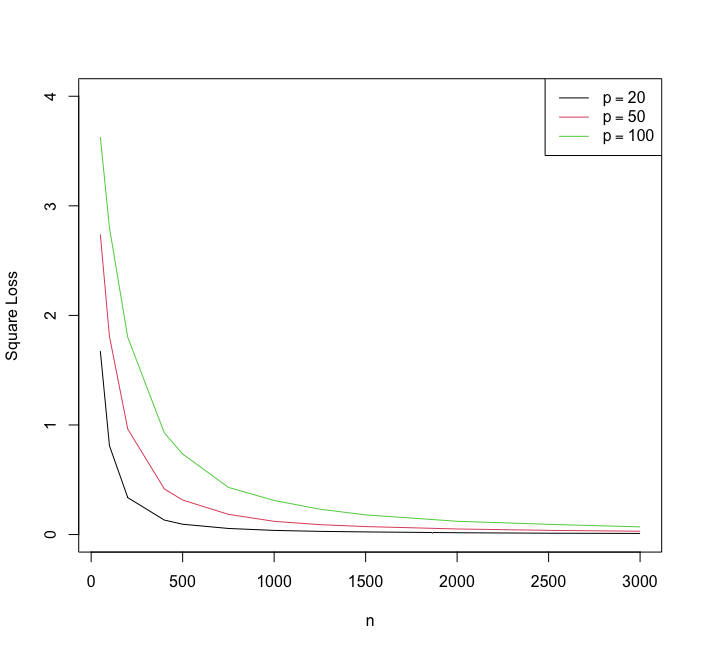}
	\includegraphics[scale=0.3]{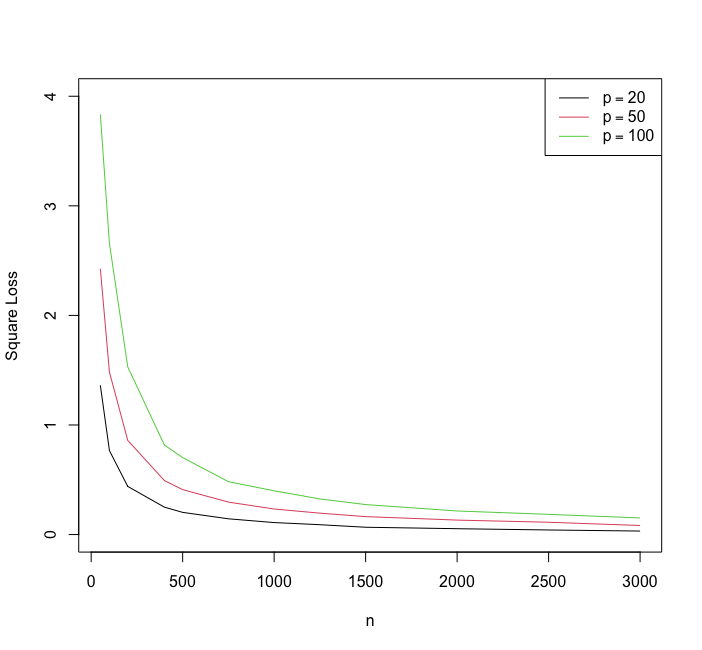}
	\caption{
    Square Loss against sample size for three different dimensions ($p=20,50,100$). 
    Left panel: CRE. Right panel: WME. The results confirm that both estimators consistently estimate the true coefficients.}
	\label{fig:consistency_cat}
\end{figure}

The results of this simulation study are shown in Figure~\ref{fig:consistency_cat}.

\subsection{Stability against the randomness of synthetic data}

We will provide a rigorous statement and proof for the result stated in Theorem~\ref{thm:stability_CR}  in the main text. 

Recall the definition of the CRE (with finite $M$) and its population version (with $M=\infty$): 
\begin{align}
\widehat{\boldsymbol{\beta}}_{CR,\tau}& =\arg\max_{\bbeta}\left[ Q_n(\bbeta)+ \frac{\tau}{M}\sum_{i\leq M} \left(\bX_i^{*\top}\bbeta-h_0^+ Y_i^*\exp(\bX_i^{*\top}\bbeta) \right)\right], \label{obj:M_CR}\\
    \widehat{\boldsymbol{\beta}}_{\infty}& =\arg\max_{\bbeta} \left[Q_n(\bbeta)+ \tau\mathbb{E}\left( \bX^{*\top}\bbeta-h_0^+ Y^*\exp(\bX^{*\top}\bbeta)\right)\right], \label{obj:infty_CR}   
\end{align}
where $Q_n(\bbeta)$ is partial likelihood and the expectation is taken over the random synthetic data. 
For simplicity, we use $ \widehat{\boldsymbol{\beta}}_{M}$ to denote $ \widehat{\boldsymbol{\beta}}_{CR,\tau}$. Theorem \ref{thm:stability_finite_M} provides a upper bound on $\|\widehat{\boldsymbol{\beta}}_{M}-\widehat{\boldsymbol{\beta}}_{\infty}\|^2$. 

\begin{theorem}\label{thm:stability_finite_M}
Suppose that $\tau>0$ and Condition \ref{conditions:synthetic_X_Y} holds. 
If there is some constant $L\geq 1$ such that    $\widehat{\boldsymbol{\beta}}_{\infty}$ lie in $\mathcal B_L=\{\bbeta:\|\bbeta\|_2\leq L\}$, then for any $\eps\in (0,1)$, there exist two positive constants $\tilde B$ and $M_{\eps,p}$ such that for all $M>M_{\eps,p}$, it holds that
$$\mathbb P\left(\|\widehat{\boldsymbol{\beta}}_{M}-\widehat{\boldsymbol{\beta}}_{\infty}\|\leq \sqrt{\frac{\tilde B p}{M}} \right)\geq 1-2\eps.$$
 In particular,   we have $\|\widehat{\boldsymbol{\beta}}_{M}-\widehat{\boldsymbol{\beta}}_{\infty}\|^{2} = O_p 
\left( \frac{p}{M }\right)$. 

\end{theorem}

The proof of this theorem relies on a careful analysis of the Cox catalytic prior density with random synthetic covariates. The following is a useful lemma to capture the curvature of the log density evaluated at $\boldsymbol{\beta}_\infty$.  

\begin{lemma} \label{lemma:smallest_eigen_cat_population_hessian}
Suppose Condition~\ref{conditions:synthetic_X_Y} holds and  $L$ is any positive number. For any $\bbeta \in \mathbb R^p$ with bounded norm $\|\bbeta\|_2\leq L$, it holds that 
$$\mathbb E\left(\exp(\bX^{*\top}\bbeta)\bX^*\bX^{*\top}   \right)\succcurlyeq C_L   \mathbb \mathbb{I}_p ,$$
where $C_L$ is a positive constant that depends on $L$ and the constant $B_1$ in Condition~\ref{conditions:synthetic_X_Y}.
\end{lemma}
The proof of Lemma~\ref{lemma:smallest_eigen_cat_population_hessian} is deferred to the end of this section. We now present the proof of Theorem~\ref{thm:stability_finite_M}. 

\begin{proof}[Proof of Theorem \ref{thm:stability_finite_M}]

To simplify the expressions of the two objective functions in \eqref{obj:M_CR} and \eqref{obj:infty_CR}, we define 
$$
\begin{aligned}
g_1(\bbeta)&=Q_n(\bbeta),  \\
g_2(\bbeta)&=\frac{1}{M}\sum_{i\leq M} \left( \bX_i^{*\top}\bbeta-h_0^+ Y_i^*\exp(\bX_i^{*\top}\bbeta)\right), \\
g_3(\bbeta)&=\mathbb{E}\left( \bX^{*\top}\bbeta-h_0^+ Y^*\exp(\bX^{*\top}\bbeta)\right).   
\end{aligned}
$$

We write the gradient of the objective function in \eqref{obj:M_CR} and \eqref{obj:infty_CR} as 	
	\begin{align*}
		&\nabla F(\boldsymbol{\beta})=\nabla Q_n(\bbeta)+ \tau\nabla g_2(\bbeta).\\
		&\nabla G(\boldsymbol{\beta})=\nabla Q_n(\bbeta)+ \tau\nabla g_3(\bbeta).
	\end{align*}
	
	Then the point estimator  $\widehat{\boldsymbol{\beta}}_{M}$ is the root of $\nabla F(\boldsymbol{\beta})=\boldsymbol{0}$ and $\widehat{\boldsymbol{\beta}}_{\infty}$ is the root of $\nabla G(\boldsymbol{\beta})=\boldsymbol{0}$. Based on \citet[Result 6.3.4]{ortega1970iterative},  it suffices to show that  for any $\epsilon>0$, there is some constant $\tilde B>0$ with $\gamma= \sqrt{\tilde Bp/M}$ such that with probability $1-2\epsilon$, the inequality $\boldsymbol{u} ^{\top} \nabla F(\widehat{\boldsymbol{\beta}}_{\infty}+\gamma\bu)<0$ holds for every unit vector $\bu$.	

We start with the expression for the gradient  $\nabla F$ evaluated at $\widehat{\boldsymbol{\beta}}_{\infty} + \gamma \bu$:
$$
\begin{aligned}
\left\langle \bu,  \nabla F\left(\widehat{\boldsymbol{\beta}}_{\infty} + \gamma\bu\right)\right \rangle = &  \left\langle \bu, \nabla G\left(\widehat{\boldsymbol{\beta}}_{\infty} + \gamma\bu\right) \right \rangle+ \tau \left\langle \bu,\left(\nabla g_{2} - \nabla g_{3}\right)\left(\widehat{\boldsymbol{\beta}}_{\infty} + \gamma\bu\right)\right \rangle \\
&=\underbrace{ \left\langle \bu, \nabla G\left(\widehat{\boldsymbol{\beta}}_{\infty} + \gamma\bu\right)- \nabla G\left(\widehat{\boldsymbol{\beta}}_{\infty}\right) \right \rangle  }_{R_1(\bu) }+  \underbrace{\tau\left\langle \bu, \left(\nabla g_{2} - \nabla g_{3}\right)\left(\widehat{\boldsymbol{\beta}}_{\infty}\right)\right \rangle }_{R_2(\bu) } \\
	& \quad \quad + \underbrace{\tau\left\langle \bu, \left(\nabla g_{2} - \nabla g_{3}\right)\left(\widehat{\boldsymbol{\beta}}_{\infty} + \gamma \bu\right) -  \left(\nabla g_{2} - \nabla g_{3}\right)\left(\widehat{\boldsymbol{\beta}}_{\infty} \right) \right \rangle }_{R_3(\bu) }.
\end{aligned}
$$
We immediately have
$$\sup_{\|\bu\|_2=1} \left\langle \bu,  \nabla F\left(\widehat{\boldsymbol{\beta}}_{\infty} + \gamma\bu\right)\right \rangle \leq \sup_{\|\bu\|_2=1} R_1(\bu)+ \sup_{\|\bu\|_2=1} R_2(\bu)+ \sup_{\|\bu\|_2=1} R_3(\bu).$$

We aim to upper bound each term on the right hand side of the above inequality and show the sum of those upper bounds is negative. 
In the rest of the proof, we assume $M/p^2$ is sufficiently large such that $\gamma \sqrt{p}B_1\leq 1$.

\textbf{For $R_1$}:

By the convexity of $-Q_n(\bbeta)$, it holds that $\nabla^2 Q_n\leq 0$. 
For any $\bu$ with $\|\bu\|_2=1$, we can apply the Taylor's theorem to see that for some $\bar{\bbeta}$  lies between $\widehat{\bbeta}_{\infty}+\gamma \bu$ and $\widehat{\bbeta}_{\infty}$, it holds that 
\begin{align*}
	R_1(\bu)&=\left\langle \bu, \gamma \nabla^2 G\left(\bar{\bbeta}\right)\bu \right \rangle  \\
	&\leq - \gamma\tau h_0^+ \xi_L C_{2L} ,\\
\end{align*}
where the inequality follows from the fact that $\|\bar{\bbeta}\|_2\leq 1+L$ and
Lemma~\ref{lemma:smallest_eigen_cat_population_hessian}. 
It follows that 
\begin{equation}\label{eq:sup_Q1_bound}
	\sup_{\|\bu\|_2=1} R_1(\bu)\leq -\sqrt{\frac{\tilde B p}{M}} \tau h_0^+ \xi_L C_{2L}  .
\end{equation}
 \textbf{For $R_2$}:  Since $\widehat{\bbeta}_{\infty}$ does not depend on the realization of synthetic data, we view $\widehat{\bbeta}_{\infty}$ as fixed in this part. By Cauchy-Schwarz inequality, we have 
 \begin{align*}
 	\sup_{\|\bu\|_2=1}R_2(\bu)&\leq \tau h_0^+ \left\|\frac{1}{M}\sum_{i\leq M} Y_i^*\exp(\bX_i^{*\top}\widehat{\bbeta}_{\infty})\bX_i^* - \mathbb{E}\left(  Y^*\exp(\bX^{*\top}\widehat{\bbeta}_{\infty})\bX^{*}\right)  \right\|_2+\tau \left\|\frac{1}{M}\sum_{i=1}^M \bX_i^*  \right\|.
 \end{align*}

Let $t_{\epsilon}=  4\xi^2B_1^2 \exp\left(\frac{B_1^2 L}{2}\right)/\varepsilon$. 
By Chebyshev's inequality, we have
\begin{align*}
   & \mathbb P\left(\left\|\frac{1}{M}\sum_{i\leq M} Y_i^*\exp(\bX_i^{*\top}\widehat{\bbeta}_{\infty})\bX_i^* - \mathbb{E}\left(  Y^*\exp(\bX^{*\top}\widehat{\bbeta}_{\infty})\bX^{*}\right)  \right\|^2\geq \frac{p}{M}t_{\epsilon}\right)\\
&\leq   \frac{ \sum_{j=1}^p \operatorname{Var}(Y^*\exp(\bX^{*\top}\widehat{\bbeta}_{\infty})\bX^*_j)}{pt_{\epsilon}}        \\
&\leq   \frac{ \sum_{j=1}^p \xi^2B_1^2 \mathbb E(2\exp(\bX^{*\top}\widehat{\bbeta}_{\infty}))}{pt_{\epsilon}}        \\
     &\leq   \frac{ 2\xi^2B_1^2 \exp\left(\frac{B_1^2 L}{2}\right)}{t_{\eps}}   \\
     &=\eps/2 .
\end{align*}
We can also use a similar argument to get $P\left(\left\|\frac{1}{M}\sum_{i=1}^M \bX_i^*  \right\|_2 >  2B_1\sqrt{p/(M\eps)}\right)\leq \epsilon/2$. 
The two bounds together show  that 
\begin{equation}\label{eq:sup_u_Q2_bound}
	\mathbb P\left(\sup_{\|\bu\|_2=1}R_2(\bu)\leq \tau h_0^+ \sqrt{t_{\eps}p/M}+ 2 \tau B_1\sqrt{p/(M\eps)} \right)\geq 1-\eps.
\end{equation}

 \textbf{For $R_3$}: 
 We should derive an upper bound on 
\begin{align*}
	\sup_{\|\bu\|_2=1} R_3(\bu)=& \tau\sup_{\|\bu\|_2=1} \frac{1}{M}\sum_{i\leq M} Y_i^* (\bu^\top \bX_i^*)  \left[1-\exp\left(\gamma \bX_i^{*\top}\bu \right) \right]        \exp\left(\bX_i^{*\top}\widehat{\bbeta}_{\infty}\right) \\
 &\quad \quad  - \mathbb{E}\left(  Y^*(\bu^\top \bX^*)\left[1-\exp\left(\gamma \bX^{*\top}\bu  \right)\right]\exp\left(\bX^{*\top}\widehat{\bbeta}_{\infty} \right)\right) .
\end{align*}
Such an upper bound is given in Lemma \ref{lemma:bound_Q3}. We apply Lemma~\ref{lemma:bound_Q3} with $c_1=.5 h_0^+\xi_L C_{2L}$ to show that there exist two positive constants $M_{\eps,p}>0$ and $C_1$ such that if $\tilde{B}>C_1$, the following holds for all $M>M_{\eps,p}$: 
\begin{equation}\label{eq:sup_Q3_bound}
	\mathbb P\left(\sup_{\|\bu\|_2=1} R_3(\bu)\leq .5 h_0^+\xi_L C_{2L} \tau \sqrt{\frac{\tilde B p}{M}}\right)\geq 1-\eps .
\end{equation}
Combining the inequalities in \eqref{eq:sup_Q1_bound}, \eqref{eq:sup_u_Q2_bound}, and \eqref{eq:sup_Q3_bound}, we have 
$$\mathbb P\left(\sup_{\|\bu\|_2=1} \left\langle \bu,  \nabla F\left(\widehat{\boldsymbol{\beta}}_{\infty} + \gamma\bu\right)\right \rangle<0 \right)\geq 1-2\eps,$$
given that  
$\sqrt{\tilde B} > \frac{2 h_0^+\sqrt{t_{\eps}}+4 B_1/\sqrt{\eps}}{h_0^+\xi_L C_{2L}}$.  
The last condition can always be satisfied by choosing a large value of $\tilde B$. Therefore, we complete the proof of Theorem \ref{thm:stability_finite_M}. 

\end{proof}

The rest of this section devotes to proving the lemmas used in the proof of Theorem~\ref{thm:stability_finite_M}. 

\begin{lemma}\label{lemma:bound_Q3}
Suppose $\{Z_i=(\bX_i,Y_i)\}_{i=1}^n$ are independent and identically distributed random vectors, where $\bX_i$ is p-dimensional vector satisfying that (1)  $\mathbb{E} X_{ij}=0$, $\operatorname{Var}\left(X_{ij}\right)=1$, and $\left|X_{ij}\right| \leq B_1$,  , for $j=1, \cdots, p$;
     (2) $X_{i1}, \cdots, X_{ip}$ are independent. Furthermore, there exist two positive constants $\xi_L$ and $\xi_U$ such that $\xi_L\leq Y_i\leq \xi_U$.  
Suppose $\gamma=\sqrt{\tilde{B} p/n}$ where $\tilde{B}$ is a constant. 
Let $\bv_0\in\mathbb R^p$ be a fixed vector with $\|\bv_0\|_2\leq L$. 
Define
       $$h_u(Z_i) \triangleq  Y_i\left(\bX_i^{\top} \bu\right)  \left(1-e^{\gamma \bX_i^{\top} \bu}\right)  \exp \left( \bX_i^{\top}\bv_0 \right).$$
For any positive number $c_1$, there exist constants $C_1, C_2,C_3,C_4$ that only depend on $c_1$, $\xi_U$, $B_1$, and $L$ such that if $\tilde{B}>C_1$ and  $n>C_2p(\log n)^{C_3}$, then 
     $$\mathbb P\left(\sup_{\|\bu\|_2=1}\frac{1}{n}\sum_{i=1}^n \left[h_u(Z_i)-\mathbb E(h_u(Z_i))\right]\leq c_1 \gamma \right)\geq 1-2\exp\left(-\frac{n}{ C_4 (\log n)^{2}} \right).$$
 \end{lemma}

\begin{proof}[Proof of Lemma \ref{lemma:bound_Q3}]
	
We begin with defining a truncation of $h_u$:

$$
\begin{aligned}
& k_u(Z_i) \triangleq  Y_i\left(\bX_i^{\top} \bu\right)\left(1-e^{\gamma \bX_i^{\top} \bu}\right)   \exp \left( \bX_i^{\top}\bv_0 \wedge D\right),
\end{aligned}
$$
where $a\wedge b:=\min(a,b)$ and the value of $D$ will be determined later.
We want to upper bound $\sup_{\bu} \frac{1}{n} \sum_{i\leq n} h_u(Z_i)-\mathbb E(h_u(Z_i))$, which can be decomposed into three parts:
\begin{align*}
	\sup_{\bu}  \frac{1}{n} \sum_{i\leq n} h_u(Z_i)-\mathbb E(h_u(Z_i))&\leq \sup_{\bu} \frac{1}{n} \sum_{i\leq n} \left[h_u(Z_i)-k_u(Z_i)\right] \\
	& + \sup_{\bu} \frac{1}{n} \sum_{i\leq n} \left[k_u(Z_i)-\mathbb E(k_u(Z_i))\right]\\
	& + \sup_{\bu} \frac{1}{n} \sum_{i\leq n} \left[\mathbb E(k_u(Z_i))-\mathbb E(h_u(Z_i))\right].
\end{align*}
Accordingly, our analysis is divided into three steps:
\begin{itemize}
	\item First step: Upper bound $h_u\left(Z_i\right)-k_u\left(Z_i\right)$. Since $x(1-e^{\gamma x})\leq 0$ for any $x\in \mathbb{R}$ any $\gamma>0$ and $\exp \left( \bX_i^{\top}\bv_0 \wedge D\right)\leq \exp \left( \bX_i^{\top}\bv_0 \right)$, we can upper bound this term by 0.
	\item Second step: Upper bound the difference between $\mathbb{E}(k_u(Z))$ and $\mathbb{E}(h_u(Z))$.
    \item Third step: Use the $\eps$-net argument to control the distance between $\frac{1}{n} \sum_{i\leq n}  k_u(Z_i)$ and its expectation for all $\bu$ satisfying $\|\bu\|_2=1$.
 \end{itemize}

We detail the second and the third steps below.

\textbf{Second step:}  To bound the difference in expectation, we first introduce 
a lemma regarding two quantities used later.

\begin{lemma}\label{lemma:expectation_difference_two_quantities}
Let $\bu$ and $\bv$ be two fixed vectors that satisfy $\|\bu\|_2=1$ and $\|\bv\|_2\leq L$. 
Let $\bX$ be a $p$-dimensional random vector with independent entries of mean zero and variance 1 and $|X_i|\leq B_1$. The followings hold: 
\begin{enumerate}
	\item $\mathbb E(\bX^\top \bu)^4\leq B_1^2+3$.
	\item Denote  	$D_0=2B_1 L\sqrt{ \log(\log n  ) }$ and $D=2 B_1^2 L^2+D_0$. It holds that
	 \begin{align*}
	 	&\mathbb E\left[\left(e^{\bX^\top\bv}-e^{D} \right)^2 I\left\{\bX^\top\bv>D\right\}\right]\leq
        \frac{6 B_1 L e^{2 B_1^2 L^2} }{(\log n)^2}.
	 \end{align*}
\end{enumerate}

\end{lemma}

We defer the proof of Lemma \ref{lemma:expectation_difference_two_quantities}. Recall $\xi_l\leq Y_i\leq \xi_U$ and $\gamma \sqrt{p}B_1\leq 1$, then we have
\begin{align*}
	\mathbb E(k_u(Z_i))-\mathbb E(h_u(Z_i))&=  \mathbb E\left(\gamma Y_i\left(\bX_i^{\top} \bu\right)^2 \frac{1-e^{\gamma \bX_i^{\top} \bu}}{\gamma \bX_i^{\top} \bu}\left[\exp \left( \bX_i^{\top}\bv_0 \wedge D\right)-\exp \left( \bX_i^{\top}\bv_0 \right)  \right] \right)\\
    &\leq 2 \gamma \xi_{U} \mathbb E\left( \left(\bX_i^{\top} \bu\right)^2 \left|\exp \left( \bX_i^{\top}\bv_0 \wedge D\right)-\exp \left( \bX_i^{\top}\bv_0 \right)  \right| \right)\\
	&\leq 2\gamma \xi_U   \sqrt{\mathbb  E\left(\bX_i^{\top} \bu\right)^4} \sqrt{\mathbb E\left(\left(e^{\bX_i^{\top}\bv_0}-e^{\bX_i^{\top}\bv_0 \wedge D}\right)^2 I\left\{\bX_i^{\top}\bv_0>D\right\}\right)} \\
	&\leq 2\gamma \xi_U \sqrt{B_1^2+3}\frac{\sqrt{6 B_1 L }\exp(B_1^2L^2)}{\log n},
\end{align*}
where the first inequality is due to the elementary inequality that $\sup _{-1<\theta<1}\left|\frac{e^\theta-1}{\theta}\right|<2$, the second inequality follows from Cauchy-Schwarz inequality, and the third inequality follows from Lemma \ref{lemma:expectation_difference_two_quantities}.
This completes the second step.

\textbf{Third step:} We begin with with a symmetrization argument. To proceed, we first need to show that for each fixed $\bu$, the variance of $k_u(Z_i)$ is bounded by some constant. By definition of $k_u$ and the fact that $\sup _{-1<\theta<1}\left|\frac{e^\theta-1}{\theta}\right|<2$, we have 
\begin{align*}
	 \mathbb E(k^2_u(Z_i))& \leqslant 4 \xi_U^2 \mathbb E\left[\left(\bX_i^\top\bu\right)^4 \exp \left(2 \bX_i^\top \bv_0\right)\right] \\ 
	 & \leqslant 4 \xi_U^2 \sqrt{\mathbb E\left[\left(\bX_i^\top\bu\right)^8\right] \mathbb E \exp \left(4 \bX_i^\top \bv_0\right)} \\ 
	 & \leqslant 4 \xi_U^2 \cdot \sqrt{C_8 \left(1+B_1^8\right)} \cdot \sqrt{e^{8 B_1^2 L^2}} \\ 
	 & = 4 \xi_U^2 \sqrt{C_8 \left(1+B_1^8\right)} e^{4 B_1^2 L^2} \triangleq R^2,
\end{align*}
where the second inequality is due to Cauchy-Schwarz inequality and the third inequality follows from  Rosenthal's inequality stated in Lemma \ref{lemma:Rosenthal}. We can use this bound on the variance to derive a bound on the tail. 
Denote $\delta_n=2 R \sqrt{2/n}$. 
For any $\bu$ with $\|\bu\|_2=1$ and any $t\geq \delta_n$, Chebyshev's inequality implies that 
$$\mathbb P\left(\left|\frac{1}{n}\sum_{i=1}^nk_u(Z_i)-\mathbb E(k_u(Z_i))\right|  \geq t/2\right)\leq \frac{4 R^2 }{n t^2}\leq \frac{1}{2}.$$

This bound justifies the condition for applying the symmetrization theorem stated in Lemma \ref{lemma:symmetrization}.
Let $(Z_1^\prime,\ldots Z_n^\prime)$ be an independent copy of $(Z_1\ldots,Z_n)$. Based on Lemma \ref{lemma:symmetrization}, the following holds for any $t\geq \delta_n$: 
\begin{equation}\label{eq:symmetrization}
	\mathbb P\left(\sup_{\|\bu\|_2=1}\left|\frac{1}{n}\sum_{i=1}^n k_u(Z_i)-\mathbb E(k_u(Z_i))\right|  \geq t \right)\leq 2 \mathbb P\left(\sup_{\|\bu\|_2=1}\left|\frac{1}{n}\sum_{i=1}^n k_u(Z_i)-k_u(Z_i^\prime)\right|  \geq t /2\right) .
\end{equation}

We will use the $\eps$-net argument to control the right hand side of \eqref{eq:symmetrization} together with an exponentially small tail probability for any fixed $\bu$. To get the tail probability, we first verify that $W_i(\bu):= k_u(Z_i)/\gamma- k_u(Z_i^\prime)/\gamma$ is a sub-exponential random variable. 
Obviously, we have $\mathbb E(W_i(\bu))=0$. For any $t>0$,  
\begin{equation}\label{eq: sub-exponential W_i(u)}
\begin{aligned}
\mathbb P\left(|W_i(\bu)|>t\right) & \leq 2 \mathbb P\left(- k_u\left(Z_i\right)/\gamma>t\right) \\
& \leq 2\mathbb P\left(\left(\bX_i^{\top} \bu\right)^2>\frac{t}{2 e^D \xi_U}\right) \\
& \leq 2\exp \left(-\frac{\frac{t}{2 e^D \xi_U}}{2 B_1^2  }\right) \\
& =2\exp \left(-\frac{2 t}{8 e^D \xi_U B_1^2  }\right),
\end{aligned}
\end{equation}
where the first inequality is due to the union bound, the second inequality is again due to the fact that $\sup _{-1<\theta<1}\left|\frac{e^\theta-1}{\theta}\right|<2$, and the third inequality is due to Lemma~\ref{lem: weighted Hoeffding}. We conclude that $W_i(\bu)$'s are sub-exponential random variables and the condition of Bernstein's inequality as stated in Lemma \ref{lemma: Bernsteins} holds with $\lambda=8 e^D \xi_U B_1^2$. 

We are now ready to apply the $\eps$-net argument. Define
$$
\bar\eps =3 \cdot \exp\left(-\frac{n}{4096 \cdot p\cdot \exp(2D) \xi_U^2 B_1^4\cdot(\log n)^{2}}\right).
$$

Let $\mathcal N$ be a $\bar \eps$-net of $\{\bu\in \mathbb R^p:\|\bu\|_2=1\}$. We have $|\mathcal N|\leq (3/\bar \eps)^p$. 
Furthermore, we have
\begin{equation}\label{eq: stability bound on net}
\begin{aligned}
	&~ \quad \mathbb P\left(\gamma^{-1}\sup_{\bu\in\mathcal N}  \left|\frac{1}{n}\sum_{i=1}^n k_u(Z_i)-k_u(Z_i^\prime)\right|\geq \frac{1}{\log n}   \right)\\
    &\leq \sum_{\bu\in\mathcal N} \mathbb P \left(\frac{1}{n}\sum_{i=1}^n W_i(\bu) \geq  \frac{1}{\log n}\right)\\
    &\leq \exp\left(p\log \left(\frac{3}{\bar \eps}\right)-\frac{n}{ 2048 (\log n)^2 e^{2D}\xi_U^2 B_1^4 } \right) \\
    &= \exp\left(-\frac{n}{ 4096 (\log n)^2 e^{2D}\xi_U^2 B_1^4 } \right),
\end{aligned}
\end{equation}
where the first inequality follows from union bound and the second inequality follows from Lemma \ref{lemma: Bernsteins} with the case that $\left(\frac{t^2}{16\lambda^2} \wedge \frac{t}{4\lambda}\right)=\frac{t^2}{16\lambda^2}$
because $\frac{1}{\log n}\leq 4 \lambda=32e^D\xi_U B_1^2$ holds when $n$ is sufficiently large.

We choose $t=c_1 \gamma/2$ in \eqref{eq:symmetrization}. Let $C_1=32 R^2/c_1^2$. For $\tilde{B}\geq C_1$, we have $t\geq \delta_n$. 
Furthermore, for $\log n > 4/c_1$, we have $\gamma/\log n < c_1 \gamma/4$. 
Consequently, we can combine 
\eqref{eq:symmetrization} and \eqref{eq: stability bound on net} together to see that 
\begin{equation}\label{eq:eps_net_concentration}
\mathbb P\left( \sup_{\bu\in\mathcal N}  \left|\frac{1}{n}\sum_{i=1}^n k_u(Z_i)-\mathbb E(k_u(Z_i))\right|\leq c_1\gamma/2  \right)\geq 1-2\exp\left( -\frac{n}{4096(\log n)^2 e^{2D}\xi_U^2B_1^4 } \right).
\end{equation}
Now suppose the event in the probability of \eqref{eq:eps_net_concentration} holds. 
For any $\bu_1$ with $\|\bu_1\|_2=1$, we can find a $\bu_2 \in \mathcal N$ such that $\|\bu_1-\bu_2\|_2\leq \bar \eps$. 
It then follows that $|\bX_i^{\top} \bu_1-\bX_i^{\top} \bu_2|\leq \sqrt{p} B_1 \eps$. 
Recall that  $\gamma \sqrt{p}\leq 1$. 
We have
\begin{align}
	|k_{u_1}(Z_i)-k_{u_1}(Z_i)|&\leq |Y_i \exp(\bX_i^\top \bv_0 \wedge D )| \cdot \left|\left(\bX_i^{\top} \bu_1\right)\left(1-e^{\gamma \bX_i^{\top} \bu_1}\right)-\left(\bX_i^{\top} \bu_2\right)\left(1-e^{\gamma \bX_i^{\top} \bu_2}\right)  \right| \nonumber \\
    &\leq \xi_U e^D \cdot 3 \gamma B_1^2 e^{B_1} \sqrt{p} \left|\bX_i^{\top} \bu_1-\bX_i^{\top} \bu_2   \right|\nonumber \\
	& \leq 3\gamma \xi_U  B_1^2 e^{B_1}  e^D p \bar\eps\nonumber \\
	&= 9 \gamma \xi_U  B_1^2 e^{B_1+D}  p \exp\left(-\frac{n}{4096 \cdot p\cdot \exp(2D) \xi_U^2 B_1^4\cdot(\log n)^{2}}\right),   
    \label{eq:u1_u2_bound_eps_net}
\end{align}
where the second inequality follows from the following application of the mean value theorem with $a=\bX_i^{\top} \bu_1$ and $b=\bX_i^{\top} \bu_2$: 
 
\begin{align*}
\left|a\left(e^{\gamma a}-1\right)-b\left(e^{\gamma b}-1\right)\right|
&= \left|\int_b^a\left[e^{\gamma x}(1+\gamma x)-1\right] d x \right| \\
&= \left|\int_b^a\left[e^{\gamma x}(1+\gamma x)-1\right] d x \right| \\
&= \left|\int_b^a\left[\int_{0}^{\gamma x}  [2e^y+ye^y] d y\right] d x \right| \\
&\leq 3 B_1 e^{B_1} \gamma |b^2-a^2|/2\\
&\leq 3 \gamma \sqrt{p} B_1^2 e^{B_1} |b-a|. 
\end{align*}

The right hand side of \eqref{eq:u1_u2_bound_eps_net} can be bounded by $c_1 \gamma/2$ if 
$$
\log\left(18 c_1^{-1} \xi_U  B_1^2\right)+ B_1+D +\log p \leq   \frac{n}{4096 \cdot p\cdot \exp(2D) \xi_U^2 B_1^4\cdot(\log n)^{2}}, 
$$
which is guaranteed to hold if $\frac{n}{p\log(n)^3}$ is sufficiently large. 
Based on \eqref{eq:eps_net_concentration} and \eqref{eq:u1_u2_bound_eps_net}, we have  
\begin{equation}\label{eq:sup_u_concentration}
	\mathbb P\left( \sup_{\|\bu\|_2=1 }  \left|\frac{1}{n}\sum_{i=1}^n k_u(Z_i)-\mathbb E(k_u(Z_i))\right|\leq c_1 \gamma  \right)\geq 1-2\exp\left( -\frac{n}{4096 (\log n)^2 e^{2D}\xi_U^2B_1^4 } \right).
\end{equation}

We complete the proof of Lemma~\ref{lemma:bound_Q3}
\end{proof}

\begin{proof}[Proof of Lemma \ref{lemma:expectation_difference_two_quantities}]
By properties of $\bX$, we have
\begin{align*}	\mathbb{E}\left[\left(\mathbf{X}^{\top} \mathbf{u}\right)^4\right]&=\sum_{i=1}^d \mathbb{E}\left[X_i^4\right] u_i^4+6 \sum_{i<j} u_i^2 u_j^2\\
	&\leq B_1^2 \sum_{i=1}^d u_i^4+3\left(1-\sum_{i=1}^d u_i^4\right) \\
	&\leq B_1^2+3.
\end{align*} 	
Furthermore, Fubini's theorem implies that 
	\begin{align*}
E\left(\left(e^{\bX_i^{\top}\bv_0}-e^{ D}\right)^2 1_{\bX_i^{\top}\bv_0>D}\right)&=\mathbb E\left(\int_0^{\infty} 2\left(e^t-e^D\right) e^t 1_{D<t<\bX_i^{\top}\bv_0} d t 1_{\bX_i^{\top}\bv_0>D}\right)\\
		& =\int_D^{\infty} 2\left(e^t-e^D\right) e^t P(\bX_i^{\top}\bv_0>t, \bX_i^{\top}\bv_0>D) d t \\
& \leqslant 2 \int_D^{\infty} e^{2 t} e^{-\frac{t^2}{2 B_1^2 L^2}} d t \\
& =2 e^{2 B_1^2 L^2} \int_{D-2 B_1^2 L^2}^{\infty} e^{-\frac{\left[\left(t-2B_1^2L^2\right) /\left(B_1 L\right)\right]^2}{2}} d t \\
& = 2 e^{2 B_1^2 L^2} B_1 L \int_{\frac{D_0}{B_1 L}}^{\infty} e^{-t^2 / 2} d t \\
& \leqslant 2\sqrt{2\pi} e^{2 B_1^2 L^2} B_1 L e^{-D_0^2 / 2 B_1^2 L^2} \\
&= \frac{2\sqrt{2\pi} B_1 L e^{2 B_1^2 L^2} }{(\log n)^2},
\end{align*}
where the first inequality is due to Lemma~\ref{lem: weighted Hoeffding} with $\bu=\bv_0/\|\bv_0\|$ and the fact that $\|\bv_0\|\leq L$,  the third equality is based on $\frac{t^2}{B^2 c^2}-4 t=\left(\frac{t}{B_1 c}-2 B_1 c\right)^2-4 B^2 c^2$, and the last inequality is due to the standard upper bound of the tail probability of the standard normal distribution.
\end{proof}

\begin{lemma}\label{lemma: Bernsteins}
(Bernstein's inequality, Theorem 1.13 in \cite{rigollet2023high})
Let $X_1, \ldots, X_n$ be independent random variables such that $\mathbb{E}\left(X_i\right)=0$ and for $t>0$, $\mathbb P(|X_i|>t)\leq 2\exp(-2t/\lambda)$. Define $ \bar{X}=\frac{1}{n} \sum_{i=1}^n X_i $. Then we have
$$
\mathbb{P}(\bar{X}>t) \vee \mathbb{P}(\bar{X}<-t) \leq \exp \left[-\frac{n}{2}\left(\frac{t^2}{16\lambda^2} \wedge \frac{t}{4\lambda}\right)\right] .
$$
\end{lemma}

\begin{lemma}\label{lemma:Rosenthal}(Rosenthal's inequality, \cite{rosenthal1970subspaces})
     Let $Z_1, Z_2, \ldots, Z_d$ be independent random variables with mean zero.  For any $p \geq 2$, the following inequality holds:
$$
E\left|\sum_{i=1}^d Z_i\right|^p \leq C_p\left(\sum_{i=1}^d E\left|Z_i\right|^p+\left(\sum_{i=1}^d E\left|Z_i\right|^2\right)^{p / 2}\right),
$$
where $C_p$ is a constant depending only on $p$.
\end{lemma}

\begin{lemma}\label{lemma:symmetrization}
(Symmetrization theorem, Lemma 3.3 in  \cite{van2000empirical}). Let $X_1, \ldots, X_n$, $X_1^{\prime}, \ldots, X_n^{\prime}$ be i.i.d. with distribution $P$.
	Suppose for all $f \in \mathcal{F}$,
	$$\mathbb{P}\left(  \left|\frac{1}{n} \sum_{i=1}^n\left(f\left(X_i\right)-\mathbb{E} f\left(X_i\right)\right)\right| \geq \delta/2 \right)\leq \frac{1}{2} . $$
	Then
		$$
\mathbb{P}\left(\sup _{f \in \mathcal{F}}\left|\frac{1}{n} \sum_{i=1}^n\left(f\left(X_i\right)-\mathbb{E} f\left(X_i\right)\right)\right| \geq \delta\right) \leq 2 \mathbb{P}\left(\sup _{f \in \mathcal{F}}\left|\frac{1}{n} \sum_{i=1}^n\left(f\left(X_i\right)- f\left(X_i^{\prime}\right)\right)\right| \geq \delta/2\right) .
$$
\end{lemma}

\begin{lemma}\label{lem: weighted Hoeffding}
    Let $X=\left(X_1, \ldots, X_p\right)^{\top}$ be a vector of independent random variables such that $E\left[X_i\right]=0$, $\operatorname{Var}\left(X_i\right)=1$, and $\left|X_i\right| \leq B_1$ almost surely. Let $u$ be a unit vector in $\mathbb{R}^p$. Then, for any $t>0$, it holds that
$$
P\left(X^{\top} u \geq t\right) \leq \exp \left(-\frac{t^2}{2 B_1^2}\right).
$$
\end{lemma}
\begin{proof}
Let $S = X^\top u = \sum_{i=1}^p u_i X_i$ and $Y_i = u_i X_i$.
By Hoeffding's lemma, the moment-generating function for $Y_i$ satisfies
$E\left[e^{\lambda Y_i}\right] \leq \exp\left(\frac{\lambda^2 B_1^2 u_i^2}{2}\right)$. 
Since $Y_i$ are independent and $\|u\|^2 = 1$, it holds that
$$
E\left[e^{\lambda S}\right] \leq \exp\left(\frac{\lambda^2 B_1^2}{2} \sum_{i=1}^p u_i^2\right) = \exp\left(\frac{\lambda^2 B_1^2}{2}\right).
$$
Markov's inequality implies that
$$
P(S \geq t) \leq \exp\left(\frac{\lambda^2 B_1^2}{2} - \lambda t\right), \forall \lambda>0.
$$
In particular, for $\lambda = \frac{t}{B_1^2}$, we have $P(S \geq t) \leq \exp\left(-\frac{t^2}{2B_1^2}\right)$.
\end{proof}

\begin{proof}[Proof of Lemma~\ref{lemma:smallest_eigen_cat_population_hessian}]

It suffices to show that for any $\|\bv\|_2=1$, $\bv^\top \mathbb E\left(\exp(\bX^{*\top}\bbeta)\bX^*\bX^{*\top}   \right) \bv\geq C_L$. 

Under Condition~\ref{conditions:synthetic_X_Y}, we have two conclusions: (1) $\operatorname{Var}(\bX^{*\top}\bbeta)\leq L^2$, and (2) there exist two positive constants $\eta_0$ and $\rho_0$ such that $\mathbb P(|\bX^{*\top}\bv|> \eta_0) \geq \rho_0$. The second statement follows from \citet[Theorem 5.7]{huang_catalytic_2020}. 
By Chebyshev's inequality, we have $\mathbb P(|\bX^{*\top }\bbeta|> L/\sqrt{\rho_0/2})\leq \rho_0\operatorname{Var}(\bX^{*\top}\bbeta)/(2L^2)\leq \rho_0/2$ using the first statement. We can then derive as follows:

\begin{align*}
&\quad 	\bv^\top \mathbb E\left(\exp(\bX^{*\top}\bbeta)\bX^*\bX^{*\top}   \right) \bv\\
 & = \mathbb E\left(\exp(\bX^{*\top}\bbeta)(\bX^{*\top}\bv)^2   \right) \\
	&\geq \mathbb E\left(\exp(\bX^{*\top}\bbeta)(\bX^{*\top}\bv)^2 \mathbf{1}\{|\bX^{*\top }\bbeta|\leq L/\sqrt{\rho_0/2}, |\bX^{*\top}\bv|>\eta_0  \}  \right)\\
&\geq \exp(-L/\sqrt{\rho_0/2})\eta_0^2 \mathbb P\left(|\bX^{*\top }\bbeta|\leq L/\sqrt{\rho_0/2}, |\bX^{*\top}\bv|>\eta_0 \right)\\
&\geq \exp(-L/\sqrt{\rho_0/2})\eta_0^2 \left[\mathbb P(|\bX^{*\top}\bv|> \eta_0 )- \mathbb P(|\bX^{*\top }\bbeta|> L/\sqrt{\rho_0/2})\right]\\
&\geq \exp(-L/\sqrt{\rho_0/2})\eta_0^2 \frac{\rho_0}{2},
\end{align*}
where the second inequality is because  $\exp(\cdot)$ is increasing on $[0, \infty)$, the third inequality is because $\mathbb P(A\cap B) = \mathbb P(B)-\mathbb P(B\cap A^c)\geq \mathbb P(B)-\mathbb P(A^c)$. 
By taking $C_L=\frac{\rho_0 \eta_0^2 }{2}\exp(-L/\sqrt{\rho_0/2})$, we proved Lemma~\ref{lemma:smallest_eigen_cat_population_hessian}.
\end{proof}

\section{Implementation of Simulation}
\label{supp:estimator_computation_detail}

\subsection{Synthetic data}

The synthetic data for specifying Cox catalytic priors are generated as follows. 
For each entry $X_{j}^{*}$ of a synthetic covariate vector $\boldsymbol{X}^*$, $X_{j}^{*}$ is sampled from the marginal empirical distribution of observed $\{X_{i,j}\}_{i=1}^{n}$. 
To accommodate the highly unbalanced binary covariate ($j=1$) in our simulations, half of the sampled $X_{1}^{*}$ will be replaced by i.i.d. random variables drawn from Bernoulli($p=0.5$); this is following the \textit{flattening} strategy proposed in the supplementary material of \cite{huang_catalytic_2020}. 
To accommodate the skewness in continuous covariate ($j\geq 2$), half of sampled $X_{j}^{*}$ will be replaced by i.i.d random variables drawn from a normal distribution with median and interquartile range (IQR) matching to those of the observed covariates. Specific, the normal distribution is $N(\mu_j,\sigma_j^2)$, where $\mu_j$ is the sample median of $\{X_{i,j}\}_{i=1}^{n}$ and $\sigma_j$ is chosen properly such that
$1/4=\Phi\left( - {IQR_j}/{(2 \sigma_j)} \right)$, 
where $IQR_j$ is the IQR of observed $\{X_{i,j}\}_{i=1}^{n}$ and $\Phi$ is the cumulative distribution function of standard normal.

We generate synthetic survival times according to the simple example in Section \ref{sec:specification}. 
The synthetic sample size is fixed at $M=1000$, a sufficiently large number relative to the dimension $p$ in our simulation settings. 
We set the surrogate component $h_0^+$ to be the MLE $\hat{\psi}$ for the following likelihood.
$$L\left( \psi \mid \left\{\left(\boldsymbol{X}_i, Y_i, \delta_i\right)\right\}_{i=1}^n\right) = \prod_{i=1}^n \psi^{\delta_i}\exp \left\{- \psi Y_i \right\}, \quad \psi>0.$$

 The detailed procedure to implement the estimation, along with the tuning via cross-validation, is provided in the supplementary material. 
\subsection{Implementation of estimator}

The MPLE and the WME in \eqref{estimator_wm} can be computed using standard statistical software; for example, one can use the function \texttt{coxph($\cdot$)} in the $\textbf{R}$ package \texttt{survival}. 
The CRE in \eqref{estimator_penalty} can be obtained using any efficient convex optimization method, such as the classical Newton-Raphson method. Furthermore, we consider Bayesian estimation based on posterior sampling. Specifically, we place a Cox catalytic prior on the coefficients and a Gamma process prior on the cumulative baseline hazard function, collect samples from the posterior distribution, and calculate the posterior mean for the coefficients.  
For the purpose of comparison, we also consider placing a multivariate Gaussian prior on the coefficients as an alternative. 

In addition to the Cox catalytic prior on $\boldsymbol{\beta}$, we also consider the Cox adaptive catalytic prior on $(\boldsymbol{\beta}, \tau)$ defined in \eqref{joint_prior} with hyperparameters $(\alpha, \gamma)$ set to be $(2,1)$. Unlike the other methods, the Cox adaptive catalytic prior has the advantage of being free of tuning parameters. 

The Ridge estimator is defined as follows:
\begin{equation}
	\label{cox_ridge}
	\hat{\boldsymbol{\beta}}_{\lambda}=\arg \max_{\boldsymbol{\beta}\in \mathbb{R}^p} \left\{\sum_{i=1}^n \delta_i \left({\boldsymbol{X}_i}^{\top} \boldsymbol{\beta}-\log \sum_{j\in \mathcal{R}_i} \theta_j\right) -\lambda f(\boldsymbol{\beta})\right\},
\end{equation}
 where $\theta_j=\exp(\boldsymbol{X}_j^{\top}\boldsymbol{\beta})$ and the penalty term $f(\boldsymbol{\beta})=\|\boldsymbol{\beta}\|^2$, assuming that each entry of the covariates $\boldsymbol{X}_{i}$'s has been standardized to have zero mean and unit variance.  
The Lasso estimator is defined in a similar way as in \eqref{cox_ridge} but with $f(\boldsymbol{\beta})=\|\boldsymbol{\beta}\|_1$. Both the Ridge and the Lasso estimates can be computed using the \texttt{glmnet} package in \textbf{R} \citep{simon_regularization_2011}.

\subsection{Selection of tuning parameters}\label{supp:tune}
 
Some of the estimation methods under consideration involve selecting the tuning parameters, including $\tau$ for the estimation based on Cox catalytic prior and $\lambda$ for the penalized estimation using \eqref{cox_ridge}. 
To make a fair comparison, we employed the $K$-fold cross-validation proposed in \citet{van2006cross} to select the tuning parameter for each method. 
To illustrate, we outline the cross-validation procedure for an estimator with tuning parameter $\tau$ and the procedure for $\lambda$ is the same. 
Given the number $K$ of folds, the original data $\boldsymbol{D}$ is first randomly split into $K$ folds of roughly equal sizes. 
Then, for each $i$ out of $K$, the $i$-th fold of data is left out and the regression estimate $\widehat{\boldsymbol{\beta}}_{-i}(\tau)$ with tuning parameter $\tau$ is computed based on the remaining $(K-1)$ folds of data. 
Unlike the standard procedure of cross-validation, where the estimate is used to make prediction on the left-out data, 
we subtract $\log PL_{-i}\left(\widehat{\boldsymbol{\beta}}_{-i}(\tau)\right)$ from $\log PL\left(\widehat{\boldsymbol{\beta}}_{-i}(\tau)\right)$, where $\log PL$ is the partial log-likelihood for all $K$ folds and $\log PL_{-i}$ is the partial log-likelihood for all folds except the $i$-th. 
The cross-validated partial log-likelihood is then defined as 
\begin{equation}\label{cvpl}
 	\text{CVPL}(\tau)= \sum_{i=1}^{K}\left[ \log PL\left(\widehat{\boldsymbol{\beta}}_{-i}(\tau)\right)-\log PL_{-i}\left(\widehat{\boldsymbol{\beta}}_{-i}(\tau)\right) \right]. 
 \end{equation}
Then the tuning parameter is set to be the value that maximizes the cross-validated partial log-likelihood. 
For CRE \eqref{estimator_penalty} and WME \eqref{estimator_wm}, the synthetic data remains the same across different folds.

For CRE, WME, Ridge estimator, Lasso estimator, the tuning parameters are selected by 10-folds cross-validation. 
For estimation based on Bayesian posterior, the cross-validation requires intensive sampling from different posterior distributions, making it computationally demanding. 
To reduce the computational burden, a simple approach is to set the total weight parameter equal to the dimension of the regression coefficients, that is, $\tau=p$, as suggested in \citep{huang2022catalytic} for logistic regression. 
Another strategy is to utilize the fact that CRE is an approximation to the posterior mode of $\boldsymbol{\beta}$ and can be computed efficiently. 
Specifically, we first conduct cross-validation to obtain $\hat{\tau}_{CR,cv}$ for CRE and then use its value to specify the hyperparameter $\tau$ for Bayesian posterior estimation.  
Extending this strategy to Bayesian inference with a normal prior on each coefficient, we first conduct cross-validation to determine $\hat{\lambda}_{2,cv}$ for the Ridge estimator, and set the variance of each normal prior to be $(2\hat{\lambda}_{2,cv})^{-1}$  (assuming that each entry of the observed covariates has been standardized).

\section{Extra simulation}
\label{supp:extra_simulation}
This section presents additional simulations that extend the numerical study presented in Section~\ref{sec:simulation}, in which a censoring rate of 20\% was assumed. Specifically, two additional settings are considered, with censoring rates of 10\% and 40\%.

\renewcommand{\arraystretch}{0.5}
\begin{table}[htbp]
\centering
\caption{Estimation and prediction performance of the MPLE, Ridge, Lasso, CRE, WME, posterior mean of different Bayesian methods when the censoring rate is 10\%. Standard error is presented in parentheses. We indicate the best method for each scenario in bold. Shorthand notations: 
``CPM", ``APM", and ``GPM" stand for Bayesian posterior mean resulting from using a Cox catalytic prior, a Cox adaptive catalytic prior, and  a Gaussian prior respectively; 
``CV" in parentheses indicates the selection of tuning parameters through cross-validation.}
% {\tiny
$$
\begin{array}{clccc}
\hline
p & \text{Methods} & \|\hat{\boldsymbol{\beta}}-\boldsymbol{\beta}_0 \|^2 & \text { Predictive deviance }  \\
\hline
\hline
 & \text{MPLE} & 0.84(0.05) & 19.35(1.19) \\
& \text{CRE (CV)} & 0.57(0.03) & 13.25(0.74) \\
& \text{WME (CV)} & \textbf{0.48}(0.02) & \textbf{12.83}(0.71) \\
20& \text{CPM (CV)} & 1.07(0.08) & 23.14(0.69) \\
& \text{APM} & 0.83(0.01) & 25.22(0.70) \\
& \text{GPM (CV)} & 0.70(0.02) & 16.74(0.64) \\
& \text{Ridge (CV)} & 0.56(0.03) & 13.50(0.65) \\
& \text{Lasso (CV)} & 0.71(0.04) & 13.99(0.67) \\
\hline\hline
 & \text{MPLE} & 3.39(0.15) & 92.77(4.43) \\
& \text{CRE (CV)} & 0.77(0.02) & 24.66(0.97) \\
& \text{WME (CV)} & \textbf{0.74}(0.02) & \textbf{23.88}(0.89) \\
40& \text{CPM (CV)} & 0.81(0.02) & 26.59(0.83) \\
& \text{APM} & 0.77(0.01) & 25.35(0.76) \\
& \text{GPM (CV)} & 0.94(0.03) & 26.27(0.91) \\
& \text{Ridge (CV)} & 0.90(0.03) & 25.18(0.95) \\
& \text{Lasso (CV)} & 1.18(0.03) & 25.50(0.88) \\
\hline\hline
 & \text{MPLE} & 17.60(0.80) & 380.56(14.53) \\
& \text{CRE (CV)} & 0.98(0.02) & 32.38(0.85) \\
& \text{WME (CV)} & 0.94(0.02) & 31.37(0.85) \\
60& \text{CPM (CV)} & 1.02(0.02) & 32.02(0.79) \\
& \text{APM} & \textbf{0.91}(0.01) & 29.64(0.74) \\
& \text{GPM (CV)} & 1.09(0.02) & 32.52(0.78) \\
& \text{Ridge (CV)} & 1.08(0.02) & 32.43(0.79) \\
& \text{Lasso (CV)} & 1.29(0.02) & \textbf{29.36}(0.75) \\
\hline
\end{array}
$$
\label{table:censor1}
\end{table}

\renewcommand{\arraystretch}{1}

\renewcommand{\arraystretch}{0.5}
\begin{table}[htbp]
\centering
\caption{Estimation and prediction performance of the MPLE, Ridge, Lasso, CRE, WME, posterior mean of different Bayesian methods when the censoring rate is 40\%. Standard error is presented in parentheses. We indicate the best method for each scenario in bold. See the caption of Table~\ref{table:censor1} for the meaning of shorthand notations. }
$$
\begin{array}{clccc}
\hline
p & \text{Methods} & \|\hat{\boldsymbol{\beta}}-\boldsymbol{\beta}_0 \|^2 & \text { Predictive deviance }  \\
\hline
\hline
 & \text{MPLE} & 1.56(0.11) & 23.08(1.50) \\
& \text{CRE (CV)} & 0.83(0.05) & 13.05(0.81) \\
& \text{WME (CV)} & \textbf{0.69}(0.03) & \textbf{12.59}(0.77) \\
20& \text{CPM (CV)} & 0.76(0.05) & 14.08(0.67) \\
& \text{APM } & 0.74(0.02) & 15.08(0.60) \\
&  \text{GPM (CV)} & 0.73(0.04) & 12.97(0.70) \\
& \text{Ridge (CV)} & 0.79(0.04) & 13.60(0.73) \\
& \text{Lasso (CV)} & 1.01(0.04) & 13.25(0.66) \\
\hline\hline
 & \text{MPLE} & 10.08(0.67) & 148.25(8.46) \\
& \text{CRE (CV)} & 1.03(0.03) & 22.91(0.78) \\
& \text{WME (CV)} & 1.00(0.03) & 22.51(0.75) \\
40& \text{CPM (CV)} & 1.05(0.02) & 21.78(0.75) \\
& \text{APM } & \textbf{0.99}(0.02) & 20.62(0.72) \\
&  \text{GPM (CV)} & 1.10(0.03) & 23.67(0.74) \\
& \text{Ridge (CV)} & 1.13(0.03) & 23.17(0.75) \\
& \text{Lasso (CV)} & 1.31(0.03) & \textbf{21.46}(0.73) \\
\hline\hline
 & \text{MPLE} & 683.79(159.72) & 1811.07(200.55) \\
& \text{CRE (CV)} & 1.16(0.02) & 27.05(0.73) \\
& \text{WME (CV)} & 1.11(0.02) & 25.91(0.72) \\
60& \text{CPM (CV)} & 1.25(0.02) & 25.89(0.73) \\
& \text{APM } & 1.21(0.02) & 24.49(0.78) \\
&  \text{GPM (CV)} & 1.17(0.02) & 25.95(0.70) \\
& \text{Ridge (CV)} & 1.19(0.02) & 26.08(0.73) \\
& \text{Lasso (CV)} & 1.34(0.02) & \textbf{22.84}(0.77) \\
\hline
\end{array}
$$
\label{table:censor4}
\end{table}

\renewcommand{\arraystretch}{1}

\section{Inforamtion about PBC data}
\label{supp:pbc_detail}
The dataset originates from a randomized clinical trial conducted by Mayo Clinic between 1974 and 1984. 
Its primary objective was to assess the efficacy of a pharmaceutical intervention in the treatment of primary biliary cirrhosis of the liver, which is a rare yet life-threatening disease. 
The dataset is appealing because it contains a variety of covariates, including both numerical and categorical variables. 
Researchers have utilized this dataset for building survival models for PBC patients and for investigating the impact of different types of variables \citep{hoeting1999bayesian}.

Our analysis is based on a publicly available version of the dataset, which is included in the \texttt{R} package \textbf{survival}. 
This dataset includes the following covariates:
\begin{itemize}
\item Categorical variables (6 in total): DPCA treatment (2 levels), gender (2 levels), presence of ascites (2 levels), edema (3 levels), hepatomegaly (2 levels), blood vessel malformations in the skin (2 levels);
    \item Numeric variables (11 in total): age, serum albumin, serum bilirubin, 
    serum cholesterol, alkaline phosphatase, aspartate aminotransferase, urine copper, platelet count, standardized blood clotting time, triglycerides, Histologic stage. 
\end{itemize}
In data preprocessing, we standardized continuous variables and converted categorical variables into binary dummy variables with value 0 or 1. 
In total, the dimension of a covariate vector is $p=18$ in our analysis. 
The patient status at the end of the study was coded as 1 for deceased patients and 0 for those still alive. 
The survival time was defined as the duration in years between the date of registration and the earliest of death, transplantation, or the study's analysis date in July 1986. 
After removing subjects with missing data, the PBC dataset consists of $n=276$ observations. 

\section{Source code information}
\label{supp:Sec:source code}
The R code for implementing the aforementioned estimators is available on Github:

\url{https://github.com/liweihaoPaul/Catalytic-Cox-Regrerssion}.

\newpage

%% file: all-bibliography.bib
@article{rosenthal1970subspaces,
  title={On the subspaces of L p (p> 2) spanned by sequences of independent random variables},
  author={Rosenthal, Haskell P},
  journal={Israel Journal of Mathematics},
  volume={8},
  pages={273--303},
  year={1970},
  publisher={Springer}
}

@book{van2000empirical,
  title={Empirical Processes in M-estimation},
  author={Van de Geer, Sara A},
  volume={6},
  year={2000},
  publisher={Cambridge university press}
}

@book{carlin2000,
  title={Bayes and empirical {B}ayes methods for data analysis},
  author={Bradley P. Carlin and Thomas A. Louis},
  edition={2},
  year={2000},
  publisher={Boca Raton: Chapman and Hall–CRC}
}

@article{rigollet2023high,
  title={High-dimensional statistics},
  author={Rigollet, Philippe and H{\"u}tter, Jan-Christian},
  journal={arXiv preprint arXiv:2310.19244},
  year={2023}
}

@book{ortega1970iterative,
  title={Iterative Solution of Nonlinear Equations in Several Variables},
  author={Ortega, JM and Rheinboldt, WC},
  volume={30},
  year={1970},
  publisher={SIAM}
}

@article{cai2005variable,
  title={Variable selection for multivariate failure time data},
  author={Cai, Jianwen and Fan, Jianqing and Li, Runze and Zhou, Haibo},
  journal={Biometrika},
  volume={92},
  number={2},
  pages={303--316},
  year={2005},
  publisher={Oxford University Press}
}

@article{bradic2011regularization,
  title={Regularization for Cox’s proportional hazards model with NP-dimensionality},
  author={Bradic, Jelena and Fan, Jianqing and Jiang, Jiancheng},
  journal={Annals of statistics},
  volume={39},
  number={6},
  pages={3092},
  year={2011},
  publisher={NIH Public Access}
}

@article{Box:1964ev,
  title={An analysis of transformations},
  author={Box, George EP and Cox, David R},
  journal={Journal of the Royal Statistical Society: Series B (Methodological)},
  volume={26},
  number={2},
  pages={211--243},
  year={1964},
  publisher={Wiley Online Library}
}

@article{beca2021impact,
	author = {Beca, Jaclyn M and Chan, Kelvin KW and Naimark, David MJ and Pechlivanoglou, Petros},
	journal = {BMC Medical Research Methodology},
	number = {1},
	pages = {1--12},
	title = {Impact of limited sample size and follow-up on single event survival extrapolation for health technology assessment: a simulation study},
	volume = {21},
	year = {2021}}

@article{gilks1992adaptive,
	author = {Gilks, Walter R and Wild, Pascal},
	journal = {Journal of the Royal Statistical Society: Series C (Applied Statistics)},
	number = {2},
	pages = {337--348},
	title = {Adaptive rejection sampling for Gibbs sampling},
	volume = {41},
	year = {1992}}

@article{saumard2014log,
	author = {Saumard, Adrien and Wellner, Jon A},
	journal = {Statistics surveys},
	pages = {45},
	title = {Log-concavity and strong log-concavity: a review},
	volume = {8},
	year = {2014}}

@inproceedings{lovasz2006fast,
	author = {Lov{\'a}sz, L{\'a}szl{\'o} and Vempala, Santosh},
	booktitle = {2006 47th Annual IEEE Symposium on Foundations of Computer Science (FOCS'06)},
	organization = {IEEE},
	pages = {57--68},
	title = {Fast algorithms for logconcave functions: Sampling, rounding, integration and optimization},
	year = {2006}}

@article{lovasz2007geometry,
	author = {Lov{\'a}sz, L{\'a}szl{\'o} and Vempala, Santosh},
	journal = {Random Structures \& Algorithms},
	number = {3},
	pages = {307--358},
	title = {The geometry of logconcave functions and sampling algorithms},
	volume = {30},
	year = {2007}}

@manual{survivalpackage,
	author = {Terry M Therneau},
	title = {A Package for Survival Analysis in R},
	year = {2023}}

@article{locke1996time,
	author = {Locke III, G Richard and Therneau, Terry M and Ludwig, Jurgen and Dickson, E Rolland and Lindor, Keith D},
	journal = {Hepatology},
	number = {1},
	pages = {52--56},
	title = {Time course of histological progression in primary biliary cirrhosis},
	volume = {23},
	year = {1996}}

@article{hoeting1999bayesian,
	author = {Hoeting, Jennifer A and Madigan, David and Raftery, Adrian E and Volinsky, Chris T},
	journal = {Statistical science},
	number = {4},
	pages = {382--417},
	title = {Bayesian model averaging: a tutorial (with comments by M. Clyde, David Draper and EI George, and a rejoinder by the authors},
	volume = {14},
	year = {1999}}

@book{therneau_modeling_2000,
	address = {New York, NY},
	author = {Therneau, Terry M. and Grambsch, Patricia M.},
	publisher = {Springer New York},
	series = {Statistics for {Biology} and {Health}},
	title = {Modeling Survival Data: Extending the Cox Model},
	year = {2000}}

@article{murphy1997maximum,
	author = {Murphy, SA and Rossini, AJ and van der Vaart, Aad W},
	journal = {Journal of the American Statistical Association},
	number = {439},
	pages = {968--976},
	title = {Maximum likelihood estimation in the proportional odds model},
	volume = {92},
	year = {1997}}

@article{wei1992accelerated,
	author = {Wei, Lee-Jen},
	journal = {Statistics in medicine},
	number = {14-15},
	pages = {1871--1879},
	title = {The accelerated failure time model: a useful alternative to the Cox regression model in survival analysis},
	volume = {11},
	year = {1992}}

@article{van2006cross,
	author = {Van Houwelingen, Hans C and Bruinsma, Tako and Hart, Augustinus AM and Van't Veer, Laura J and Wessels, Lodewyk FA},
	journal = {Statistics in medicine},
	number = {18},
	pages = {3201--3216},
	title = {Cross-validated Cox regression on microarray gene expression data},
	volume = {25},
	year = {2006}}

@article{bedrick1996new,
	author = {Bedrick, Edward J and Christensen, Ronald and Johnson, Wesley},
	journal = {Journal of the American Statistical Association},
	number = {436},
	pages = {1450--1460},
	title = {A new perspective on priors for generalized linear models},
	volume = {91},
	year = {1996}}

@article{chen2000power,
	author = {Chen, Ming-Hui and Ibrahim, Joseph G and Shao, Qi-Man},
	journal = {Journal of Statistical Planning and Inference},
	number = {1-2},
	pages = {121--137},
	title = {Power prior distributions for generalized linear models},
	volume = {84},
	year = {2000}}

@article{iwaki1997posterior,
	author = {Iwaki, Katsuaki},
	journal = {J. Econ. Asia Univ},
	pages = {105--134},
	title = {Posterior expected marginal likelihood for testing hypotheses},
	volume = {21},
	year = {1997}}

@article{bedrick1997bayesian,
	author = {Bedrick, Edward J and Christensen, Ronald and Johnson, Wesley},
	journal = {The American Statistician},
	number = {3},
	pages = {211--218},
	title = {Bayesian binomial regression: Predicting survival at a trauma center},
	volume = {51},
	year = {1997}}

@article{perez2002expected,
	author = {P{\'e}rez, Jos{\'e} M and Berger, James O},
	journal = {Biometrika},
	number = {3},
	pages = {491--512},
	title = {Expected-posterior prior distributions for model selection},
	volume = {89},
	year = {2002}}

@article{raiffa1961applied,
	author = {Raiffa, Howard and Schlaifer, Robert and others},
	title = {Applied statistical decision theory},
	year = {1961}}

@book{good1983good,
	author = {Good, Irving John},
	publisher = {U of Minnesota Press},
	title = {Good thinking: The foundations of probability and its applications},
	year = {1983}}

@article{neal2001transferring,
	author = {Neal, RM},
	journal = {Department of Statistics, University of Toronto},
	title = {Transferring Prior Information Between Models Using Imaginary Data Technical Report No. 0108},
	year = {2001}}

@article{urridge1981empirical,
	author = {Urridge, JB},
	journal = {Journal of the Royal Statistical Society: Series B (Methodological)},
	number = {1},
	pages = {65--75},
	title = {Empirical Bayes analysis of survival time data},
	volume = {43},
	year = {1981}}

@article{kalbfleisch1978non,
	author = {Kalbfleisch, John D},
	journal = {Journal of the Royal Statistical Society: Series B (Methodological)},
	number = {2},
	pages = {214--221},
	title = {Non-parametric Bayesian analysis of survival time data},
	volume = {40},
	year = {1978}}

@article{wong1986theory,
	author = {Wong, Wing Hung},
	journal = {The Annals of statistics},
	pages = {88--123},
	title = {Theory of partial likelihood},
	year = {1986}}

@article{tsiatis1981large,
	author = {Tsiatis, Anastasios A},
	journal = {The Annals of Statistics},
	number = {1},
	pages = {93--108},
	title = {A large sample study of {Cox}'s regression model},
	volume = {9},
	year = {1981}}

@article{andersen1982cox,
	author = {Andersen, Per Kragh and Gill, Richard D},
	journal = {The annals of statistics},
	pages = {1100--1120},
	title = {Cox's regression model for counting processes: a large sample study},
	year = {1982}}

@book{fleming2005counting,
	author = {Fleming, Thomas R and Harrington, David P},
	publisher = {John Wiley \& Sons},
	title = {Counting processes and survival analysis},
	year = {2005}}

@article{bovelstad2007predicting,
	author = {B{\o}velstad, Hege M and Nyg{\aa}rd, St{\aa}le and St{\o}rvold, Hege L and Aldrin, Magne and Borgan, {\O}rnulf and Frigessi, Arnoldo and Lingj{\ae}rde, Ole Christian},
	journal = {Bioinformatics},
	number = {16},
	pages = {2080--2087},
	title = {Predicting survival from microarray data{--}a comparative study},
	volume = {23},
	year = {2007}}

@article{zhang2022modern,
	author = {Zhang, Xianyang and Zhou, Huijuan and Ye, Hanxuan},
	journal = {arXiv preprint arXiv:2204.01161},
	title = {A Modern Theory for High-dimensional {Cox} Regression Models},
	year = {2022}}

@article{wu2018assessing,
	author = {Wu, Jing and de Castro, M{\'a}rio and Schifano, Elizabeth D and Chen, Ming-Hui},
	journal = {Journal of statistical theory and practice},
	pages = {23--41},
	title = {Assessing covariate effects using Jeffreys-type prior in the Cox model in the presence of a monotone partial likelihood},
	volume = {12},
	year = {2018}}

@article{Bryson1981,
	author = {Maurice C. Bryson and Mark E. Johnson},
	journal = {Technometrics},
	number = {4},
	pages = {381-383},
	title = {The Incidence of Monotone Likelihood in the Cox Model},
	volume = {23},
	year = {1981}}

@article{neuenschwander2010summarizing,
	author = {Neuenschwander, Beat and Capkun-Niggli, Gorana and Branson, Michael and Spiegelhalter, David J},
	journal = {Clinical Trials},
	number = {1},
	pages = {5--18},
	title = {Summarizing historical information on controls in clinical trials},
	volume = {7},
	year = {2010}}

@article{van2018including,
	author = {van Rosmalen, Joost and Dejardin, David and van Norden, Yvette and L{\"o}wenberg, Bob and Lesaffre, Emmanuel},
	journal = {Statistical methods in medical research},
	number = {10},
	pages = {3167--3182},
	title = {Including historical data in the analysis of clinical trials: Is it worth the effort?},
	volume = {27},
	year = {2018}}

@article{brard2017bayesian,
	author = {Brard, Caroline and Le Teuff, Gw{\'e}na{\"e}l and Le Deley, Marie-C{\'e}cile and Hampson, Lisa V},
	journal = {Clinical Trials},
	number = {1},
	pages = {78--87},
	title = {Bayesian survival analysis in clinical trials: What methods are used in practice?},
	volume = {14},
	year = {2017}}

@article{murphy2000profile,
	author = {Murphy, Susan A and Van der Vaart, Aad W},
	journal = {Journal of the American Statistical Association},
	number = {450},
	pages = {449--465},
	title = {On profile likelihood},
	volume = {95},
	year = {2000}}

@article{verweij1993cross,
	author = {Verweij, Pierre JM and Van Houwelingen, Hans C},
	journal = {Statistics in medicine},
	number = {24},
	pages = {2305--2314},
	title = {Cross-validation in survival analysis},
	volume = {12},
	year = {1993}}

@article{ibrahim1998prior,
	author = {Ibrahim, Joseph G and Chen, Ming-Hui},
	journal = {Sankhy{\=a}: The Indian Journal of Statistics, Series B},
	pages = {48--64},
	title = {Prior distributions and Bayesian computation for proportional hazards models},
	year = {1998}}

@article{cox1975partial,
	author = {Cox, David R},
	journal = {Biometrika},
	number = {2},
	pages = {269--276},
	title = {Partial likelihood},
	volume = {62},
	year = {1975}}

@article{simon_regularization_2011,
	author = {Simon, Noah and Friedman, Jerome and Hastie, Trevor and Tibshirani, Rob},
	journal = {Journal of Statistical Software},
	number = {5},
	pages = {1--13},
	title = {Regularization {Paths} for {Cox}'s {Proportional} {Hazards} {Model} via {Coordinate} {Descent}},
	volume = {39},
	year = {2011}}

@article{aalen1989linear,
	author = {Aalen, Odd O},
	journal = {Statistics in medicine},
	number = {8},
	pages = {907--925},
	title = {A linear regression model for the analysis of life times},
	volume = {8},
	year = {1989}}

@article{hanson2013surviving,
	author = {Hanson, TIMOTHY E and Jara, ALEJANDRO},
	journal = {Bayesian theory and applications},
	pages = {593--615},
	title = {Surviving fully {Bayesian} nonparametric regression models},
	year = {2013}}

@article{huang_penalized_2002,
	author = {Huang, Jie and Harrington, David},
	journal = {Biometrics},
	number = {4},
	pages = {781--791},
	title = {Penalized partial likelihood regression for right-censored data with bootstrap selection of the penalty parameter},
	volume = {58},
	year = {2002}}

@article{sinha_bayesian_2003,
	author = {Sinha, D.},
	journal = {Biometrika},
	month = sep,
	number = {3},
	pages = {629--641},
	title = {A {Bayesian} justification of {Cox}'s partial likelihood},
	volume = {90},
	year = {2003}}

@book{chen2012monte,
	author = {Chen, Ming-Hui and Shao, Qi-Man and Ibrahim, Joseph G},
	publisher = {Springer Science \& Business Media},
	series = {Springer {Series} in {Statistics}},
	title = {{Monte Carlo methods in Bayesian computation}},
	year = {2012}}

@book{ibrahim_bayesian_2001,
	address = {New York, NY},
	author = {Ibrahim, Joseph G. and Chen, Ming-Hui and Sinha, Debajyoti},
	publisher = {Springer New York},
	series = {Springer {Series} in {Statistics}},
	title = {Bayesian {Survival} {Analysis}},
	year = {2001}}

@article{lee2011bayesian,
	author = {Lee, Kyu Ha and Chakraborty, Sounak and Sun, Jianguo},
	journal = {The International Journal of Biostatistics},
	number = {1},
	pages = {0000102202155746791301},
	title = {Bayesian variable selection in semiparametric proportional hazards model for high dimensional survival data},
	volume = {7},
	year = {2011}}

@article{cox1972regression,
	author = {Cox, David R},
	journal = {Journal of the Royal Statistical Society: Series B (Methodological)},
	number = {2},
	pages = {187--202},
	title = {Regression models and life-tables},
	volume = {34},
	year = {1972}}

@article{huang2022catalytic,
	author = {Huang, Dongming and Wang, Feicheng and Rubin, Donald B and Kou, SC},
	journal = { arXiv:2208.14123},
	title = {Catalytic Priors: Using Synthetic Data to Specify Prior Distributions in Bayesian Analysis},
	year = {2022}}

@article{huang_catalytic_2020,
	author = {Huang, Dongming and Stein, Nathan and Rubin, Donald B. and Kou, S. C.},
	journal = {Proc. Natl. Acad. Sci. U.S.A.},
	month = jun,
	number = {22},
	pages = {12004--12010},
	title = {Catalytic prior distributions with application to generalized linear models},
	volume = {117},
	year = {2020}}

@article{tibshirani1997lasso,
	author = {Tibshirani, Robert},
	journal = {Statistics in medicine},
	number = {4},
	pages = {385--395},
	title = {The lasso method for variable selection in the Cox model},
	volume = {16},
	year = {1997}}

@article{ibrahim2000power,
	author = {Ibrahim, Joseph G and Chen, Ming-Hui},
	journal = {Statistical Science},
	pages = {46--60},
	title = {Power prior distributions for regression models},
	year = {2000}}

@article{hoerl1970ridge,
	author = {Hoerl, Arthur E and Kennard, Robert W},
	journal = {Technometrics},
	number = {1},
	pages = {55--67},
	title = {Ridge regression: Biased estimation for nonorthogonal problems},
	volume = {12},
	year = {1970}}
